\documentclass[a4paper,fleqn,usenatbib]{mnras}

\usepackage{newtxtext,newtxmath}
\usepackage{multirow}

\usepackage[T1]{fontenc}
\usepackage{ae,aecompl}

\usepackage{graphicx}	
\usepackage{amsmath}	
\usepackage{amssymb}	
\usepackage[usenames]{color}
\usepackage[normalem]{ulem}

\newcommand{\hii}    {\ion{H}{II}}
\newcommand{\mkm}    {$\mu$m}
\newcommand{\tgas}   {$T_{\rm gas}$}

\newcommand{\kms}    {km s$^{-1}$}

\newcommand{\co}     {$^{13}$CO}
\newcommand{\cvo}    {C$^{18}$O}
\newcommand{\offsets}{($\Delta \alpha, \Delta \delta$)}

\newcommand{\vexp}{$V_{\rm exp}$}

\newcommand{\nco}    {$N_{\rm ^{13}CO}$}

\title[Molecular envelope around RCW~120]{Molecular envelope around the HII region RCW~120}

\author[M. S. Kirsanova et al.]{
M. S. Kirsanova,$^{1}$\thanks{E-mail: kirsanova@inasan.ru}
Ya. N. Pavlyuchenkov$^{1}$,
D. S. Wiebe$^{1}$,
P. A. Boley$^{2,3}$,\newauthor
S. V. Salii$^{3}$,
S. V. Kalenskii$^{4}$,
A. M. Sobolev$^{3}$,
L.~D.~Anderson$^{5,6,7}$
\\
$^{1}$Institute of Astronomy, Russian Academy of Sciences, 119017, 48 Pyatnitskaya Str., Moscow, Russia\\
$^{2}$Moscow Institute of Physics and Technology, 141701, 9 Institutskiy per., Dolgoprudny, Moscow Region, Russia\\
$^{3}$ Ural Federal University, 620075, 19 Mira Str., Ekaterinburg, Russia\\
$^{4}$Astro Space Center, Lebedev Physical Institute, Russian Academy of Sciences, 117997, 84/32 Profsoyuznaya Str., Moscow, Russia\\
$^{5}$Department of Physics and Astronomy, West Virginia
University, Morgantown WV 26506, USA\\
$^{6}$Adjunct Astronomer at the Green Bank Observatory, P.O.
Box 2, Green Bank WV 24944, USA\\
$^{7}$Center for Gravitational Waves and Cosmology, West
Virginia University, Chestnut Ridge Research Building, Morgantown, WV
26505, USA
}
\date{Accepted \today. Received \today; in original form \today}

\pubyear{2015}

\begin{document}
\label{firstpage}
\pagerange{\pageref{firstpage}--\pageref{lastpage}}
\maketitle

\begin{abstract}
The \hii{} region RCW~120 is a well-known object, which is often considered as a target to verify theoretical models of gas and dust dynamics in the interstellar medium. However, the exact geometry of RCW~120 is still a matter of debate. In this work, we analyse observational data on molecular emission in RCW~120 and show that \co(2--1) and \cvo(2--1) lines are fitted by a 2D model representing a ring-like face-on structure. The changing of the \cvo(3--2) line profile from double-peaked to single-peaked from the dense molecular Condensation~1 might be a signature of stalled expansion in this direction. In order to explain a self-absorption dip of the \co(2--1) and \co(3--2) lines, we suggest that RCW~120 is surrounded by a diffuse molecular cloud, and find confirmation of this cloud on a map of interstellar extinction. Optically thick \co(2--1) emission and the infrared 8~\mkm{} PAH band form a neutral envelope of the \hii{} region resembling a ring, while the envelope breaks into separate clumps on images made with optically thin \cvo(2--1) line and far-infrared dust emission.
\end{abstract}



%
%
%
%

\section{Introduction}
\label{intro}

{\it Spitzer} and {\it WISE} maps of the Milky Way Galaxy have revealed more than five thousand ring-like structures, which are commonly referred to as infrared (IR) bubbles, around \hii{} regions excited by young massive stars, e.g. \citet{Churchwell_2006,Churchwell_2007,2010A&A...523A...6D,Anderson_2011,Anderson_2014,Simpson_2012}, and also recent studies by \citet{Topchieva_2017, Topchieva_2018}. {\it Spitzer} 8~$\mu$m and {\it WISE} 11~$\mu$m emission is commonly attributed to polycyclic aromatic hydrocarbon (PAH) molecules, which are excited by the absorption of ultraviolet photons from massive stars \citep[see, for example, classical studies by][]{Sellgren_1984, Leger_1984} in the photo-dissociation regions (PDRs) on the border between the \hii{} region and surrounding neutral material.

While the word `bubble' implies a 3D structure, the {\it Spitzer} bubbles look like rings in the plane of the sky, so their actual geometry might be different. This is an important issue, as it may shed light on the detailed structure of matter distribution in star-forming regions, on the interaction of massive stars with their parent molecular clouds, on the dust evolution in the vicinity of massive stars, and on the possibility and extent of triggered star formation. But the problem of the actual geometry of IR ring nebulae is still far from being solved.

An important study of this issue was performed by \citet{Beaumont_2010}. They utilised observations of CO(3--2) and HCO$^+$(4--3) molecular lines to study gas morphology in the vicinity of 43 {\it Spitzer} bubbles, and found that the distribution of neutral material around these bubbles lacks convincing signatures of near and far molecular walls, which would be expected if the bubbles were 3D shells. In addition, in molecular emission, the inner regions of many bubbles in their sample are fainter than in the surrounding material. Finally, the kinematic structure of the bubbles does not show any signs of expansion. \citet{Beaumont_2010} therefore concluded that the bubbles and \hii{} regions in their study are enclosed in molecular rings, rather than spherical shells, indicating that the parent molecular clouds are flattened. While a flattened shape of a molecular cloud is evident, when an \hii{} region is observed edge-on as a bipolar structure \citep[e.g.][]{2018A&A...617A..67S}, a ring-shaped region requires a more thorough analysis. Despite this fact, ring-like nebulae are often implicitly assumed to have a 3D geometry \citep[e.g.,][]{2010ApJ...713..592E,dustwave}.

In this paper we consider one of the most well-known examples of an infrared bubble, RCW~120. The \hii{} region RCW~120 is one of the closest to the Sun, at a distance of 1.34\,kpc~\citet{Russeil_03}. The parallax of the ionizing star of RCW~120, CD-38~11636, measured by Gaia~DR2 \citep{2018yCat.1345....0G}, corresponds to a distance of $0.5-1.1$\,kpc. However, this estimate has large uncertainties, as the astrometric solution for the star is poor. Therefore, in this work, we adopt 1.3\,kpc as a distance to RCW~120 to keep consistency with our previous studies~\citep{Pavlyuchenkov_2013,Akimkin_2015}. An effective temperature of CD-38~11636 is 37000~K, as found by~\cite{Martins_10}.

RCW~120 has a nearly perfect ring-like shape in {\it Spitzer} and {\it Herschel} infrared images~\citep{Deharveng_09, anderson_12}. The infrared ring is claimed to be a projection of neutral material collected during the period of expansion of the \hii{} region. \citet{Tremblin_14} analysed the column density maps from {\it Herschel} observations of the neutral environment of RCW~120, and found that it consists of a compression zone induced by the expansion of the ionized gas and surrounding turbulent medium. This apparent simplicity has made RCW~120 a template object to study various aspects of the interaction of massive stars with their parent molecular clouds, including triggered star formation.

Signatures of triggering in RCW~120 are seen in the fine region structure: the shell of neutral material breaks down into up to eight or nine dense condensations, seen in continuum observations at 1.2~mm and 870~$\rm \mu$m by \citep{Zavagno_2007} and \citep{Deharveng_09}, respectively, who argue that the condensations might have been formed as a result of fragmentation of the neutral material due to the gravitational instability during the expansion of the \hii{} region. There are numerous young stellar objects embedded in the condensations, including a massive compact source in Condensation~1, adjacent to the ionization front of RCW~120 \citep[see also][]{Zavagno_2010}. \citet{Figueira_2017} recently re-examined the {\it Herschel} data and found that 80\% of the massive compact sources in RCW~120 belong to Condensation~1. ALMA observations by \citet{Figueira_2018} reveal the importance of turbulence or magnetic fields to explain the low observed fragmentation, which is inconsistent with a pure thermal Jeans model of gravitational instability.

The geometry of RCW~120 remains a matter of debate. It has been considered as a bubble viewed in projection in series of works by \citet{Deharveng_2005, Zavagno_2007, Deharveng_09, Zavagno_2010}. However, \citet{Anderson_2010} found a lack of emission in the interior of RCW~120 at wavelengths $ \leq 250$\,$\rm \mu$m, which suggests a two dimensional ring, rather than a three dimensional bubble. 

\citet{Pavlyuchenkov_2013} tried to fit the {\it Herschel} observations using a 1D spherical model of an expanding \hii{} region, and showed that the 1D model predicts appreciable emission at 100\,$\rm \mu$m from the central part of the object and the ionizing star (the front and back molecular walls), while this emission is essentially absent in the observations. They concluded that dust envelope of RCW~120 may not be spherical, but instead has cylindrical symmetry viewed along the axis of the cylinder. In particular, the 1D model of an \hii{} region expanding into a neutral molecular cloud predicts that the maximum velocity difference should be observed between the near and far sides, centred at the ionizing star.

\citet{Zavagno_2007} analysed the \ion{H}{I} emission from the Southern Galactic Plane Survey \citep[SGPS,][]{McClure-Griffiths_2005} and found its presence in a broad area around RCW~120. They looked for the near and far expanding sides of the neutral environment of the \hii{} region and did not find it. \citet{Anderson_2015} studied the expansion of the neutral material around the \hii{} region with the $J=1-0$ transition of $^{12}$CO, $^{13}$CO and C$^{18}$O molecules using the 22-m Mopra telescope. They made no claims about the geometry, because they found no velocity difference between the suggested near and far sides. They concluded that if the \hii{} region is expanding, then the expansion velocity (\vexp) should be $\le1$\kms. We note that in the series of simulations by \citet{Pavlyuchenkov_2013,Akimkin_2015,Akimkin_2017}, where the parameters of RCW~120 were used as a basic model, the value of \vexp{} was indeed found to be about 1--1.5\,\kms. Recently, \citet{Marsh_2019} re-examined {\it Herschel} data using a Bayesian procedure and surprisingly found that spatial distribution of warm dust in the \hii{} region is consistent with the projection of a spherical shell, while colder peripheral dust has a ragged clumpy structure.

The aim of this paper is to use available data about molecular emission around RCW~120 and a 1D spherically symmetric simulation of the \hii{} region, developed by \citet{Pavlyuchenkov_2013,Akimkin_2015,Akimkin_2017}, to find a more reliable model for the RCW~120 geometry.

\section{Observations, archival data, and numerical modelling}\label{Sec:Obs}

\subsection{Molecular emission}

We observed CO molecular lines in RCW~120 at several locations in Condensation~1 around the peak of 870~$\rm \mu$m emission found by \citet{Deharveng_09} ($\alpha_{2000}=17^{\rm h} 12^{\rm m} 08^{\rm s}$ $\delta_{2000}=-38^\circ$30\arcmin\,45\arcsec{}) with the APEX telescope during 5\,hours from 7th to 9th July of 2009 in service mode (project number is O-083.F-9311A-2009) using SHeFI recievers \citep[][]{2006SPIE.6275E..0GB, 2008A&A...490.1157V}. We used position switching mode with OFF position at \offsets=(1800\arcsec,0\arcsec) relative to the reference position. Observations were done in good weather conditions, with the amount of precipitable water vapour in between 0.4 and 0.7\,mm. Our choice of APEX instrumentation and line selection is shown in Table~\ref{tab:obs}. The Fast Fourier Transform Spectrometer with 8192 spectral channels provided a spectral resolution about 0.17\,\kms. Reduction and calibration of the observations were performed by the APEX staff.

\begin{table}
\centering
\caption{Basic observational parameters.}
\begin{tabular}{ccccc}
\hline
Lines                & Reciever & Freq. range  & Beam      & rms\\
                     &          & (GHz)        & (\arcsec) & (K)\\
\hline
\co\,(2--1)         & APEX-1   & 220.0--221.0  & 29        & 0.1\\
\cvo\,(3--2)        & APEX-2   & 328.9--329.9  & 19        & 0.2\\
\co\,(3--2)         & APEX-2   & 330.1--331.1  & 19        & 0.2\\
\hline
\end{tabular}
\label{tab:obs}
\end{table}

We also use APEX archival data at a frequency of 220~GHz to reveal the large-scale distribution of molecular gas around RCW~120. The data were obtained as a part of the SEDIGISM project \citep[E-193.C-0584A-2014, PI: F. Schuller][]{Schuller_2017}. The region around RCW~120 (source G348.25+0.38) was observed twice (in 2014 and 2015). We use calibrated data from the observational session in 2015 and in the present study concentrate mainly on \co(2--1) at 220.3987~GHz and \cvo(2--1) at 219.5604~GHz. We smoothed and regridded the original archival data at 220~GHz to obtain 1-$\sigma$ noise level of 1~K on $T_{\rm mb}$ scale at 0.2~\kms, i.e. an order of magnitude higher than in our data from 2009 (see Table~\ref{tab:obs}). Since only a portion of the sky with galactic latitude $b \leq 0.5^\circ$ was covered by \citet{Schuller_2017}, there are no data for the north-west part of RCW~120.

Analysis of the data cubes was performed using GILDAS\footnote{\url{http://www.iram.fr/IRAMFR/GILDAS}}.

\subsection{Optical depth and column density}\label{lte:Eq}

We calculate the $^{13}$CO column density (\nco) using a standard LTE analysis described by \citet{Mangum_2015} (Eq.~80) and an optical depth correction factor introduced by   \citet{Goldsmith_1999}. This procedure implicitly assumes the same, constant excitation temperature for \co{} and \cvo{} along the line of sight. The $^{13}$CO-to-C$^{18}$O abundance ratio is assumed to be 7.5, based on the isotopic ratios of $^{13}$C/$^{12}$C and $^{16}$O/$^{18}$O from \citet{Wilson_1999}. To convert \nco{} to molecular hydrogen column density we use a relative $^{13}$C abundance of $2\cdot10^{-6}$.

\subsection{Dust extinction}

We use extinction maps based on the 2MASS near-infrared photometry from \citet{Juvela_2016} to study the large-scale dust distribution around RCW~120. We downloaded $A_{\rm J}$ extinction maps with a spatial resolution of 3.0\arcsec{}, produced by the NICEST method. Parameters for $R_{\rm V}=3.1$ and 5.5 given in \citet{Cardelli_1989} were used to calculate extinction in $A_{\rm V}$ from the $A_{\rm J}$ data.

\subsection{Numerical model}

Observations provide us with information on the large-scale distributions of the \co(2--1) and \cvo(2--1) emission, and this allows us to probe the geometry and kinematics of the molecular envelope around RCW~120. For this purpose, we adopt the model of an expanding \hii{} region MARION, which was developed by \citet{Kirsanova_2009} and further modified by \citet{Pavlyuchenkov_2013, Akimkin_2015, Akimkin_2017}. While many other models for RCW~120 have been proposed, invoking various physical effects such as stellar wind \citep[][]{Sanchez-Cruces_2018}, a moving star \citep{Mackey_2016}, or a radiation-driven dust wave \citep{dustwave},  we consider the `classical' picture of an expanding \hii{} region, relying on our experience in reproducing the observed intensities of dust emission from 8 to 500~$\rm \mu$m with a detailed radiative transfer model (see \citet{Pavlyuchenkov_2013} and Pavlyuchenkov~et~al., in prep.).

We performed new simulations of the expanding \hii{} region with the MARION code, using the same spherically symmetric model as \citet{Akimkin_2017}, but with modified initial parameters. We updated the stellar radiation field using the calibrations of \citet{Martins_2005} with the following star parameters: $T_{\rm eff} = 37000$~K and $\log Q_0=48.8$~s$^{-1}$. The initial hydrogen number density was raised from $n_{\rm gas}=3\cdot 10^3$~cm$^{-1}$ to $10^4$~cm$^{-1}$, since the latter value allows us to get a better fit to the observed intensities of dust emission at 100--500~$\rm \mu$m (Pavlyuchenkov et al., in prep.). An age of 590~kyr is adopted as a the age of the RCW~120 region, as this time is required for the dense molecular layer to reach the 1.2~pc radius, which is the minimum projected separation between the ionizing star and the molecular envelope, given the distance to RCW~120.

We use the two-dimensional non-LTE code URAN(IA) \citep{Pavlyuchenkov:2004,Pavlyuchenkov:2008} to calculate the emergent line profiles. The code is designed to solve the system of radiative transfer and balance equations with the accelerated-iterations method, where the mean intensity is calculated using a Monte-Carlo approach. One of the important parameters which affects the width and shape of the emission profiles is the non-thermal velocity $V_{\rm nth}$, which is associated with an uncertain turbulent field. To study the effect of $V_{\rm nth}$, we assumed a uniform non-thermal velocity distribution (micro-turbulent velocity) over the entire region and tested two values, namely, 0.3~\kms{} and 1.0~\kms.

\section{Observational perspective on neutral material around RCW~120}\label{Sec:ObsRes}

\subsection{\co{} and \cvo{} emission around the HII region}\label{subsec:largescale}

\begin{figure}
\includegraphics[width=\columnwidth]{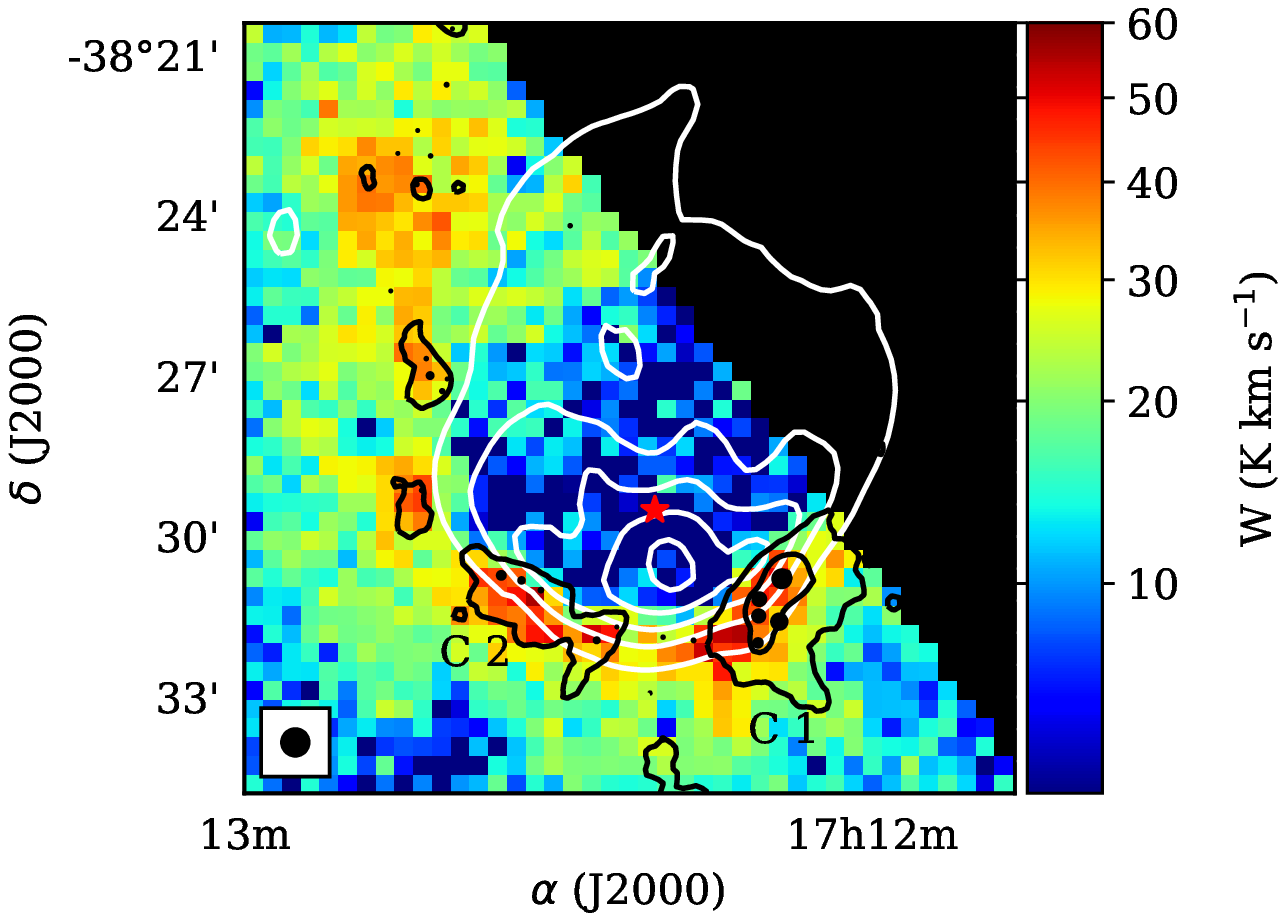}\\
\caption{\co(2--1) line intensity integrated over the $-20$ to 0~\kms{} velocity range. The original \co(2--1) data was smoothed and regridded to reach noise level of 1~K. Only pixels with the intensity higher than 6~K\,\kms{} are shown. The pixel size is 20\arcsec$\times$20\arcsec. Black circles show the locations of compact sources described by \citet{Figueira_2017} (their Table~5). The sizes of the circles depend linearly on the source masses ($M_{\rm env}$ from \citet{Figueira_2017}). The position of the ionizing star of RCW~120 is marked as a red star. The 843~MHz radio continuum emission is shown with white contours linearly spaced from from 0.02 to 0.4 Jy~beam$^{-1}$. ATLASGAL 870\,$\rm \mu$m contours for 0.4, 2.0 and 10.0 Jy~beam$^{-1}$ are shown in black. Condensations~1 and~2 are designated as C~1 and C~2, respectively.}
\label{fig:archive}
\end{figure}

Maps of the \co(2--1) line emission around RCW~120 are shown in Fig.~\ref{fig:archive}. A ring-like molecular envelope surrounds the \hii{} region from the west, south and east. The envelope contains several condensations initially distinguished by \citet{Zavagno_2007}. The two most prominent condensations, Condensation 1 and Condensation 2, are designated in Fig.~\ref{fig:archive}. Emission at 870~\mkm{} from the ATLASGAL archive \citep{Schuller_2009} is over-plotted on the \co(2--1) map to show the condensations more clearly. The integrated intensity of the \co(2--1) emission is 10 times higher at the position of the envelope, compared to the position of the ionizing star. \co(2--1) emission is observed within the \hii{} region itself only in the north-eastern part of RCW~120. The spatial distribution of the \cvo(2--1) emission is similar to that of \co(2--1); however, while the integrated line intensities are almost the same in both condensations, in Condensation~2 the peak of the \co(2--1) emission is 1.5 times higher and the lines are narrower.

The 843~MHz radio continuum map from the Sydney University Molonglo Sky Survey \citep[SUMSS,][]{Bock_1999} is also shown in Fig.~\ref{fig:archive}. The signal of the radio continuum map at about a 10-$\sigma$ level has an ellipse-like appearance, filling the \co(2--1) shell. The radio continuum emission is concentrated to the south of the ionizing star. The peak of the 1.4~GHz radio continuum emission from the NRAO VLA Sky Survey \citep[NVSS,][]{Condon_1998} is also shifted to the south of the star (not shown in Fig.~\ref{fig:archive}).

The compact sources identified by \citet{Figueira_2017} as reliable are shown in Fig.~\ref{fig:archive} as black circles. The region with the highest \co(2--1) integrated intensity in Condensation~1 corresponds to the concentration of the sources, but there is also a region in Condensation~1 with bright \co(2--1) emission and no sources with reliable SEDs. Only sources with unconstrained SEDs are observed in this part of the shell \citep[see Fig. 7 in][]{Figueira_2017}.

\subsection{Gas kinematics}

\begin{figure}
\includegraphics[width=0.98\columnwidth]{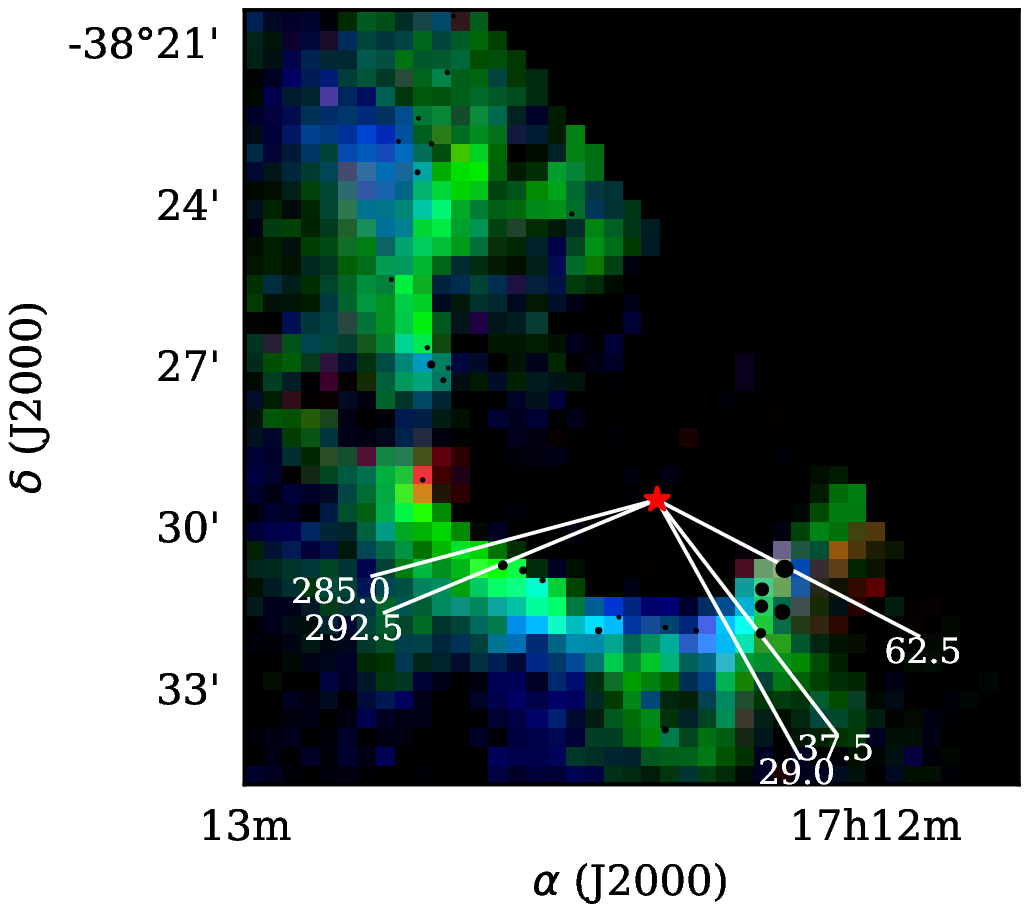}\\
\caption{Three-channel map of \co(2--1) integrated intensity: $V_{\rm lsr} > -6$\,\kms{} (red), $-8 \leq V_{\rm lsr} \leq -6$\,\kms{} (green), $V_{\rm lsr} < -8$\,\kms{} (blue). The original \co(2--1) and \cvo(2--1) data was smoothed and regridded to reach noise level of 1~K. The colorscale corresponds to the intensity range from 6 to 25~K\,\kms. The colour scale is linear in contrast with Fig.~\ref{fig:archive}. The pixel size is 20\arcsec$\times$20\arcsec. Positions of the compact sources from \citet{Figueira_2017} are shown as black circles, where the sizes of the circles depend linearly on the source masses. The position of the ionizing star of RCW~120 is marked as a red star. The cuts used for the PV diagrams in Fig.~\ref{fig:pvobscond1} are shown as white lines.}
\label{fig:channel}
\end{figure}

According to \citet{Anderson_2015}, the difference between the velocity peaks corresponding to the near and far sides of the expanding material does not exceed 2~\kms. This implies that the expansion velocity of RCW~120 cannot be higher than 1~\kms. To isolate a dense molecular shell formed due to the \hii{} region expansion, we separated all the velocity channels into several intervals, looking for an interval with the most uniform spatial distribution of line emission around the \hii{} region. Channel maps of the \co(2--1) emission are shown in Fig.~\ref{fig:channel} for three different intervals. We found that the most uniform ring-like distribution is demonstrated by the integrated emission between $-8 \leq V_{\rm lsr} < -6$~\kms, especially to the south-east and north-east of RCW~120. Blue-shifted ($V_{\rm lsr} < -8$\,\kms) and red-shifted ($V_{\rm lsr} > -6$\,\kms) \co(2--1) emission shows up only locally in the shell. The velocities and \co(2--1) line profiles are in general more regular in Condensation~2 than in Condensation~1. There are two bright areas on the red-shifted channel map in Fig.~\ref{fig:channel} around $\phi=60^\circ$ and $\phi=270^\circ$, respectively, where the azimuthal angle $\phi$ is measured from south to west, with the ionizing star at the origin. Emission of \co(2--1) in the blue-shifted velocity range is observed to the south of the molecular envelope. High-mass compact sources with constrained SEDs from \citet{Figueira_2017} in Condensations~1 and~2 are absent from the region of the blue-shifted \co(2--1).

\begin{figure}
\includegraphics[width=0.49\columnwidth]{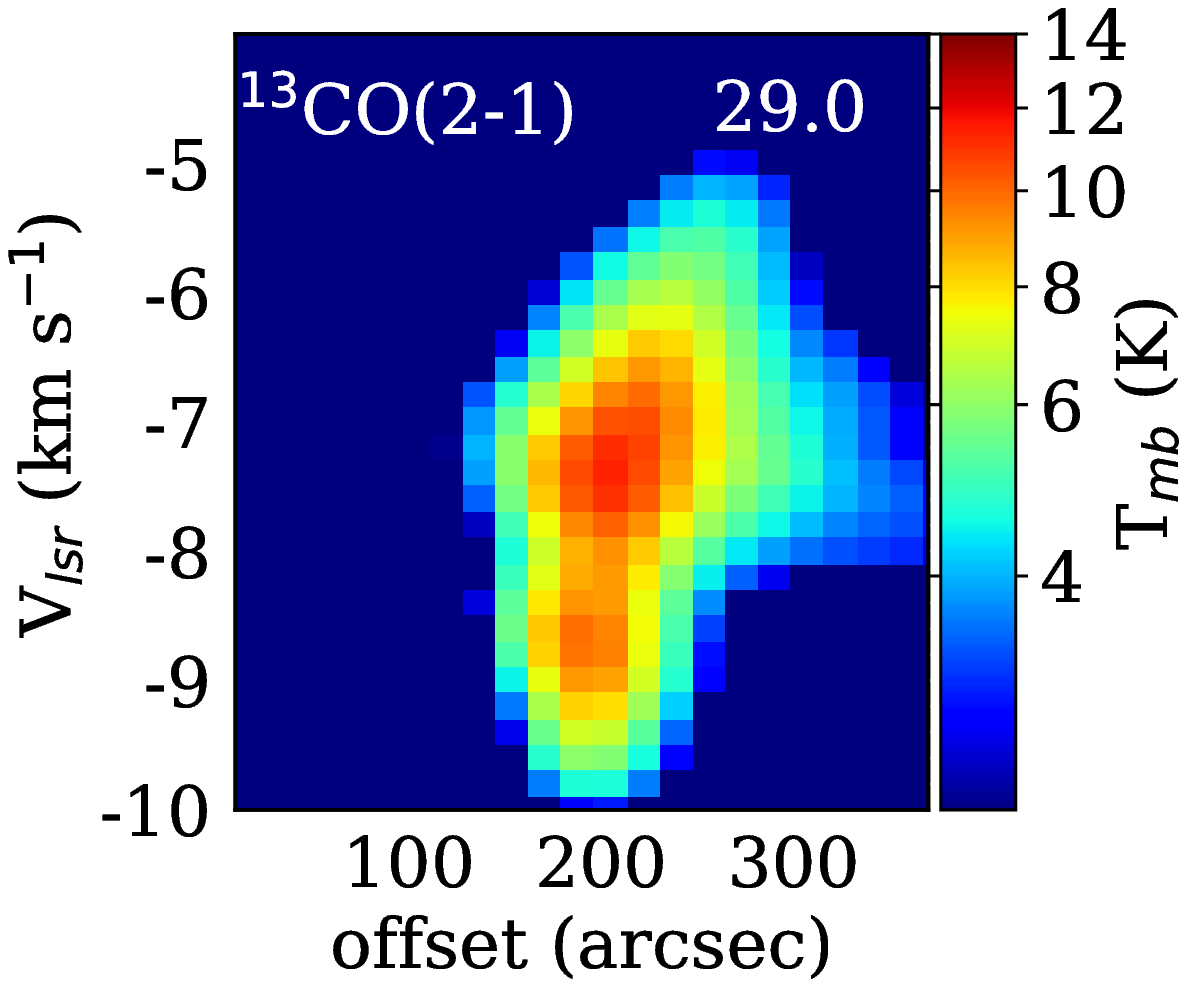}
\includegraphics[width=0.49\columnwidth]{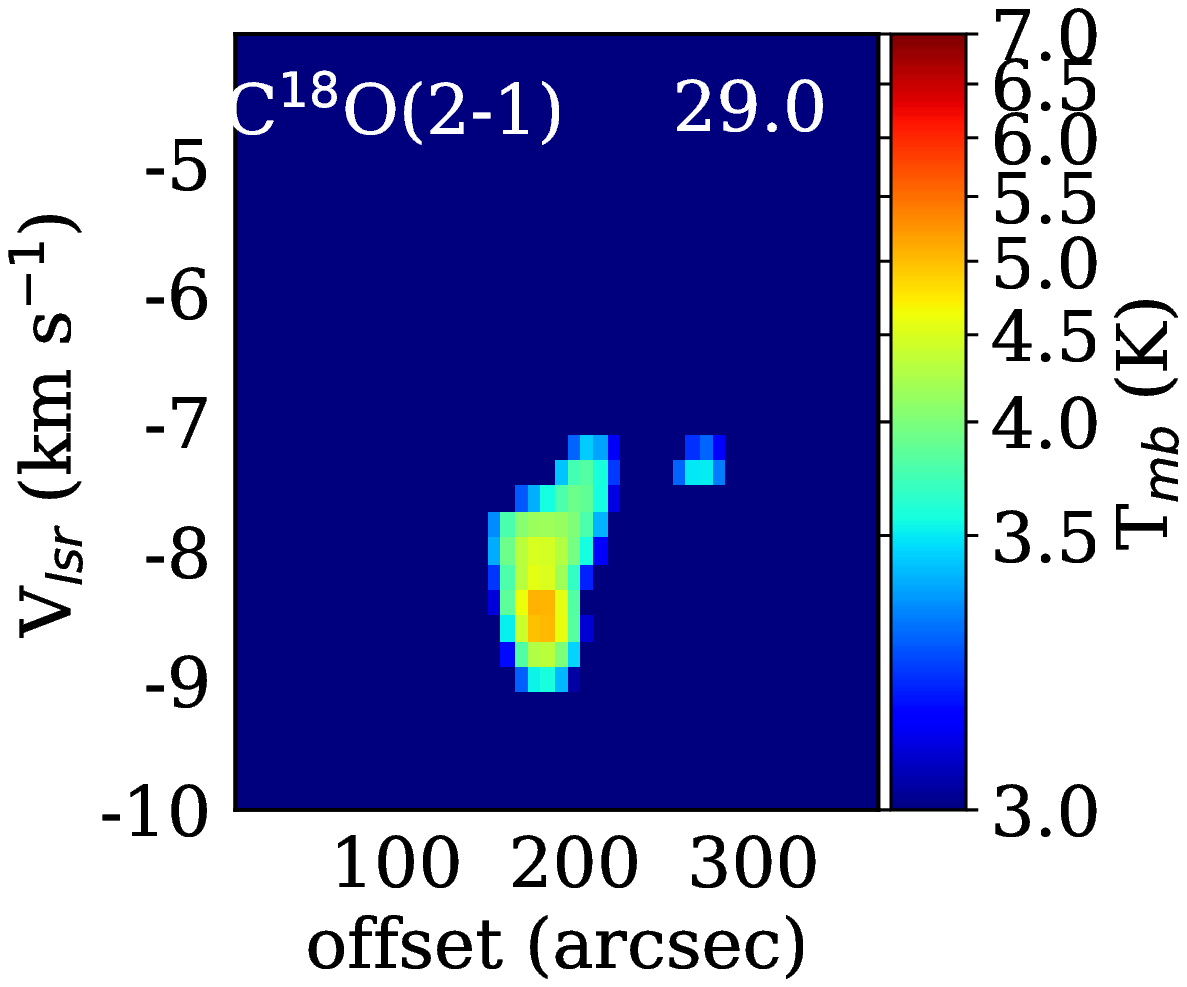}\\
\includegraphics[width=0.49\columnwidth]{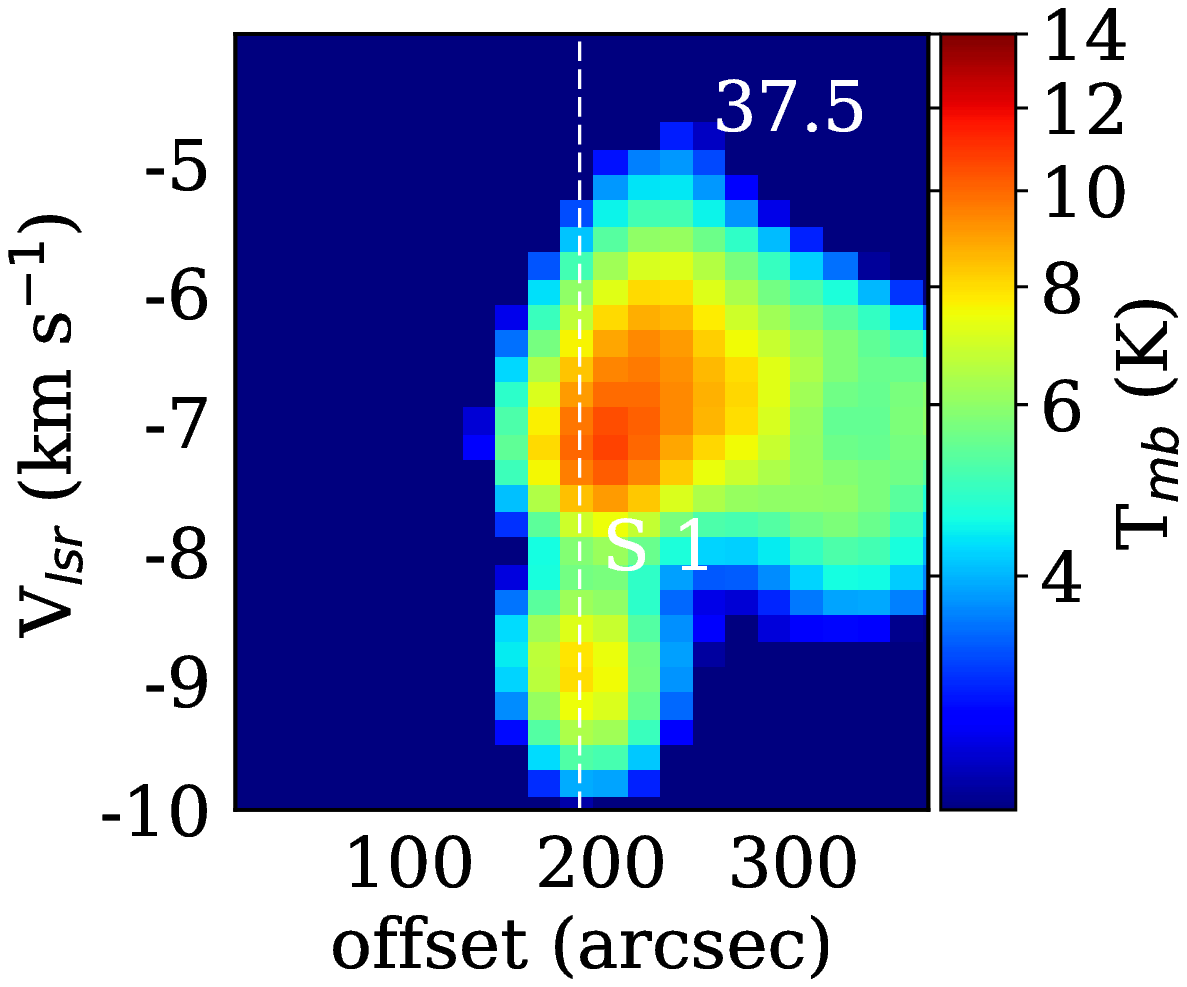}
\includegraphics[width=0.49\columnwidth]{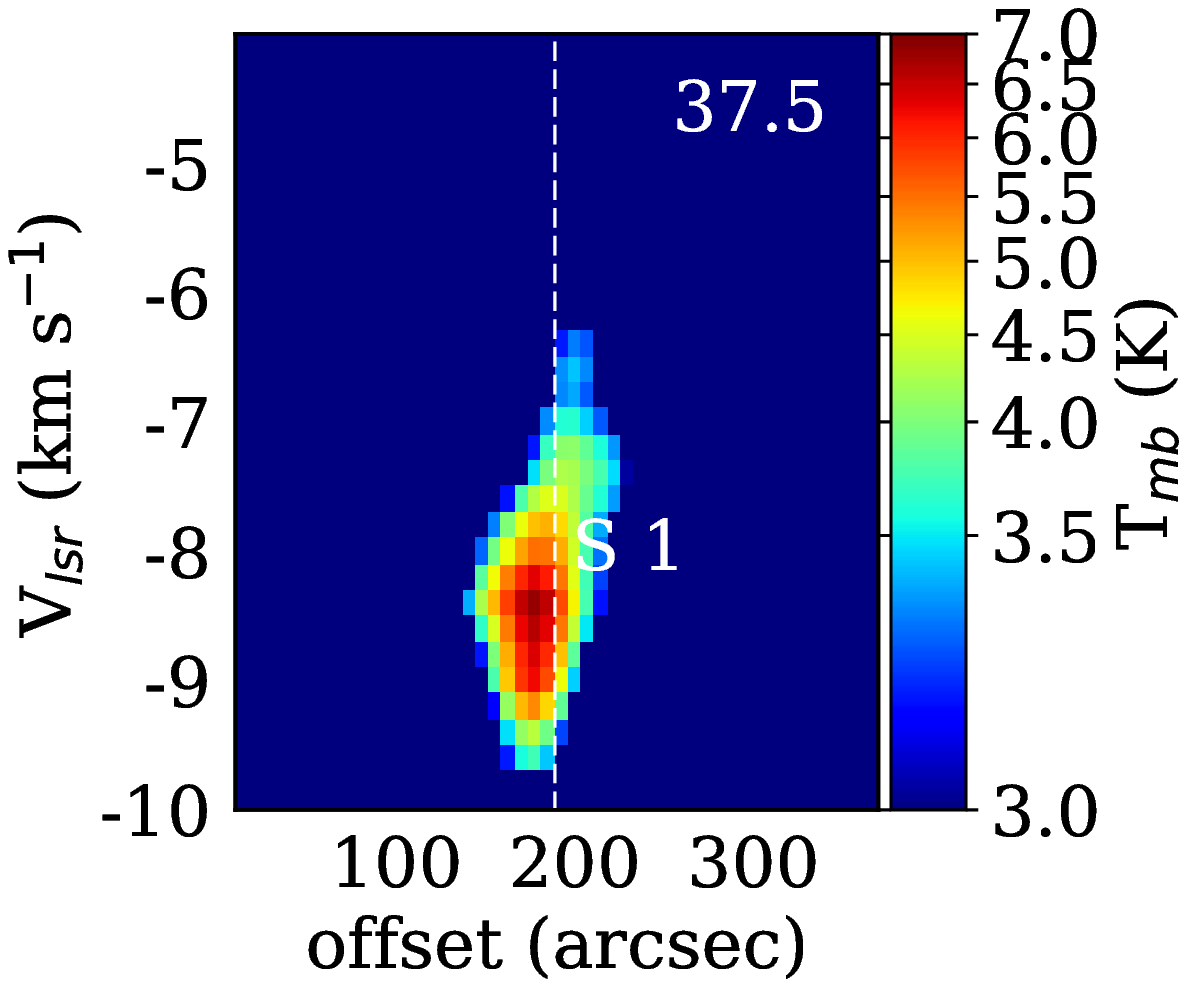}\\
\includegraphics[width=0.49\columnwidth]{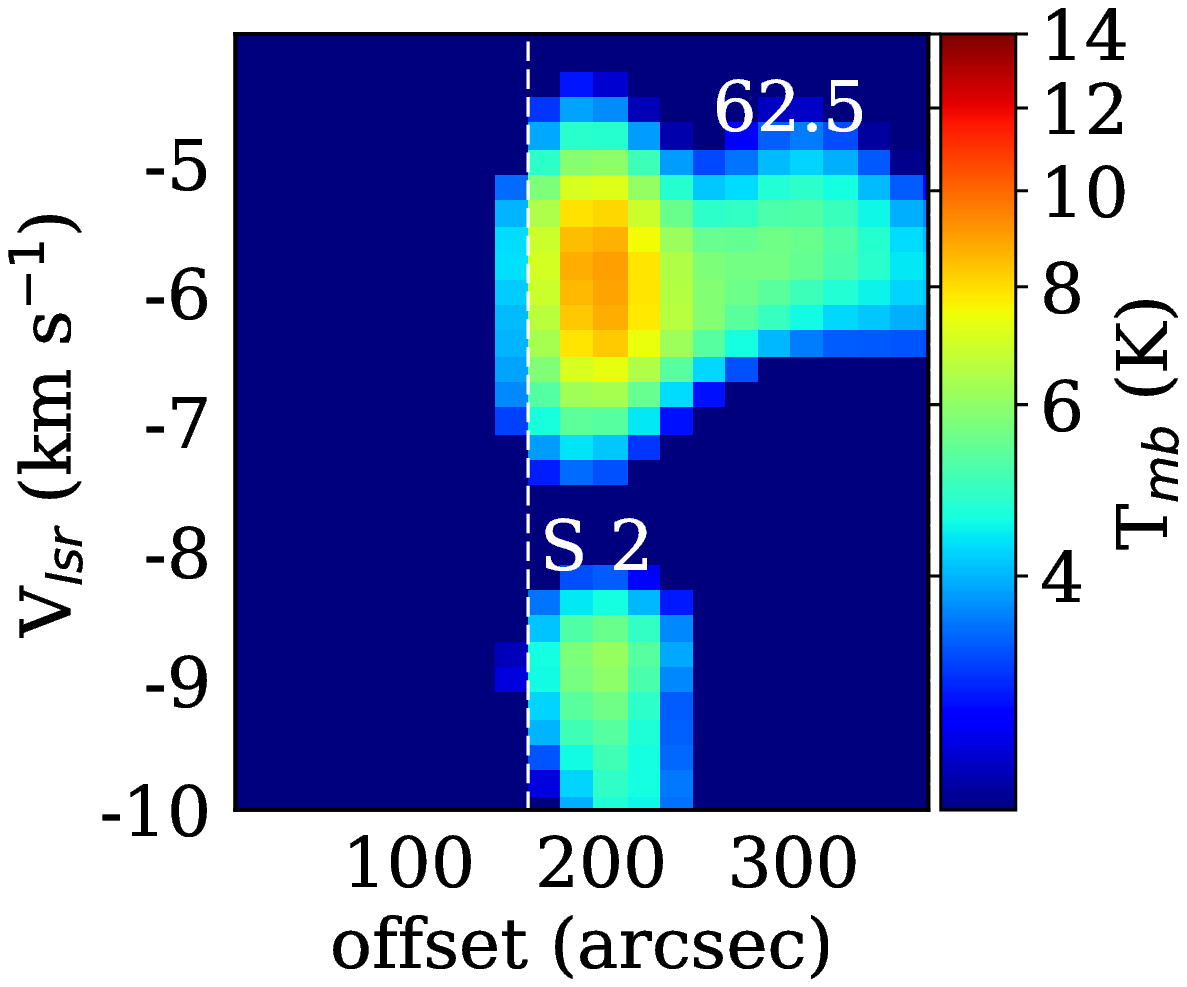}
\includegraphics[width=0.49\columnwidth]{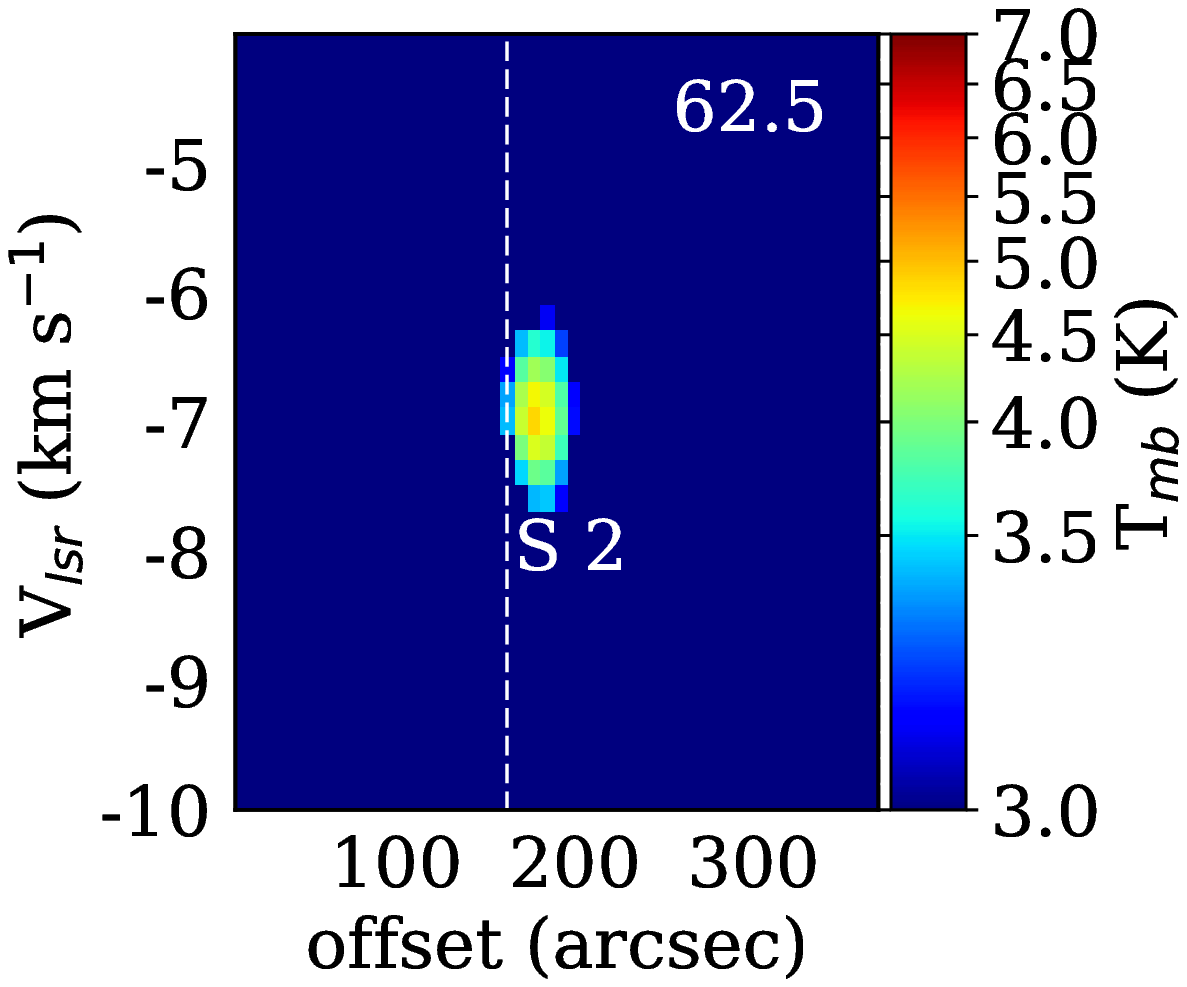}\\
\includegraphics[width=0.49\columnwidth]{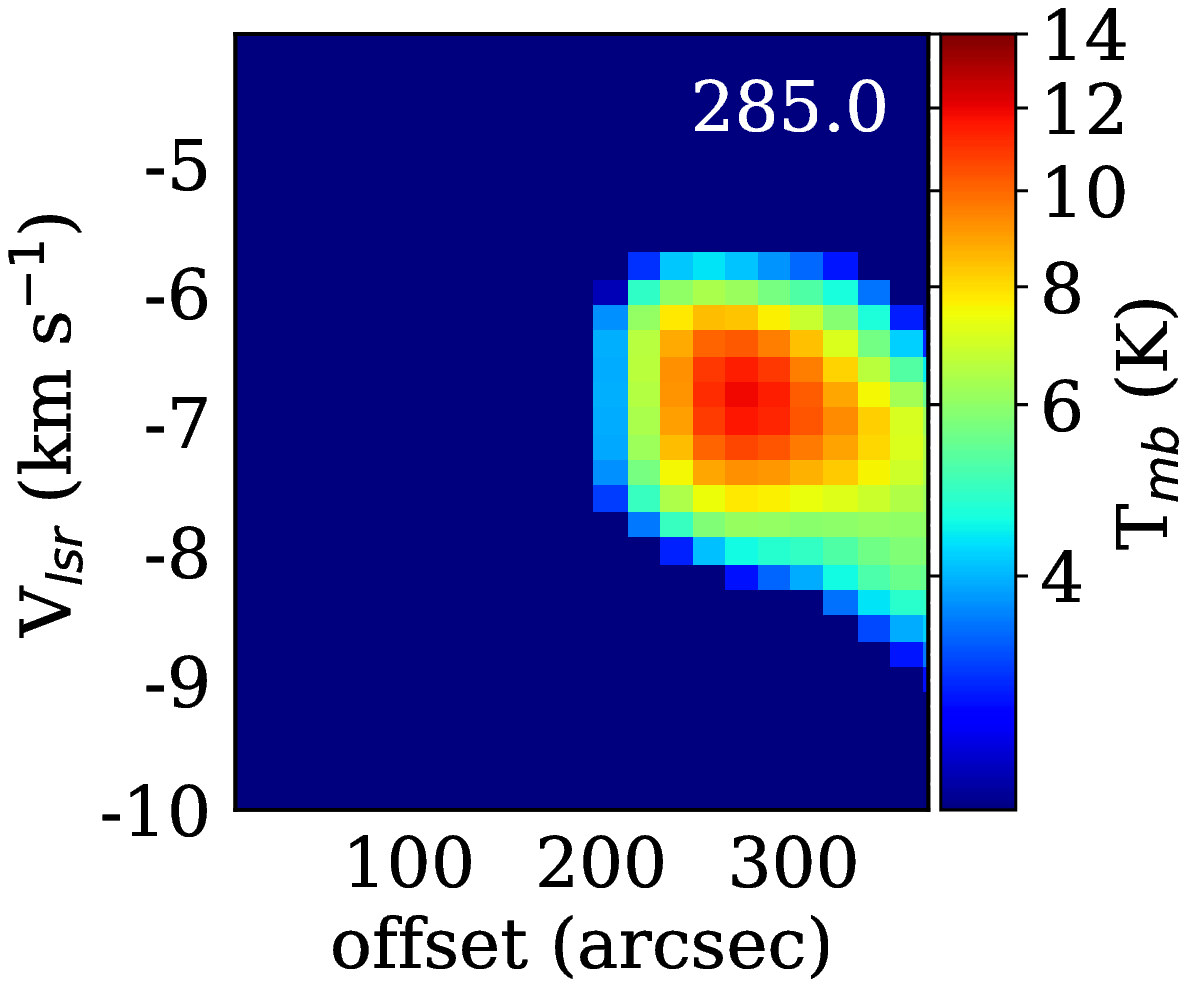}
\includegraphics[width=0.49\columnwidth]{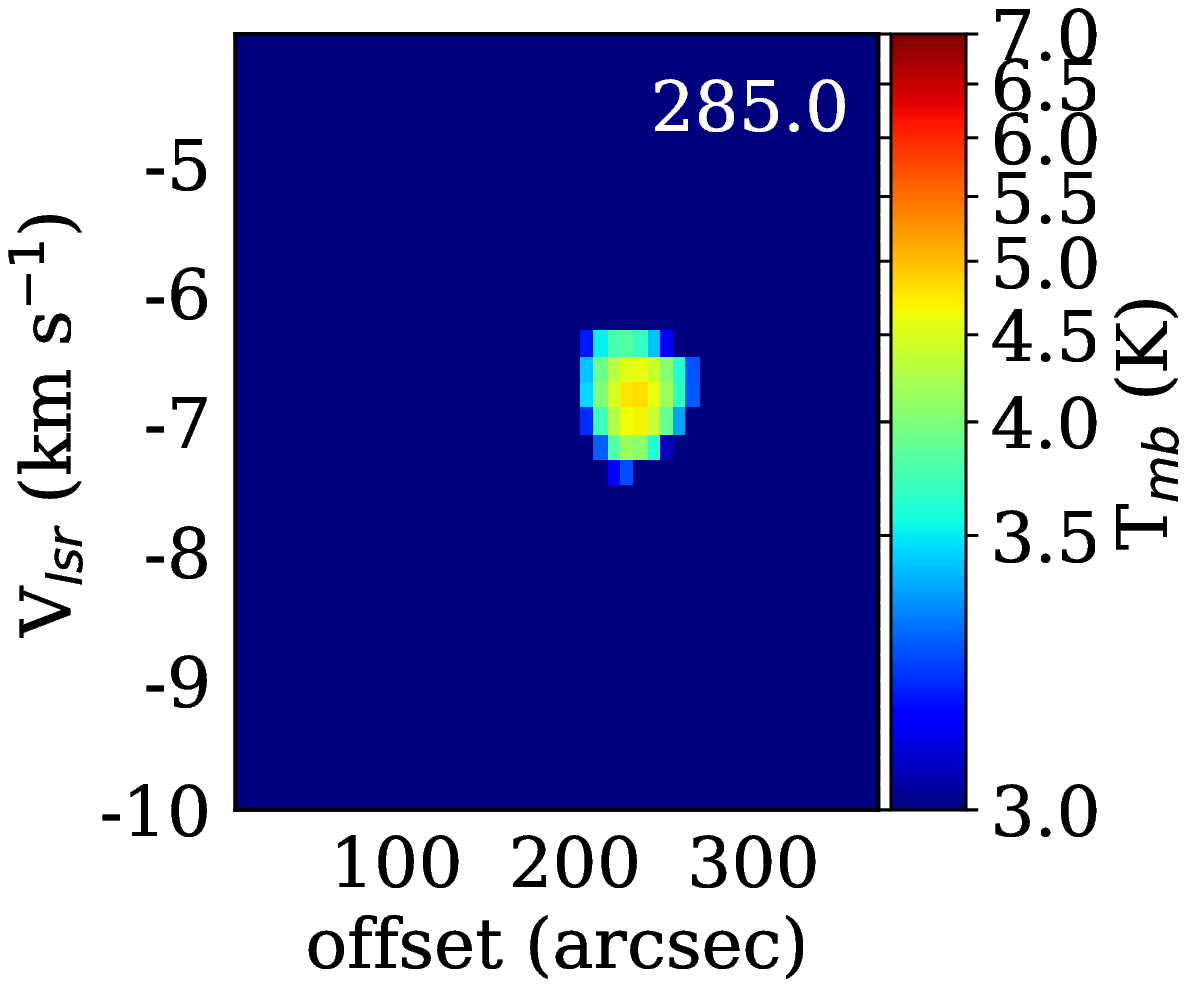}\\
\includegraphics[width=0.49\columnwidth]{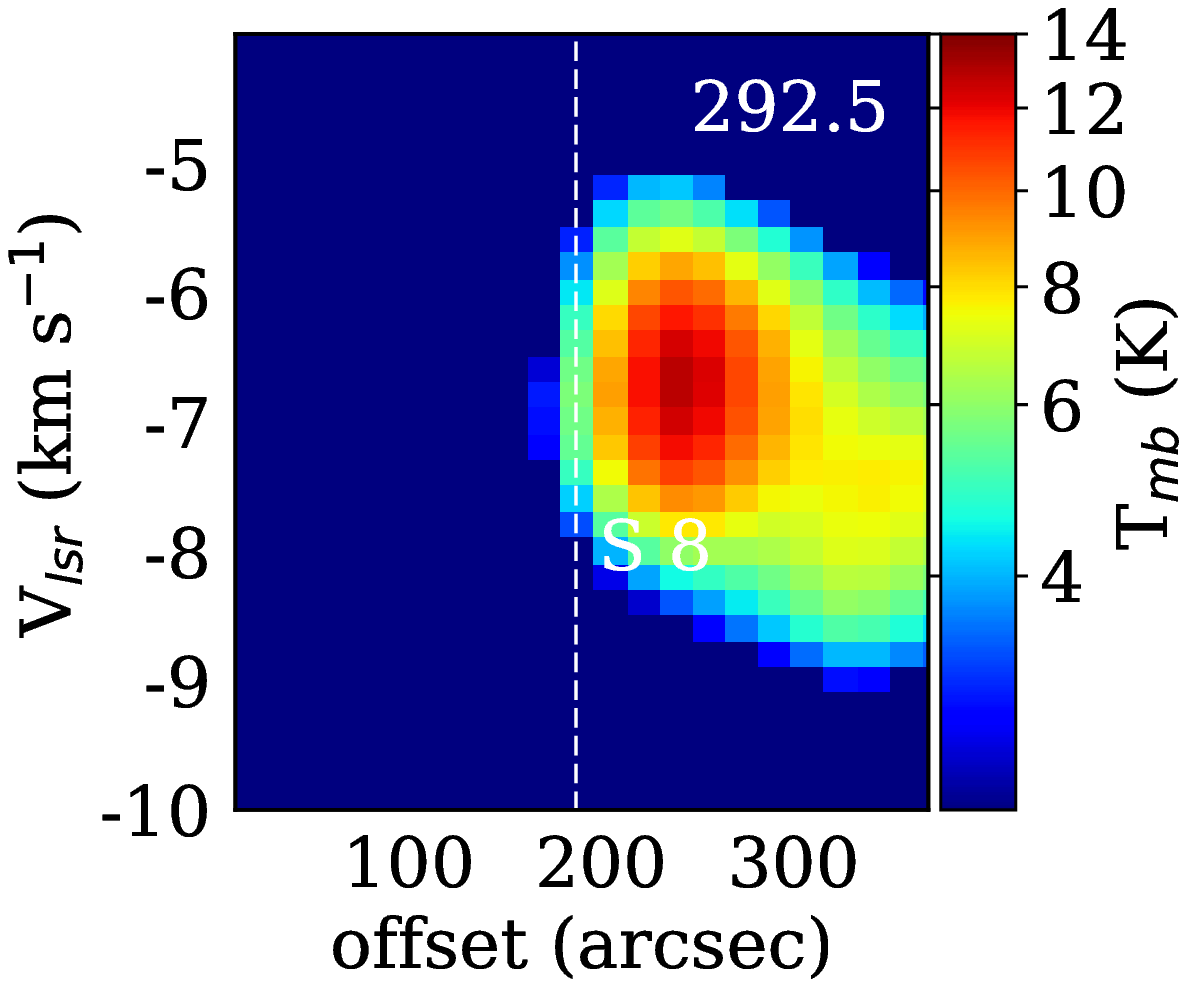}
\includegraphics[width=0.49\columnwidth]{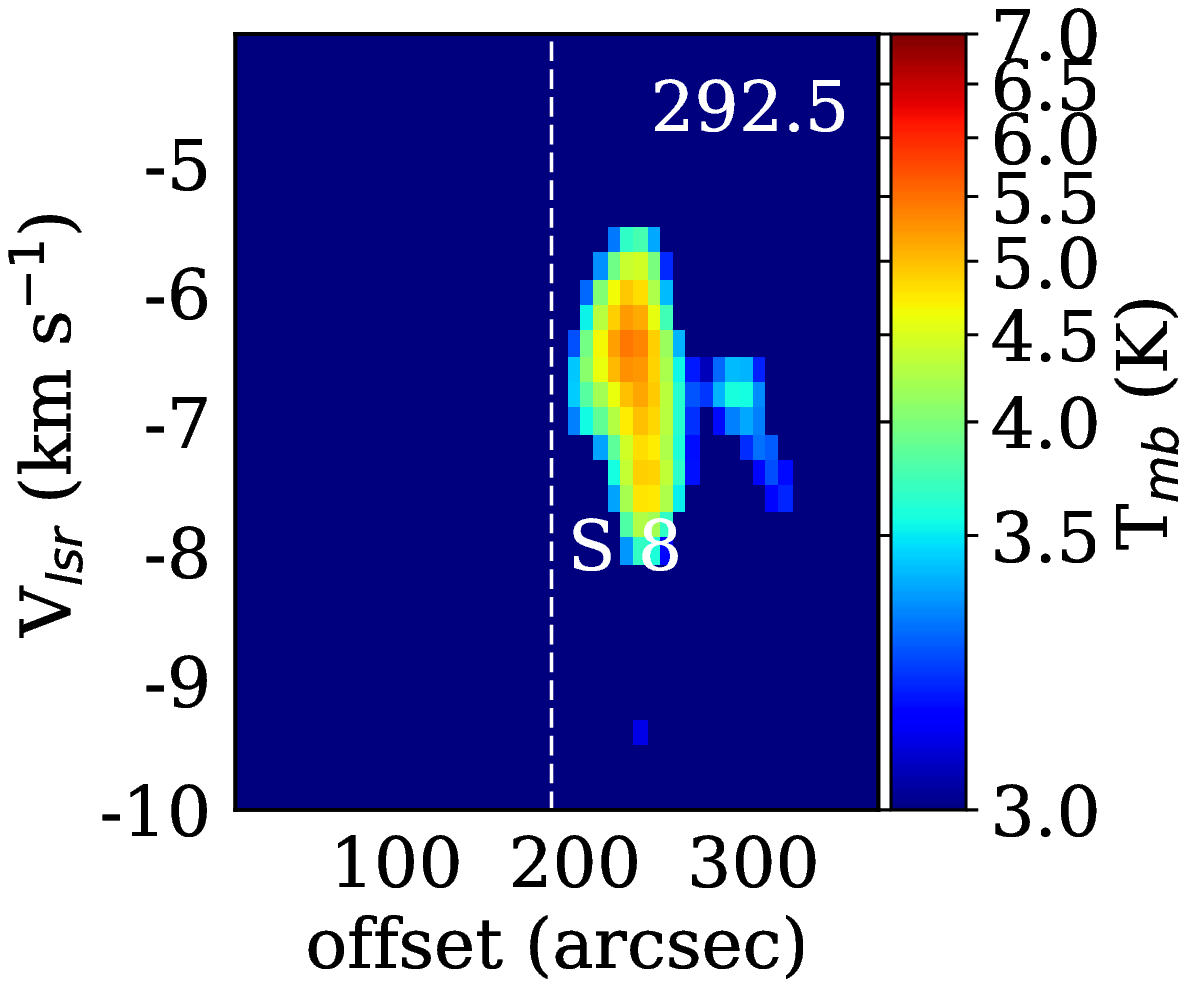}\\
\includegraphics[width=0.49\columnwidth]{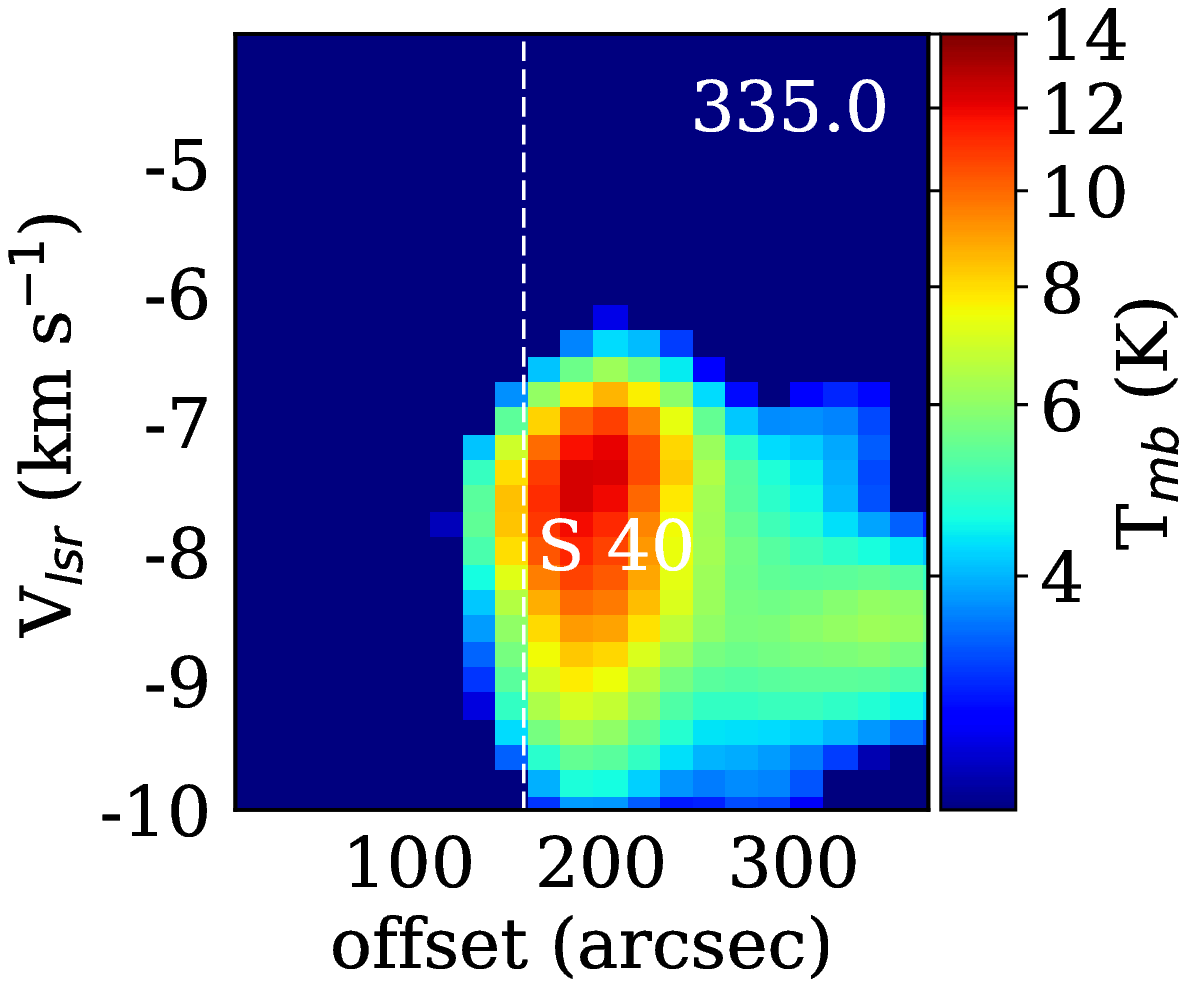}
\includegraphics[width=0.49\columnwidth]{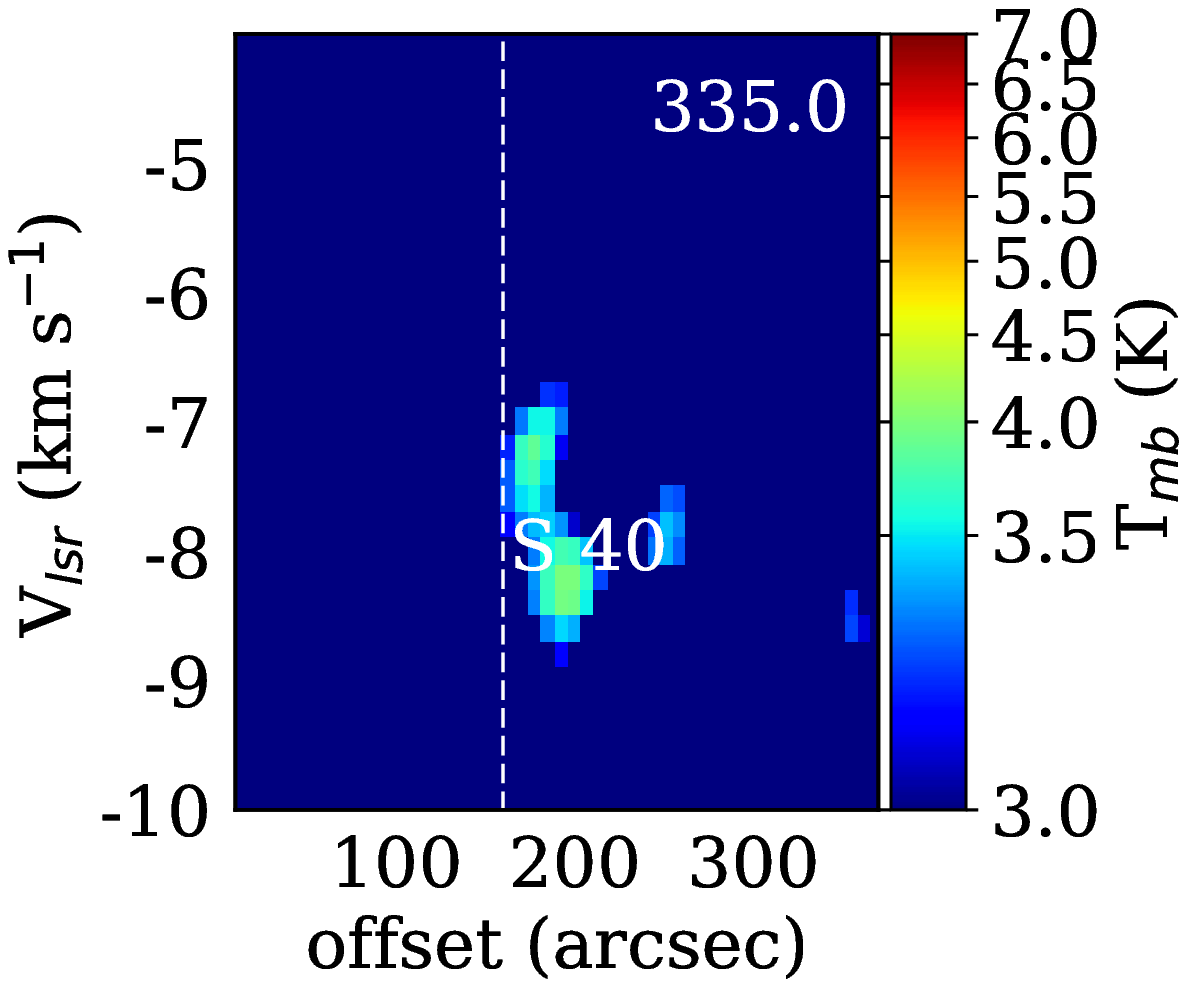}\\
\caption{PV diagrams for \co(2--1) (left) and \cvo(2--1) (right) using APEX archive data for Condensation~1 (rows 1--3) and Condensation~2 (rows 4--5). The original \co(2--1) data was smoothed and regridded to reach noise level of 1~K. The projected distance between the compact sources and the ionizing star are shown as white dashed lines (the source are designated as S1, S2 etc). Top to bottom: PV diagrams for $\varphi=29.0, 37.5, 62.5, 285.0, 292.5, 335.0^\circ$.}
\label{fig:pvobscond1}
\end{figure}

We present position-velocity (PV) diagrams for various directions from the ionizing star from the periphery of the \hii{} region, both with and without compact sources, in Fig.~\ref{fig:pvobscond1} (see Fig.~\ref{fig:channel} for the locations of the cuts with different $\phi$ for the PV diagrams). To find out whether compact sources are located inside the collected molecular layer, we plot the projected distance between the ionizing star and the compact sources on the diagrams. Lines of \co(2--1) in Condensation~1 (rows 1--3 in Fig.~\ref{fig:pvobscond1}) clearly have double-peaked profiles, with a velocity difference of $1-3$~\kms{} between the peaks. Single-peaked profiles symmetric or skewed to the red are observed in Condensation~2 (rows 4--5 in Fig.~\ref{fig:pvobscond1}). The \cvo(2--1) lines have either single- or double-peaked profiles. When they are double-peaked, the velocity difference between the \cvo(2--1) peaks is about 1~\kms. The \cvo(2--1) peaks fall in the intermediate velocity interval mentioned above. The self-absorption effect is more significant for \co(2--1): the \cvo(2--1) peaks correspond to the \co(2--1) dips.

Compact sources 2, 9, 10 and 39 in Condensation~1 \citep[source index number in Table~5 in][]{Figueira_2017} are projected closer to the ionizing star than the dense molecular shell, but source 1 is projected farther from the star than the bright \co(2--1) and \cvo(2--1) emission. All compact sources in Condensation~2 are projected closer to the star, which could mean that source 1 formed independently from the molecular shell collected by the expansion of the \hii{} region.

\subsection{Detailed study of Condensation~1}

\begin{figure}
\includegraphics[width=\columnwidth]{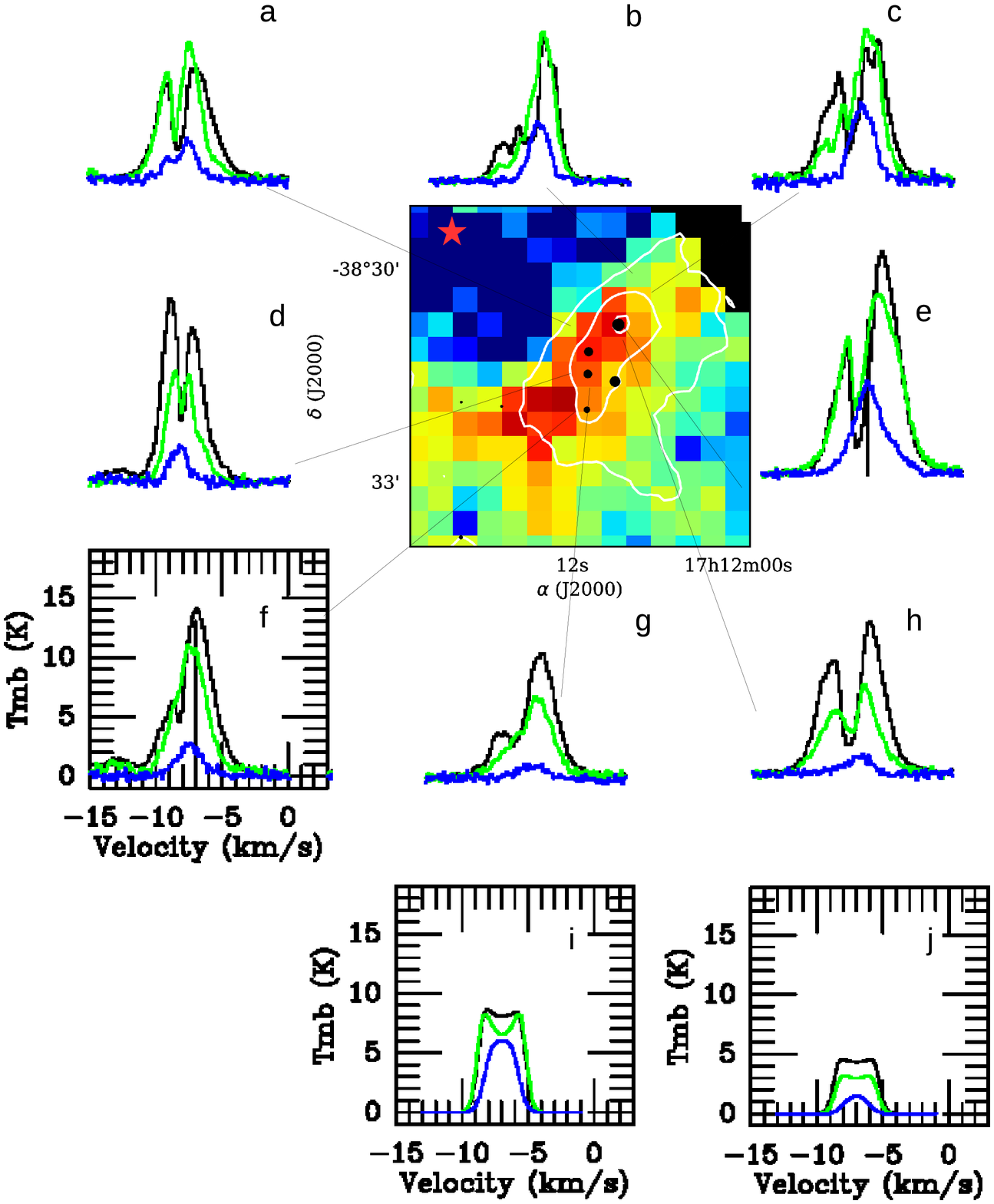}
\caption{Spectra in Condensation 1. \co(2--1), \co(3--2) and \cvo(3--2) lines are shown with black, green and blue lines, respectively. \co(2--1) integrated intensity map from Fig.~\ref{fig:archive} is shown in the center. (Positions of the compact sources from \citet{Figueira_2017} are shown as black circles, where the sizes of the circles depend linearly on the source masses. The position of the ionizing star of RCW~120 is marked as a red star.) All spectra have the same velocity and line T$_{\rm mb}$ scale. We also show typical simulated spectra for the dense shell (panel i) and the undisturbed envelope (panel j) extracted from the PV diagram of the 2D model with the microturbulent velocity $V_{\rm nth} = 1$\kms{}, see bottom panels of Fig.~\ref{fig:pv2d}.}
\label{fig:COspectramap}
\end{figure}

In order to demonstrate possible complications that may arise in the analysis of the molecular shell kinematics, we consider line observations separately in Condensation~1. Emission of \co(2--1), \co(3--2) and \cvo(3--2) was found at all observed positions in Condensation~1. Examples of spectra in selected positions are shown in Fig.~\ref{fig:COspectramap}. The \co{} lines have double-peaked profiles at almost all observed positions (confirming results obtained with lower S/N ratio from the APEX archive) which are shown in Fig.~\ref{fig:pvobscond1}, but sometimes demonstrate more complicated profiles (spectra b and c in Fig.~\ref{fig:COspectramap}, black curves). Both \co(2--1) and \co(3--2) double-peaked lines have red asymmetry at the majority of the observed positions, including the compact source~2 (spectrum e). The asymmetry is less prominent at the position of compact source~1 (spectrum f). The brightness of the \co(2--1) and (3--2) emission is almost the same, which implies a significant optical depth of the \co{} lines in a diagonal strip from the south-east to north-west along the molecular envelope of the \hii{} region. The \cvo(3--2) lines are single-peaked in almost all observed positions, but we find red and blue \cvo(3--2) line wings from the compact source~2 (spectrum e). The peaks of the \cvo(3--2) line coincide with the \co(2--1) and \co(3--2) dips, confirming self-absorption in the \co{} emission. There is a velocity gradient from --8.2~\kms{} at the south-east to --7~\kms{} at the north-west, which agrees with the gradient demonstrated on the channel maps in Fig.~\ref{fig:channel}.

Double-peaked \cvo(3--2) line profiles are observed in positions near the border between the molecular shell and the PDR (e.g. spectrum a in Fig.~\ref{fig:COspectramap}). The \cvo(3--2) dips coincide with the \co(3--2) dips in these directions with the velocity difference up to 2--3\kms{} between the peaks. We suggest that the velocity difference might be caused by the projected relative motion of different parts of the expanding shell. The \cvo(3--2) line becomes single-peaked farther from the PDR border, which means that the proposed expansion of the shell does not involve the dense material of Condensation~1.

We find blue asymmetry of the double-peaked \co{} lines in the eastern part of Condensation~1 (spectrum d) in agreement with the appearance of significant emission in the blue-shifted velocity interval ($V_{\rm lsr} < -8$\,\kms{}) visible in Fig.~\ref{fig:channel}. The \cvo(3--2) line has a single-peaked profile, with the peak corresponding to the \co{} self-absorption dip. Poor spatial coverage of Condensation~1 by single-point observations from 2008 session does not allow us to infer a reason for the change in line asymmetry. Weak \cvo(3--2) emission is observed in the outer part of Condensation~1 (spectra g and h), but the \co{} lines still show self-absorption features.

Overall, we conclude that local phenomena significantly impact the gas kinematics in Condensation~1 and hamper the analysis of the global kinematics of the entire region.

\subsection{Dust extinction around RCW~120}\label{subsec:largescale}

\citet{Beaumont_2010} noticed that many infrared bubbles are embedded in extended molecular envelopes. To verify if this is the case for RCW~120, we checked available infrared images of RCW~120, looking for additional surrounding neutral material. 
We inspected 2MASS images at 1.24, 1.66 and 2.16~$\rm \mu$m and {\it WISE}{} images at 3.6~$\rm \mu$m (see the images online through Aladin at {{\tt\url{http://cdsportal.u-strasbg.fr/?target=RCW\%20120}}), and indeed found obscuration of background stars in a broad area around RCW~120. As we move to longer wavelengths, obscuration becomes less significant and more patchy. Instead of continuous absorption, we see a number of isolated cloudlets in the 12~$\rm \mu$m {\it WISE}{} image, delineating the border of the absorbing cloud. Sharp linear absorption across the face of the \hii{} region is visible at optical wavelengths \citep[see also H$\alpha$ images by][]{Zavagno_2007, Anderson_2015} might be a part of the absorbing cloud. The extent of the cloud is 30\arcmin{} in the plane of the sky, as can be seen in the extinction map shown in Fig.~\ref{fig:av} in $A_{\rm J}$ units. The minimum $A_{\rm J} = 3-4$~mag is observed towards the ionizing star, which corresponds to $A_{\rm V} = 8-12$~mag (depending on the extinction law used). The angular size of the cloud corresponds to 11.3~pc at a distance of 1.3~kpc. RCW~120 is isolated from other star-forming regions, and situated at some distance from the galactic plane. Therefore, we propose that the absorbing cloud is physically related to the \hii{} region, but not just projected at the same line of sight. Visual extinction $A_{\rm V} = 4.36$~mag determined by \citet{Zavagno_2007} using optical and infrared photometry data of the ionizing star is less than our estimated value, suggesting that the absorbing cloud is partly in front of and partly behind the \hii{} region.

\begin{figure}
\includegraphics[width=\columnwidth]{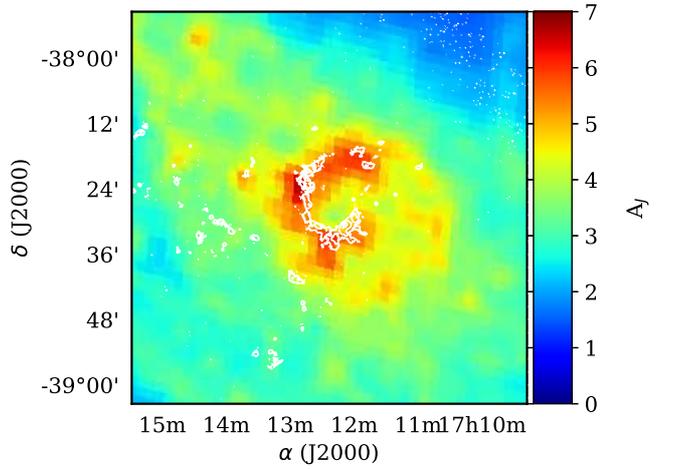}
\caption{Extinction map in A$_{\rm J}$ units around RCW~120. White contour shows ATLASGAL 870~$\rm \mu$m emission. Note the cut-off of the 870~$\rm \mu$m image on the right side of the map.}
\label{fig:av}
\end{figure}

\section{Simulated line emission in RCW~120}\label{Sec:ModelRes}

The results of a simulation of RCW~120 with the MARION code are shown in Fig.~\ref{fig:modelcalc}. Ionized gas with $n_{\rm H^+}\approx 10^2$~cm$^{-2}$ is surrounded by a dense shell of molecular hydrogen with $n_{\rm H_2}\approx 10^5$~cm$^{-2}$. The shell is truncated by a shock at $\approx 1.25$~pc preceding the ionization front, located at $\approx 1.15$~pc. The dense shell is surrounded by an undisturbed molecular gas envelope with $n_{\rm H_2} = 0.5\cdot 10^4$~cm$^{-2}$ and an outer radius of 1.6~pc. We adopt this value for the outer radius as we see molecular emission up to 300\arcsec{} offset from the ionizing star. The maximum velocity of the moving gas is about 1~\kms. We note that within the \hii{} region the carbon is double-ionized, while near the inner boundary of the dense shell it resides mainly in the form of C$^+$, and in the densest part of the shell carbon is bound to CO. Thus, the CO-bearing region is characterized by $T_{\rm gas} < 60$~K and $10^4 < n_{\rm gas} < 10^6$~cm$^{-3}$.

\begin{figure}
\includegraphics[width=0.49\columnwidth]{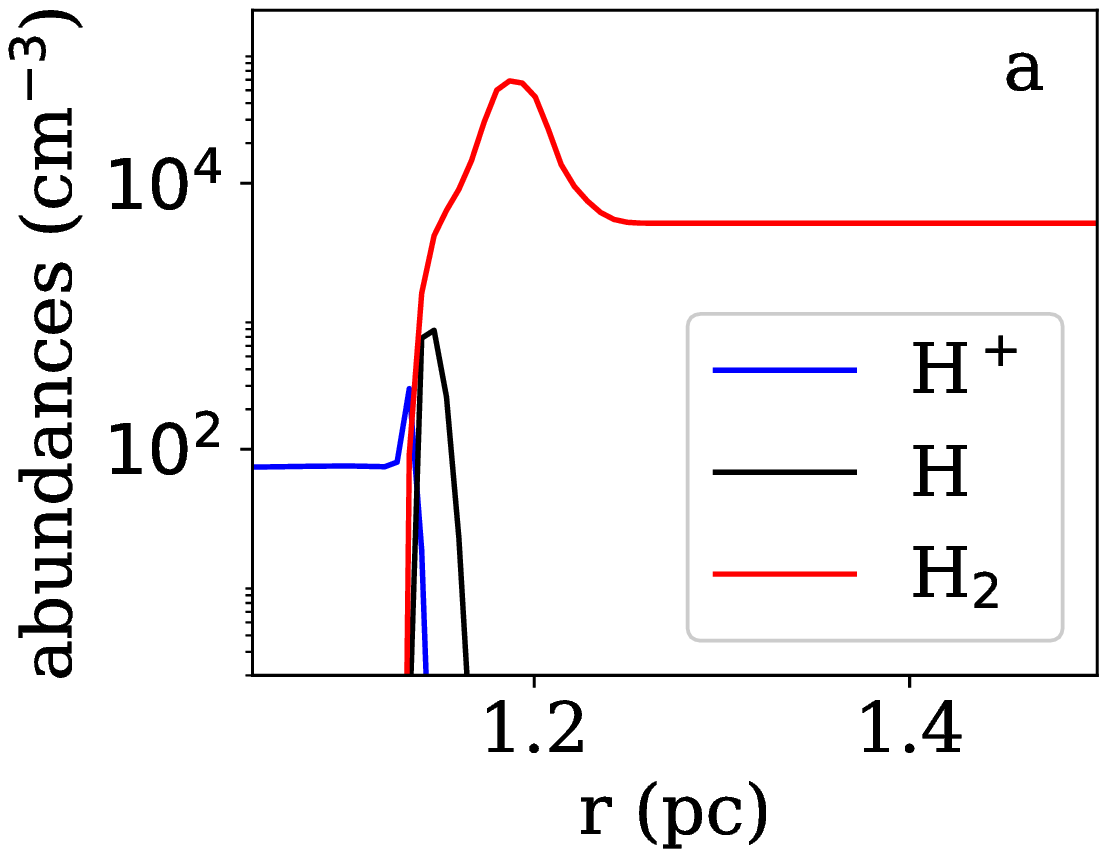}
\includegraphics[width=0.49\columnwidth]{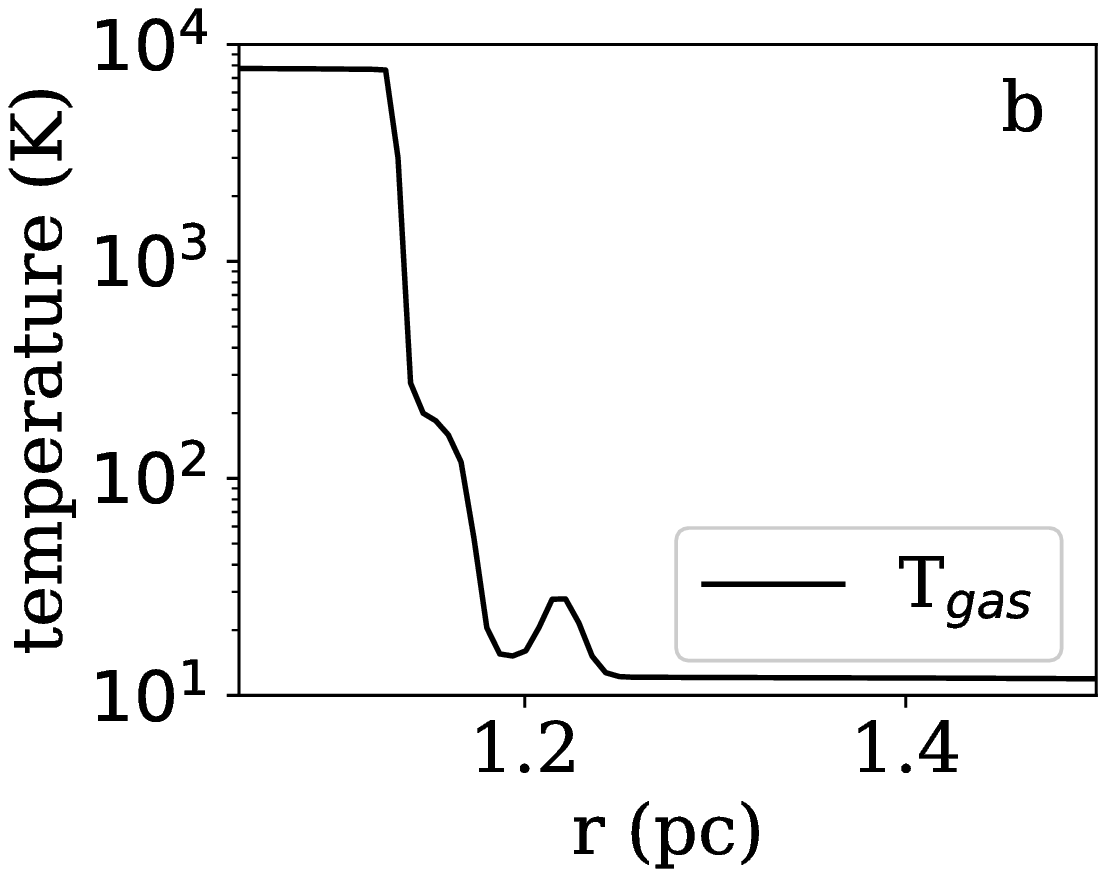}\\
\includegraphics[width=0.49\columnwidth]{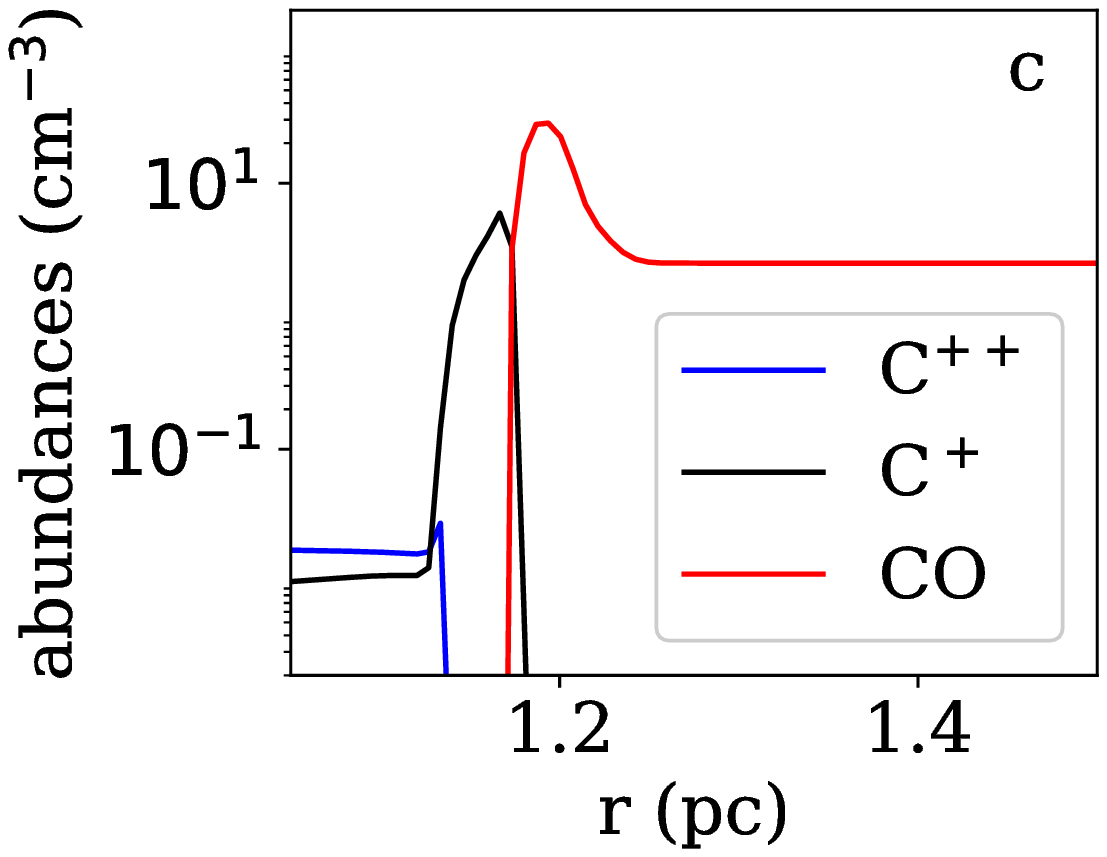}
\includegraphics[width=0.49\columnwidth]{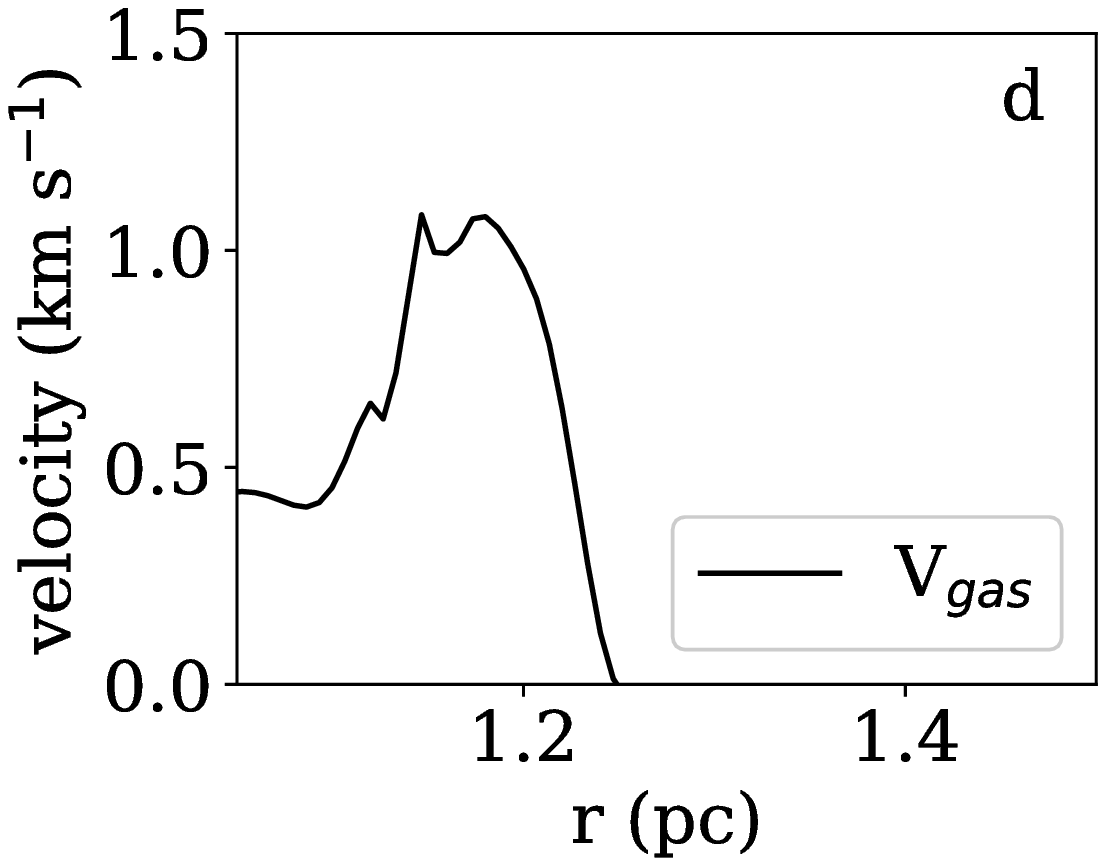}
\caption{Modelled chemical composition and physical conditions in the dense molecular shell around the \hii{} region. a) abundances of H$^+$, H and H$_2$, b) gas temperature, c) abundances of C$^{++}$, C$^+$, and CO, d) gas velocity.}
\label{fig:modelcalc}
\end{figure}

\subsection{Line emission in a spherically-symmetric molecular shell around HII region}

We start with the assumption of a 3D expanding bubble and present results for a spherically symmetric model. In Fig.~\ref{fig:pv1d}, we show PV diagrams for the \co(2--1), \cvo(2--1) and \cvo(3--2) emission along the radial direction. Dashed vertical lines on the plots show the location of the dense shell compressed due to expansion of the \hii{} region into molecular gas. The simulated intensity of \co(2--1) from the dense shell is close to the observed values (see Fig.~\ref{fig:pvobscond1}), while the \cvo(2--1) and \cvo(3--2) intensities are about 2 times higher than those observed (see also Fig.~\ref{fig:COspectramap}). The major discrepancy between the observed and simulated PV diagrams is the presence of the bright CO emission over the entire extent of the modelled \hii{} region (corresponding offsets are $<200$\arcsec), which is not observed in RCW~120. The simulated \co(2--1), \cvo(2--1), and \cvo(3--2) profiles from the inner parts of the region are strong and double-peaked, which is also not observed.

One may suppose that this discrepancy is related to a value of $n_{\rm gas}$ too high in the shell. Indeed, an overestimated initial density would result in an over-dense molecular layer, making both \co{} and \cvo{} emission optically thick over the entire extent of the \hii{} region. We checked if this is possible to resolve the inconsistency by artificially decreasing the \co{} and \cvo{} abundances. PV diagrams for the abundances decreased by a factor of 8 are shown in Fig.~\ref{fig:pv1d}. As expected, the intensities of all CO transitions become lower, which brings the \cvo(2--1) and \cvo(3--2) peak temperatures much closer to the observed values. However, the extended double-peaked emission from the \hii{} region is still significant in all three maps. Further reducing the \co{} and \cvo{} abundance does not make sense, since the model intensities would become much lower than those observed. Therefore, we conclude that a spherically symmetric model cannot reproduce the observed PV diagrams, both in terms of line shapes and line intensities, for any reasonable CO abundance in the shell.

\begin{figure*}
\hspace{-0.6cm}\includegraphics[width=0.5\columnwidth]{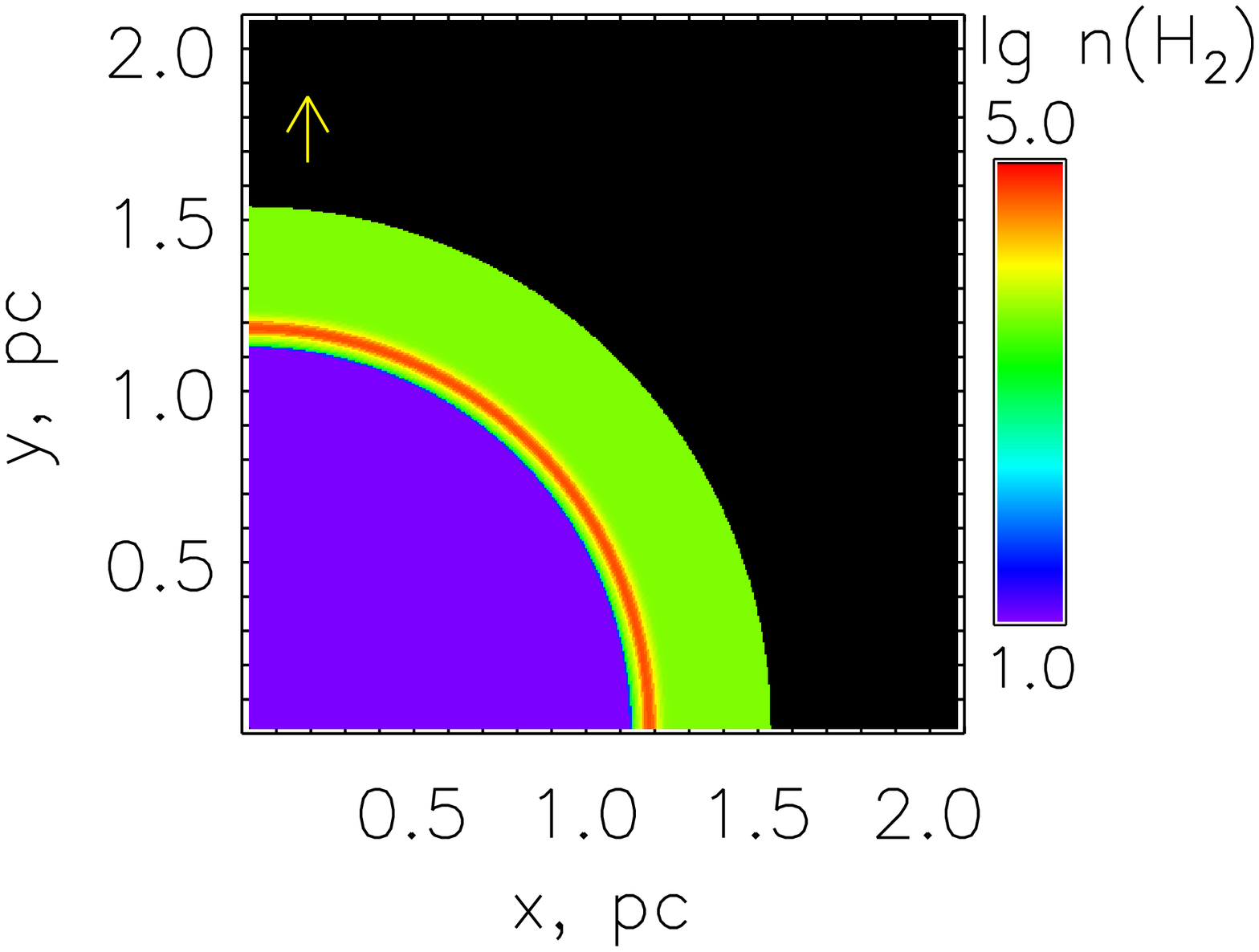}
\includegraphics[width=0.52\columnwidth]{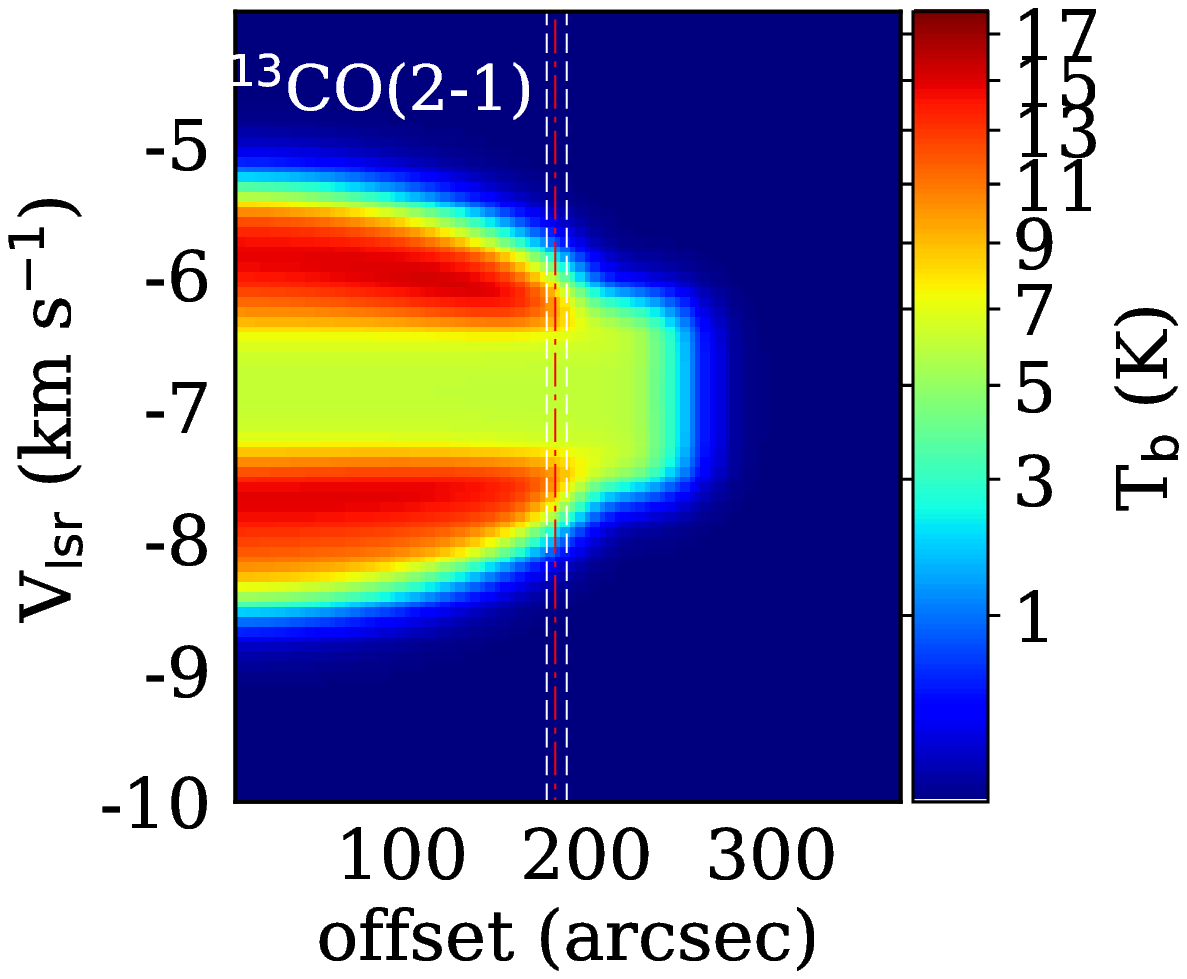}
\includegraphics[width=0.52\columnwidth]{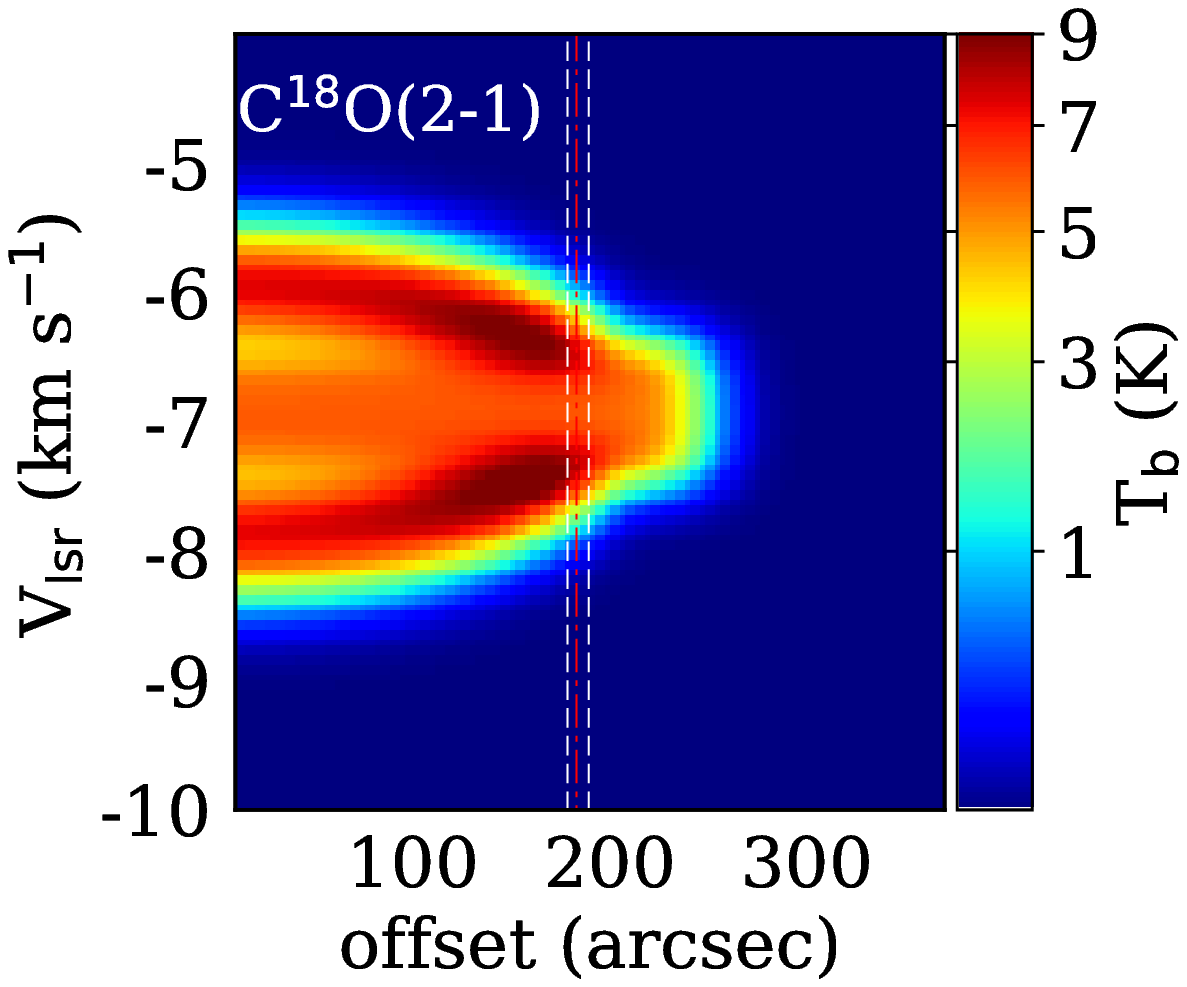}
\includegraphics[width=0.52\columnwidth]{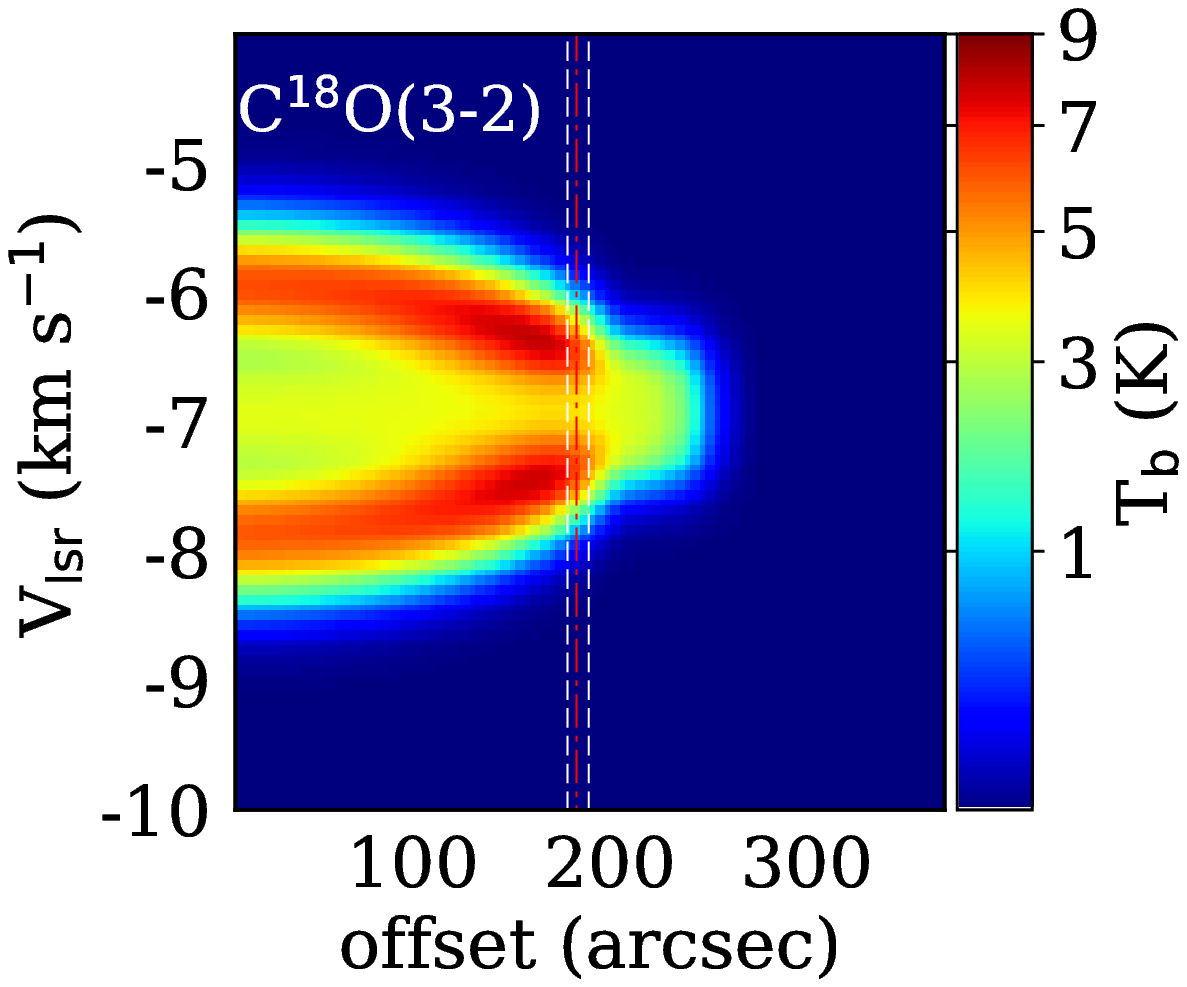}\\
\hspace{4cm}\includegraphics[width=0.52\columnwidth]{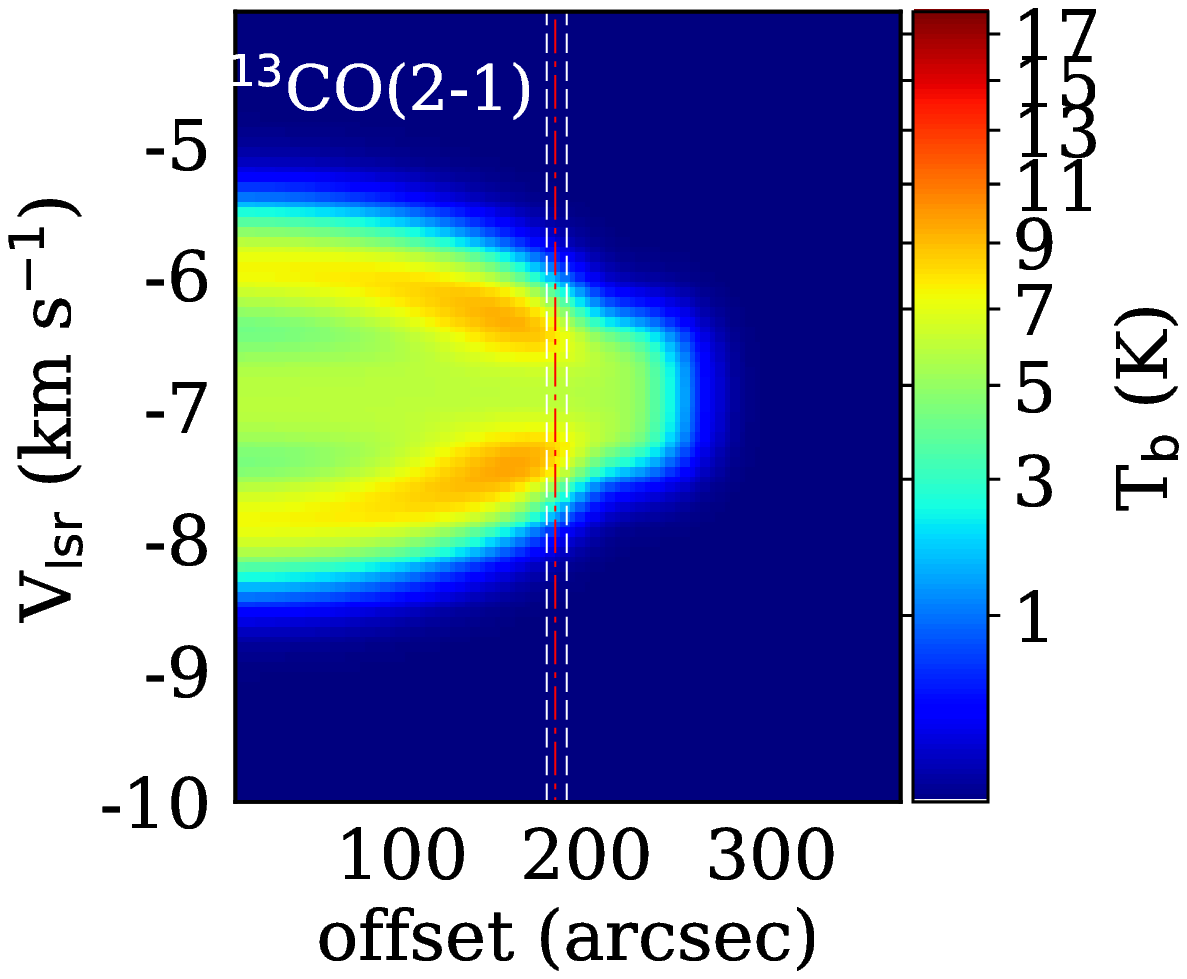}
\includegraphics[width=0.52\columnwidth]{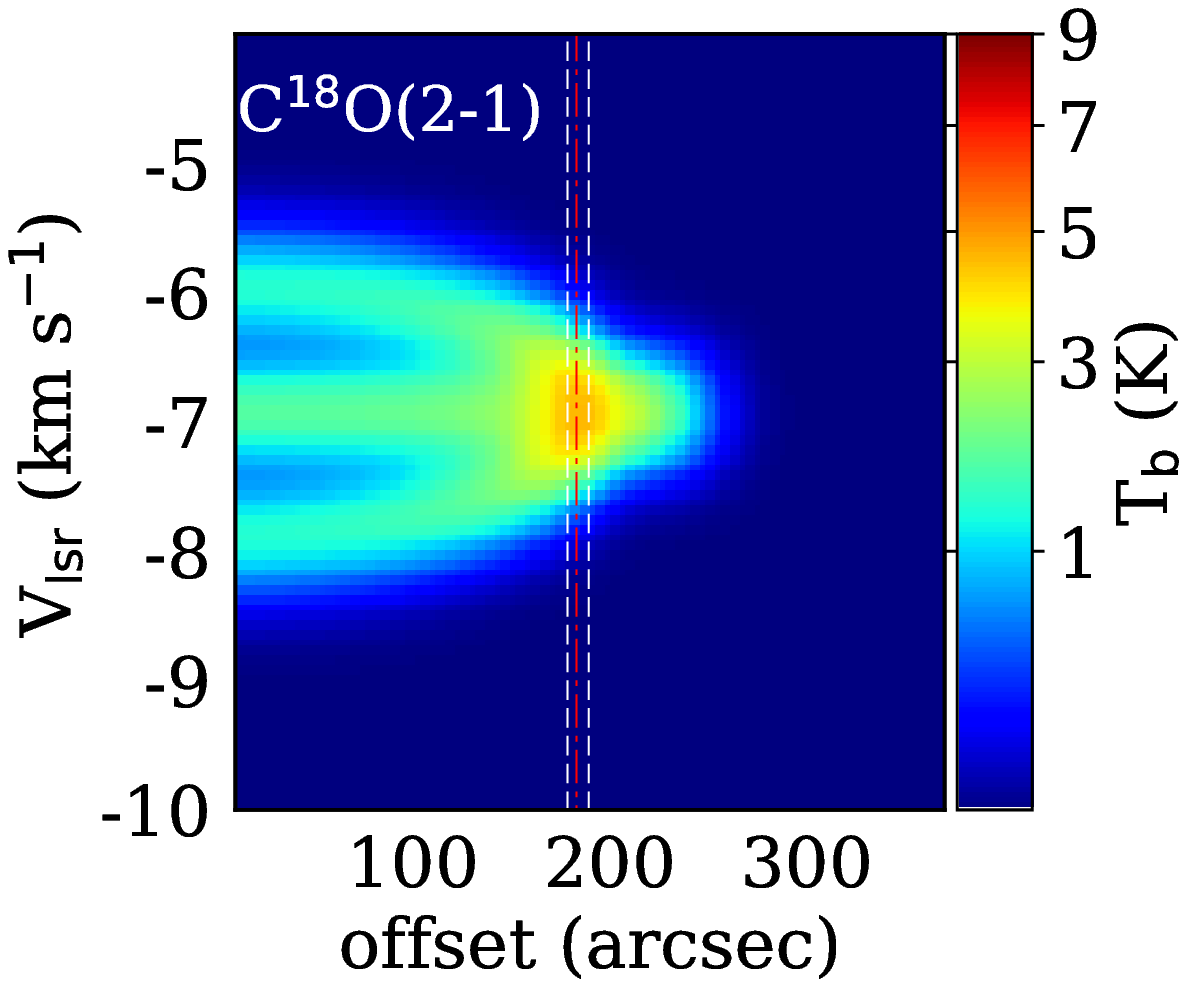}
\includegraphics[width=0.52\columnwidth]{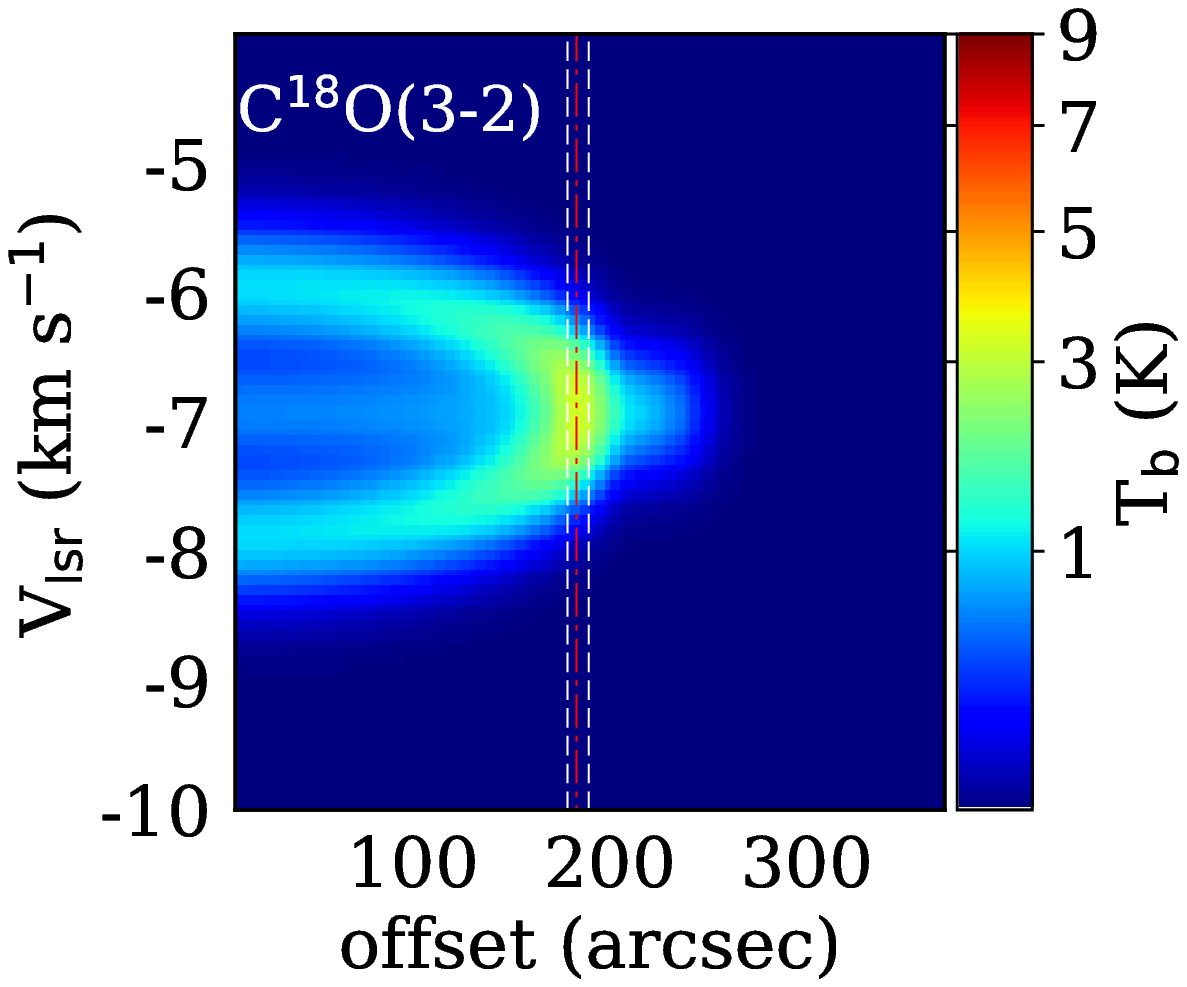}\\
\caption{PV diagrams for a spherical molecular envelope around the \hii{} region, based on the original MARION results (top row), and calculated with reduced CO abundances (bottom row). The direction to the observer is shown as a yellow arrow in the left panel. Dashed vertical lines show the location of the dense compressed shell. The red line shows the location of the density peak.}
\label{fig:pv1d}
\end{figure*}

\subsection{Position-velocity diagrams for a phenomenological 2D model}

In principle, the absence of a dense anterior part of the shell can be inferred from the presence of optical emission coming from the interior of the \hii{} region. Thus, a possible way to resolve the inconsistency between the simulated and observed PV~diagrams is to assume that RCW~120 is not a spherically symmetric object. The absence of strong \co{} and \cvo{} emission from the inner parts of the region can be explained if RCW~120 is an oblate structure seen face-on. Such a structure may arise naturally as a result of expansion of an \hii{} region in a flattened molecular cloud. The best proof of this concept would be to present synthetic PV-diagrams based on a self-consistent 2D/3D hydrodynamical model, however, here we illustrate the associated geometrical effects by manually modifying our simpler spherically symmetric model.

In Fig.~\ref{fig:pv2d}, we present PV~diagrams for the same model as in Fig.~\ref{fig:pv1d} (with unaltered \co{} and \cvo{} abundances), but with conical parts of the dense layer removed, as shown in the top left panel of Fig.~\ref{fig:pv2d}. The line of sight is parallel to the polar axis. As expected, the synthetic PV~diagrams do not reveal strong molecular emission within a circular area with a radius $200$\arcsec. Both \cvo(2--1) and \co(3--2) profiles are double-peaked, while \cvo(2--1) and \cvo(3--2) are single-peaked towards the dense shell, as is indeed observed.

While these synthetic PV~diagrams better reproduce the observed maps, the observed profiles are generally broader than modelled profiles (see Fig. \ref{fig:pvobscond1}). So we calculated another set of synthetic PV~diagrams additionally increasing the micro-turbulent velocity in the input model from $V_{\rm nth} = 0.3$~\kms{} to 1~\kms{}, as the observed CO lines have widths of several \kms{} (Sec.~\ref{Sec:ObsRes}). The corresponding PV~diagrams are shown in the bottom row in Fig.~\ref{fig:pv2d}, and the spectra extracted from the PV~diagrams in Fig.~\ref{fig:COspectramap}. The extracted spectra are shown for the direction of the intensity peak in the dense compressed shell (spectrum i for the offset 200\arcsec, at Fig.~\ref{fig:pv2d}) and in the outer part of the undisturbed molecular gas envelope (spectrum j at the offset 250\arcsec). As a result, the width of the synthetic \co(2--1) lines gets closer to the observed values, while the synthetic \cvo(2--1) profiles become broader than those observed in the direction with $\varphi=285^\circ$ in Condensation~2, corresponding to the molecular envelope with no embedded compact sources.

The presented 2D model does not include the extended rarefied molecular cloud discussed in Sec.~\ref{subsec:largescale}. We testified only one model of the cloud in order to see how the simulated line profiles changes.  Specifically, the cloud encompassing the \hii{} region is assumed to have a thickness of 1~pc, gas density of 50 cm$^{-3}$, temperature (\tgas) of 10\,K, and the same CO abundance as in the undisturbed envelope. Such a rarefied cloud produces weak extended emission over the considered maps. The signal-to-noise ratio of the large-scale \co(2--1) and \cvo(2--1) data presented in Fig.~\ref{fig:archive} and \ref{fig:pvobscond1} is too low to detect this emission. The self-absorption in the cloud additionally splits the \co(2--1) profile. To study how warm the absorbing extended envelope might be to produce a self-absorption effect on the line profiles, we calculated the excitation temperature $T_{\rm ex}$ of the \co(2--1) lines using the RADEX code \citep{radexpaper}, with $10 \leq T_{\rm gas} \leq 2000$~K (the latter value gives the pressure equilibrium between the extended rarefied cloud and the undisturbed envelope) and \nco{} from 10$^{14}$ (corresponding to the front of the rarefied cloud) to 10$^{17}$~cm$^{-2}$. We found that $T_{\rm ex}$ is always less than the observed value of $T_{\rm mb} = 14-16$~K for \co(2--1). So, using higher \tgas{} values we are still on the safe side regarding to the self-absorption effect.

Comparing the observed PV diagrams in Fig. \ref{fig:pvobscond1} with those in Figs~\ref{fig:pv1d} and~\ref{fig:pv2d}, we make the following conclusions. An expanding shell produces double-peaked line profiles with an enhanced red peak due to self-absorption effects, or single-peak profiles skewed to the red, depending on $V_{\rm exp}$ and the micro-turbulent velocity of the molecular gas. A static (non-expanding) optically thick shell would give symmetric double-peaked line profiles. On the one hand, neither in the case of a spherical expanding bubble nor in the case of an expanding torus do we see any noticeable asymmetry in the theoretical line profiles. This means that the expansion velocity in our model is too small to produce line asymmetry, even at the centre of the spherical bubble. This is even more true for the torus model, where the line of sight (presumably) passes through the molecular envelope nearly perpendicular to the expansion velocity. On another hand, the observed double-peaked line profiles are markedly asymmetrical towards Condensation~1, which could mean that even in the relatively simple case of RCW~120 that the actual line profiles are strongly affected by local kinematics. If we neglect these local effects, we can argue that the theoretical PV~diagrams resemble the observed ones for the torus model with $V_{\rm nth} = 1$~\kms{}, (compare with the directions with no embedded compact sources in Fig.~\ref{fig:pvobscond1}). 

Our model is consistent with the absence of front and back walls of the molecular shell, but we still need some molecular gas in the foreground of the \hii{} region in order to quantitatively reproduce the depth of the self-absorption dip in the double-peaked lines. Moreover, \citet{Anderson_2015} detected CO(1--0) and \co(1--0) emission at the central part of the \hii{} region, which could not be reproduced by the face-on torus envelope alone. Given the complex density and velocity structure of the shell on small spatial scales, one may argue that this self-absorption can also be local, arising not in the associated foreground cloud, but in the clumps and condensations themselves. The cloud hypothesis seems to be preferable, as we observe self-absorbed lines both within and beyond dense condensations, and we also see signs of some foreground absorbing layer in the extinction maps.

We note that the goal of these exercises is to show the most promising way to develop consistent hydrodynamical models, rather than to describe all aspects of the observations. Given this, we conclude that such a model should take into account not only the non-spherical geometry of the source, but also the extended cloud and kinematic inhomogeneities.

\begin{figure*}
\hspace{0cm}\includegraphics[width=0.41\columnwidth]{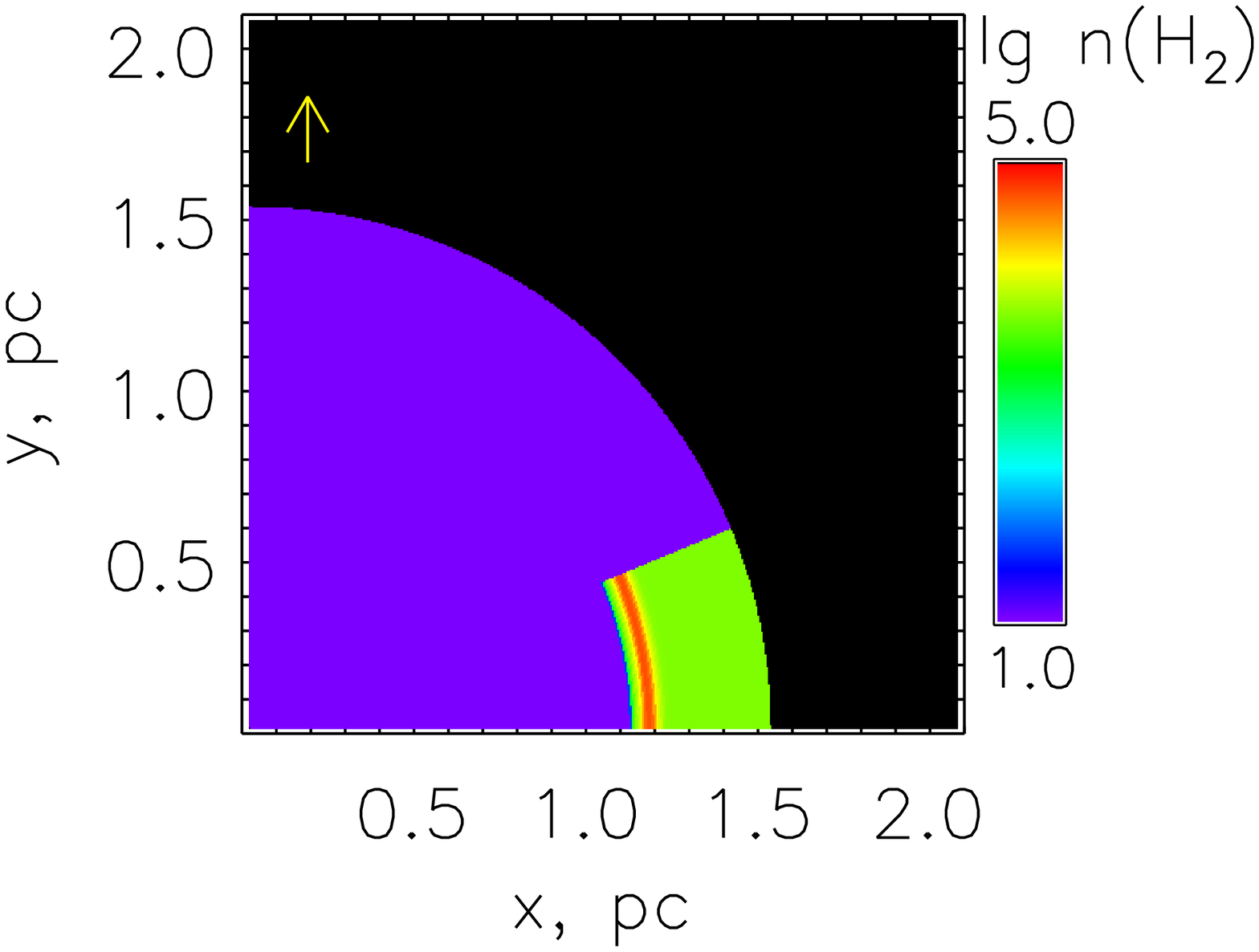}
\includegraphics[width=0.41\columnwidth]{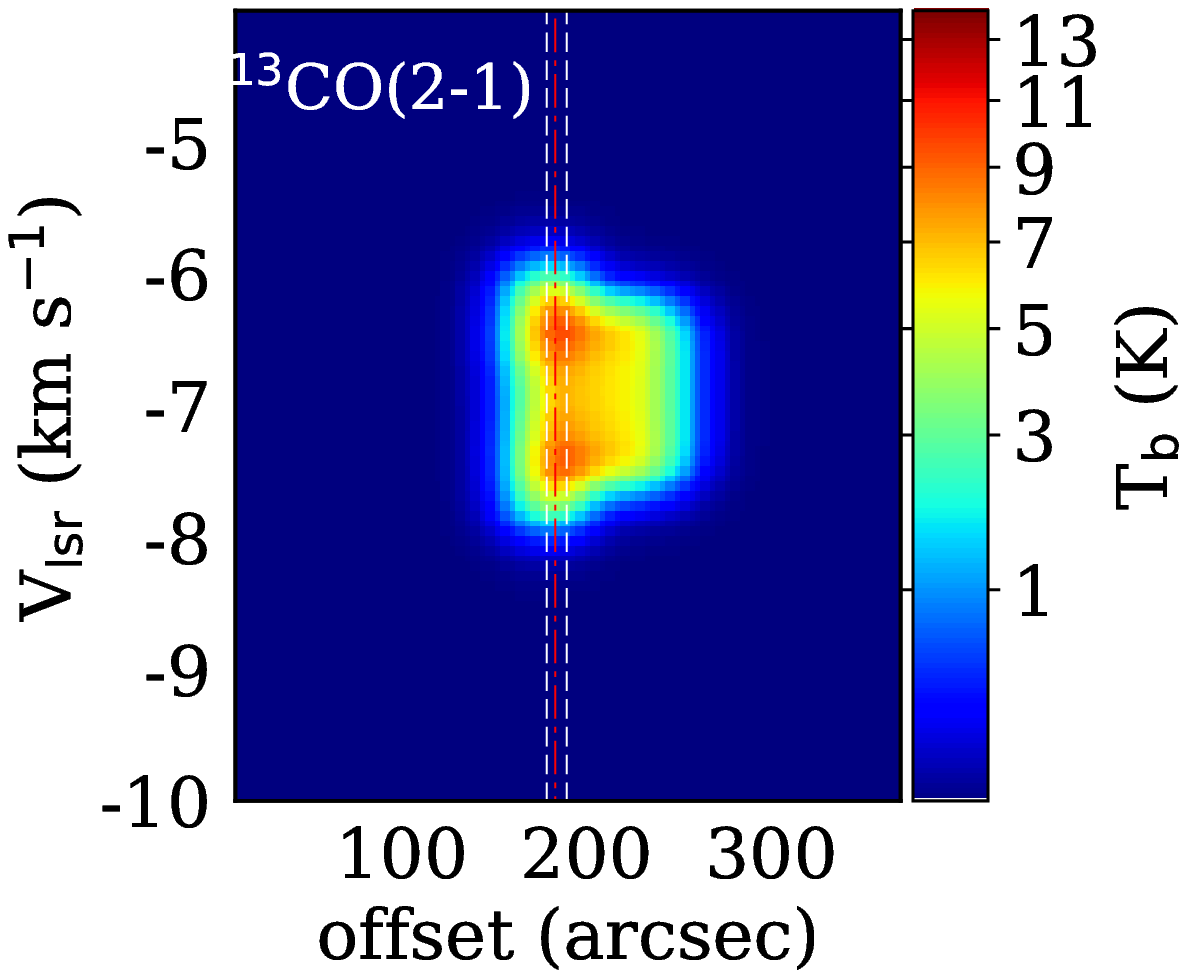}
\includegraphics[width=0.41\columnwidth]{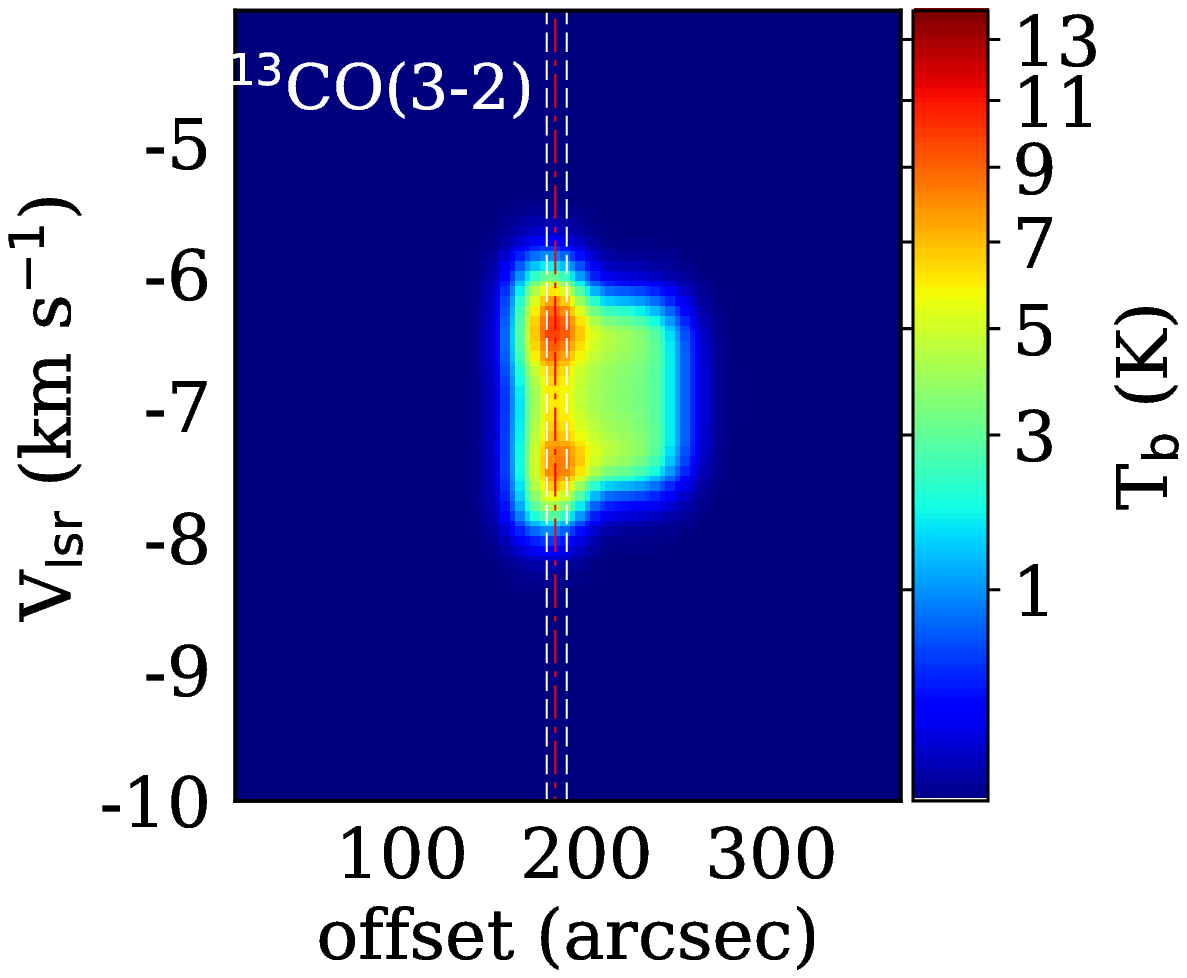}
\includegraphics[width=0.41\columnwidth]{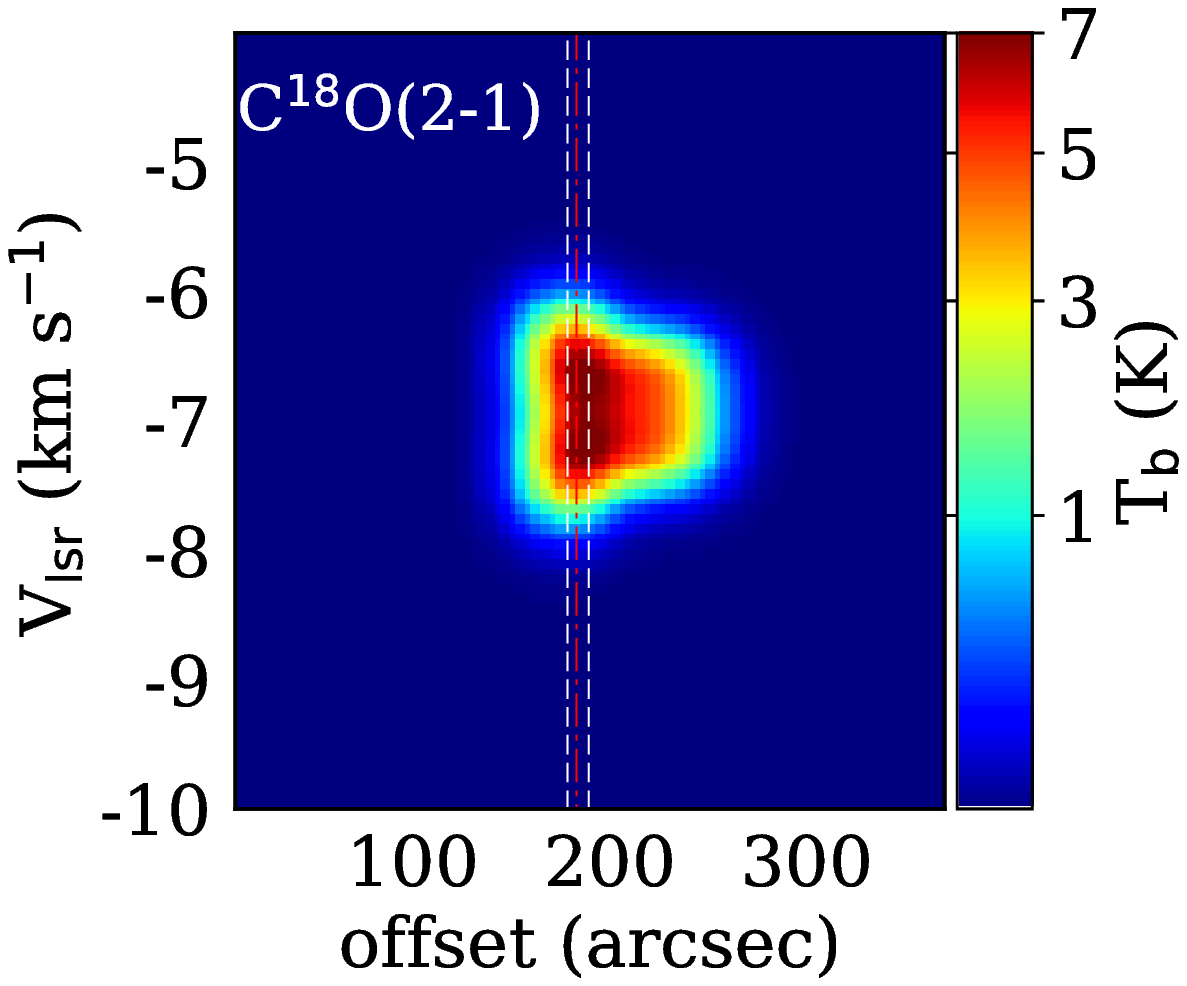}
\includegraphics[width=0.41\columnwidth]{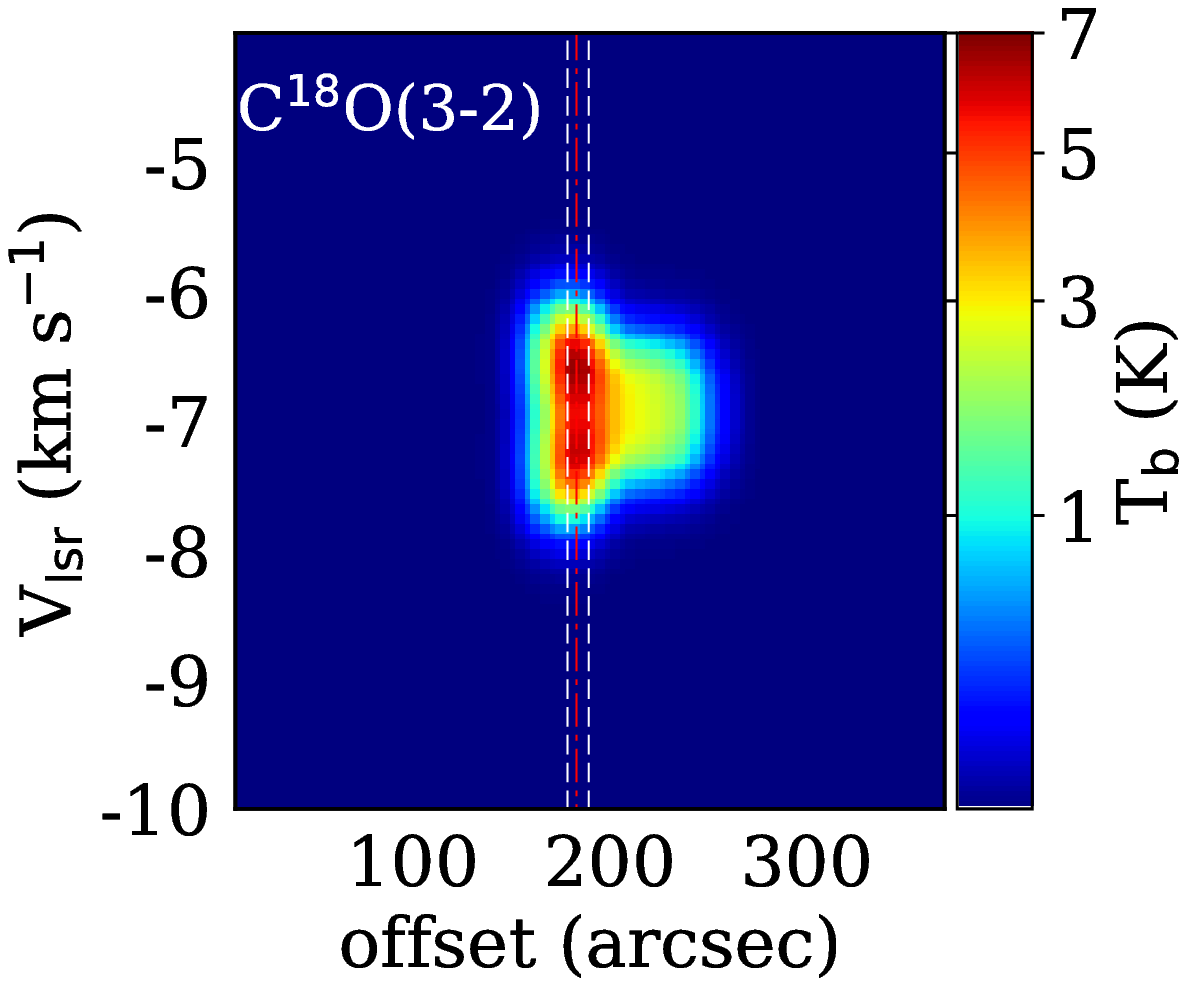}\\
\hspace{3.5cm}\includegraphics[width=0.41\columnwidth]{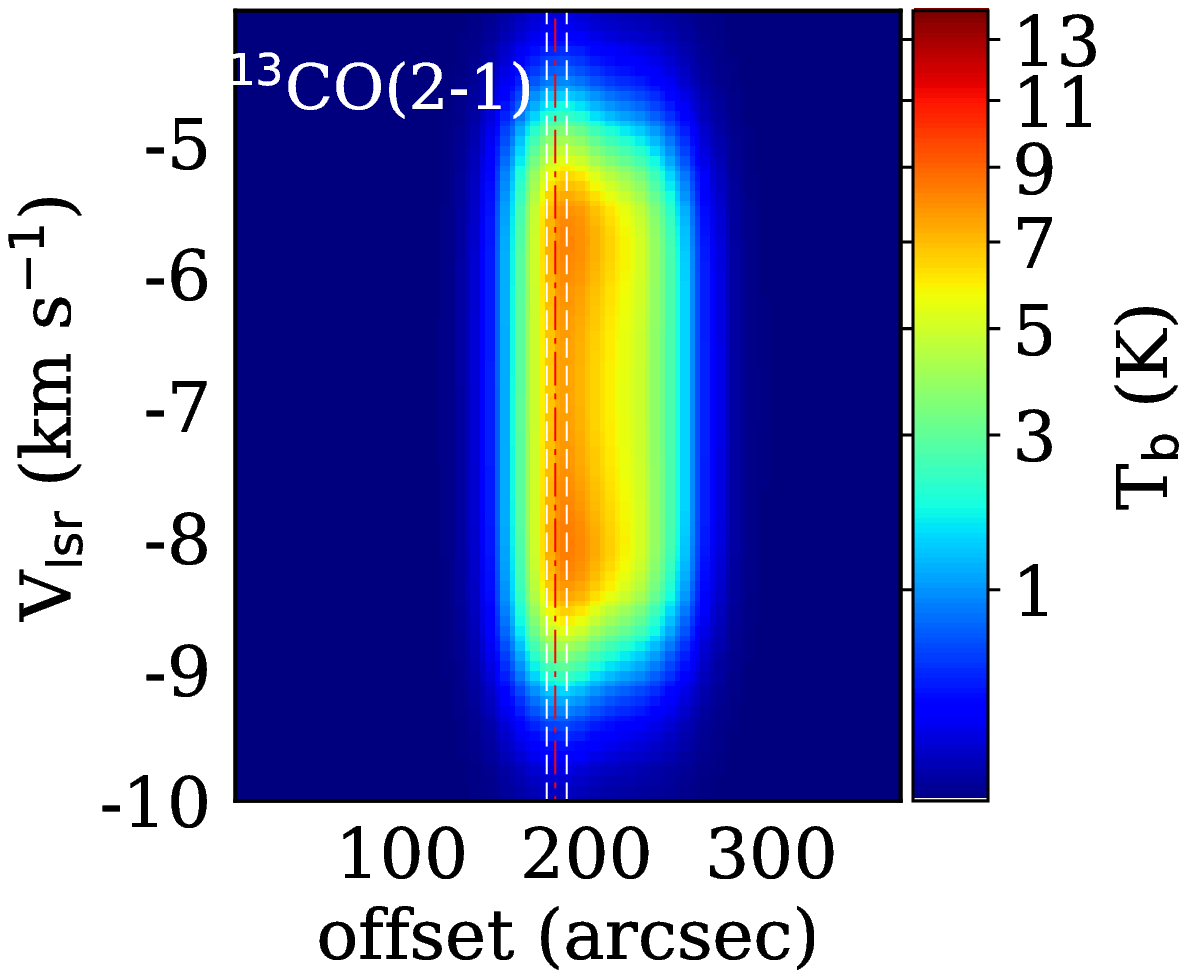}
\includegraphics[width=0.41\columnwidth]{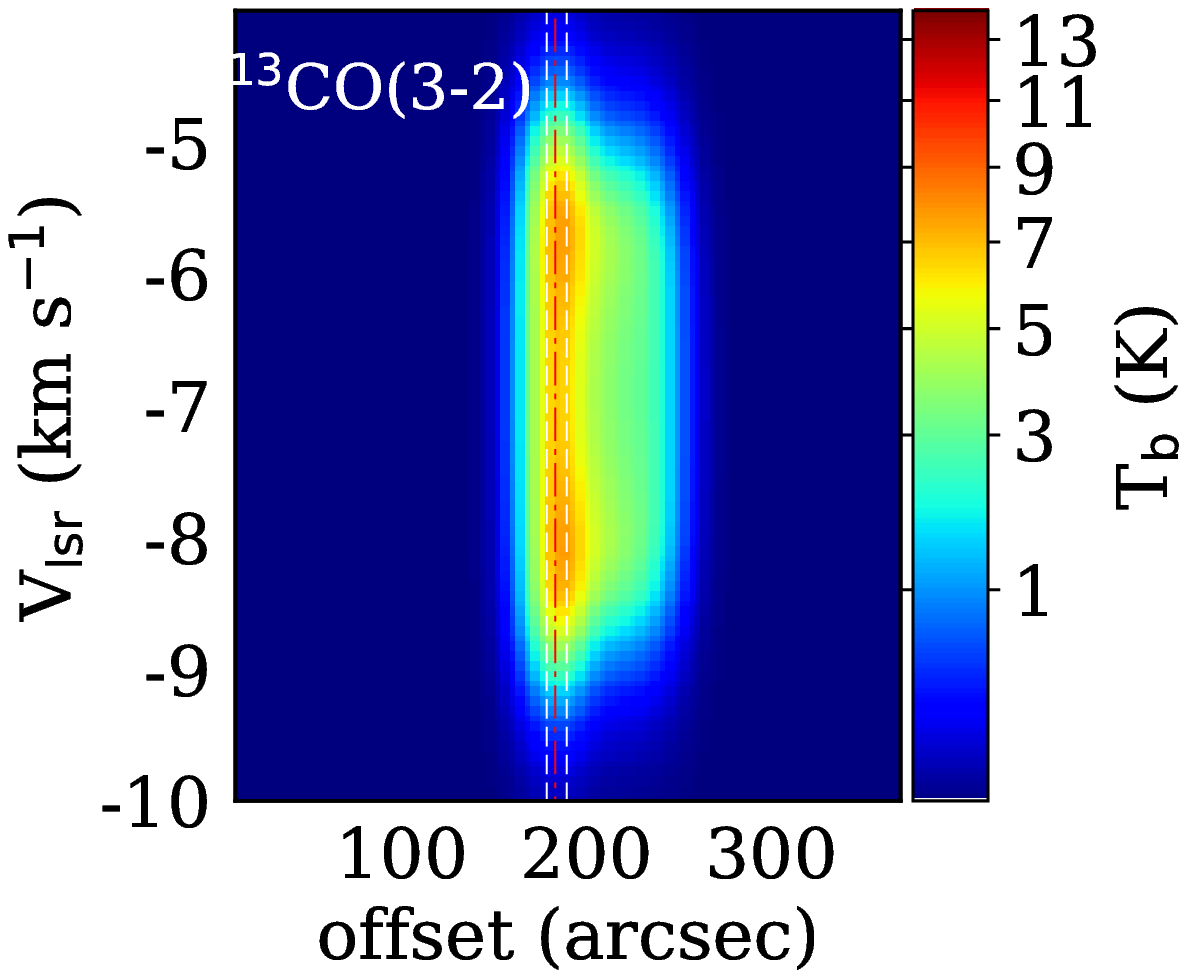}
\includegraphics[width=0.41\columnwidth]{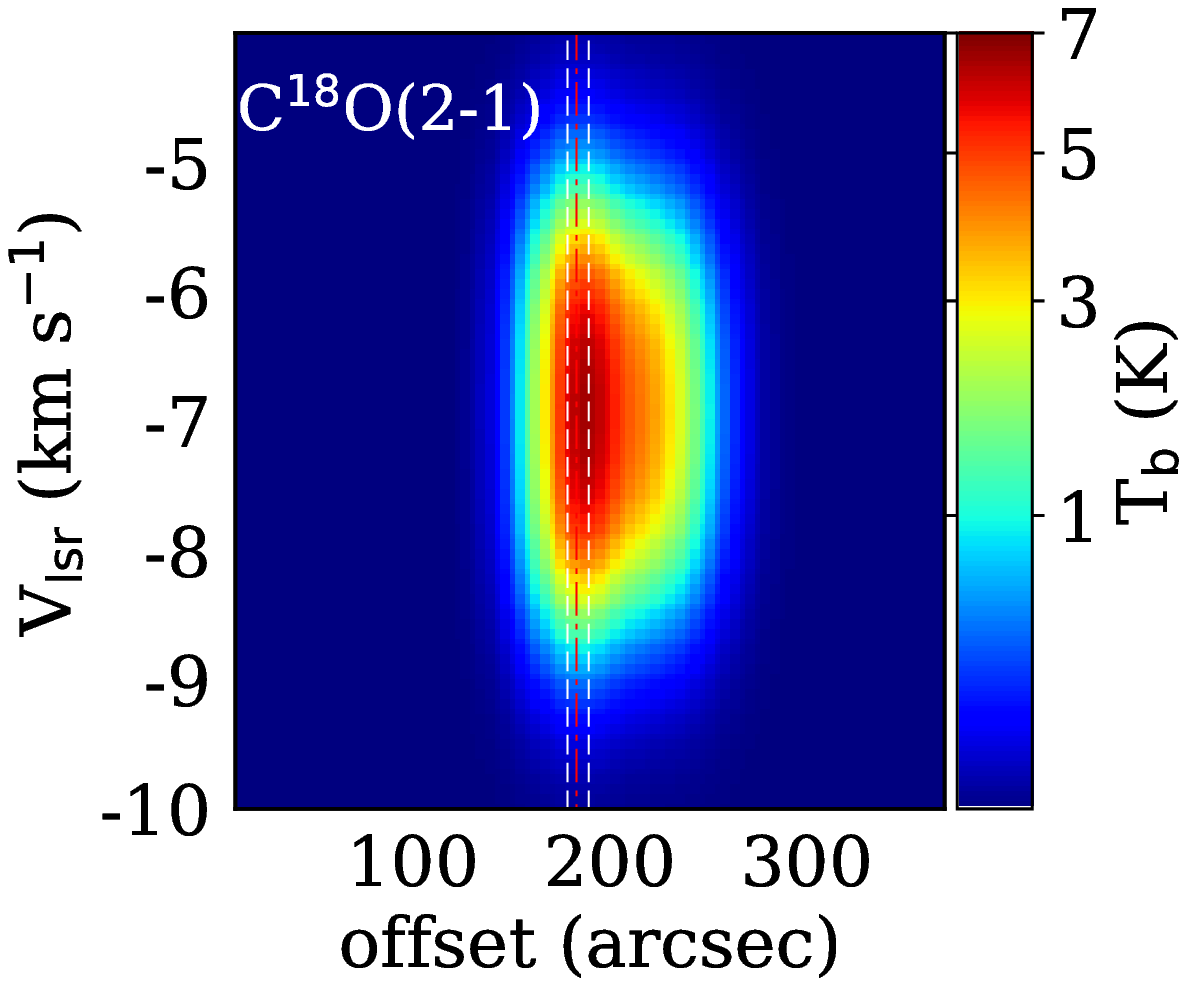}
\includegraphics[width=0.41\columnwidth]{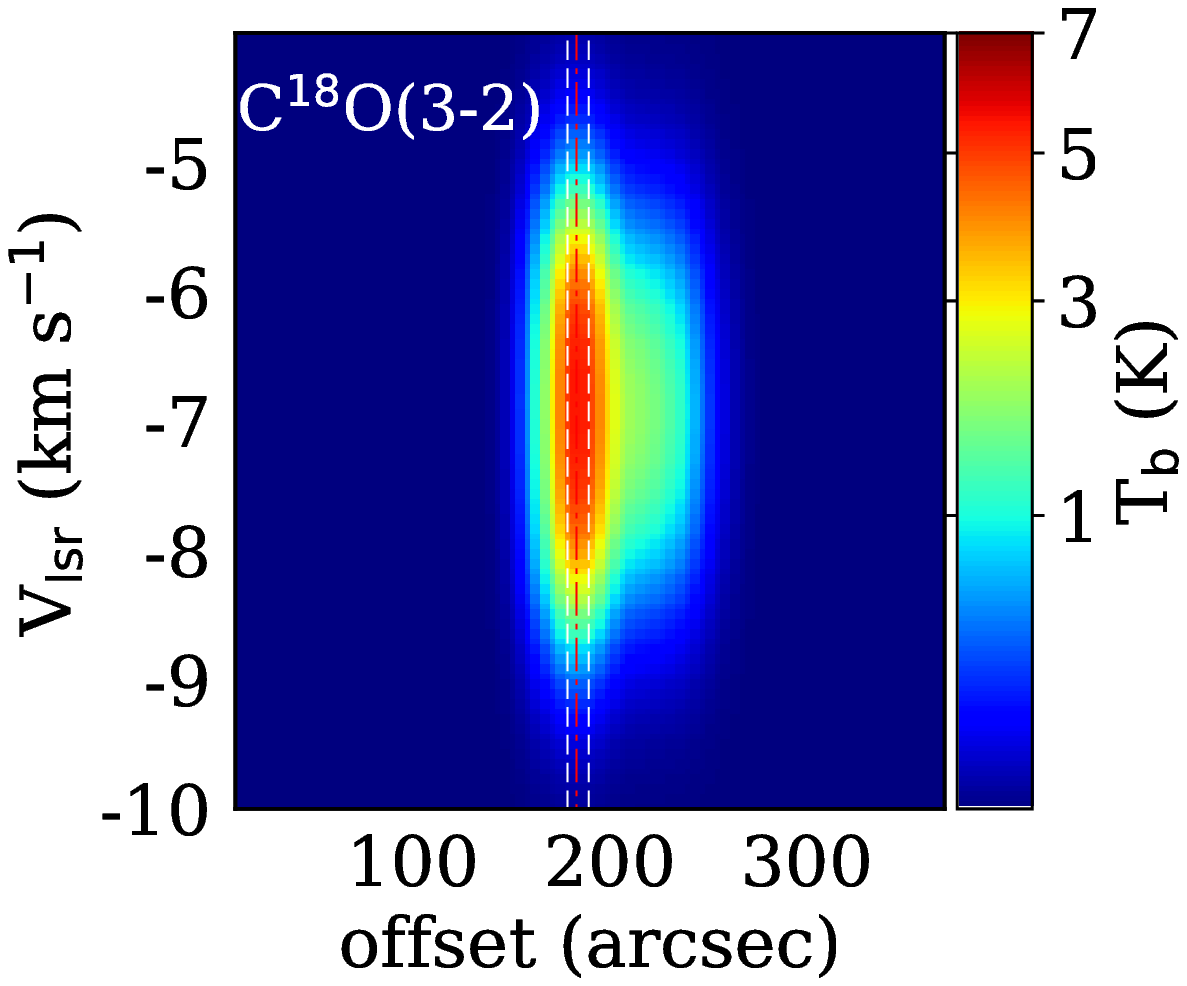}
\caption{PV diagrams for the torus envelope around the \hii{} region. The direction to the observer is shown as a yellow arrow in the left panel. Dashed vertical lines show location of the dense compressed shell. Red line shows location of the density peak. Microturbulent velocity $V_{\rm nth} = 0.3$~\kms{} for the top panel and 1~\kms{} for the bottom panel. Distributions of $n_{\rm gas}$ in the physical representations of the molecular envelope is shown in the left column.}
\label{fig:pv2d}
\end{figure*}

\section{Gas as a proxy of dust and vice versa}

So far, our results are in line with the references mentioned in Sec.~\ref{intro}, even though we have not yet constructed a real physical model of the RCW~120 and its surroundings. Anyway, RCW~120 is definitely not a classical spherical expanding \hii{} region surrounded by dense molecular gas which was collected and compressed by a shock wave preceding the ionization front as was theoretically analysed by \citet{Spitzer_1978} and \citet{Elmegreen_1977} (in the context of triggered star formation), see also \citet{Hosokawa_2006, Raga_2012, Bisbas_2015}.

\citet{Beaumont_2010} found that the thickness of the dense shells around infrared {\it Spitzer} bubbles is $\leq 20$\% of their radius. Our torus model corresponds to the thickness of the molecular shell to be 16\% of the radius of the \hii{} region (Fig.~\ref{fig:modelcalc}) in the plane of the sky, but its extent along the line of sight varies from 50\% to 90\% of the radius, depending on the impact parameter (Fig.~\ref{fig:pv2d}). While the PV~diagrams for RCW~120 indeed suggest that this object is an example of an \hii{} region in a flattened molecular cloud, the extended surrounding cloud, which we need to introduce into the model and is seen in the dust extinction map, is actually quite extended in the direction perpendicular to the plane of the sky. 
We also do not attempt to make any conclusions regarding the shape of the torus structure and the diffuse cloud from Fig.~\ref{fig:av}, because they both have a clumpy structure.

It is interesting to note how the clumpy molecular shell around the \hii{} region transforms into a more or less continuous ring with rising optical depth of the dust and gas emission. Using equations from Sec.~\ref{lte:Eq}, we find $\tau_{\rm ^{13}CO} \approx 2-7$ in the molecular envelope of RCW~120. The corresponding optical depth for the rare isotopologue $\tau_{\rm C^{18}O}$ is $\approx 0.3-1$. The molecular hydrogen column density calculated in Sec.~\ref{lte:Eq} agrees within a factor of a few with the calculations made by \citet{anderson_12} using {\it Herschel} data and a canonical dust-to-gas mass ratio of 1:100. Thus, the dust is well mixed with the gas in the molecular envelope of RCW~120. Simulations of \citet{Akimkin_2015, Akimkin_2017} show how charged dust can be expelled from an \hii{} region by radiation pressure, and estimate that the dust-to-gas mass ratio for the interior of an \hii{} region like RCW~120 can be less than 50\% of the canonical value. However, the drifting dust is stopped inside the collected dense molecular envelope of the \hii{} region due to high gas density, and dust-to-gas mass ratio there remains near the canonical of 1:100.

We find agreement between the gas column densities found by CO emission and by dust emission even along positions with $\phi{} \approx 0^\circ$ and $\phi{} \approx 270^\circ$, where \citet{anderson_12} find discontinuities in the neutral envelope of RCW~120. \citet{Anderson_2010} found stellar radiation leaking out the shell through these discontinuities, and dust heating outside the PDR. We note also that because the \co{} and \cvo{} lines have more negative radial velocities along these directions, compared to other regions of the envelope (see Fig.~\ref{fig:channel}), the radiation might leak into the line of sight of the observer in these discontinuities.

On the other hand, the bright \co(2--1) integrated intensity does not completely coincide with the bright ATLASGAL~870~\mkm{} emission. In particular, there is bright \co(2--1) towards the south and east discontinuities in the neutral envelope of the \hii{} region. Comparing the gas emission with dust emission in different bands observed by the {\it Spitzer} and {\it Herschel} telescopes, we conclude that the neutral envelope of the \hii{} region looks different in optically thin and optically thick regimes. The \co(2--1) line emission spatially coincides with the 8~\mkm{} band emission of polycyclic aromatic hydrocarbons (PAHs) around the \hii{} region, as shown in Fig.~\ref{fig:opticaldepth}. These tracers of gas and dust have significant optical depth at the south and east discontinuities ($\tau_{\rm ^{13}CO} \approx 3-4$, $\tau_{\rm 8\mu {\rm m}} \approx 1-3$; Ya. Pavlyuchenkov et al., in prep.; see also \citet{Robitaille_2012}). The distribution of optically thin dust emission in the far infrared (160-1100~\mkm{}) agrees with the \cvo(2--1){} line emission, which is also optically thin. The optically thin tracers can reveal discontinuities in the neutral envelope, where strong ultraviolet radiation leaks the ionized gas volume. Therefore, gas is a good proxy of dust, and vice versa, if we consider these tracers in the same (either thin or thick) optical regimes.

Analysing a large set of objects from a catalogue of IR ring-like nebulae around \hii{} regions embedded into molecular clouds \citep{Topchieva_2017}, and also large sample of {\it AKARI} data of IR bubbles \citep{2019PASJ...71....6H}, we conclude that the images of the \hii{} regions at 8~\mkm{} appear as quite uniform bubbles due to the significant optical depths, while their shape at longer wavelengths represents a number of clumps around massive stars. \citet{Topchieva_2018ion} constructed an evolutionary sequences of \hii{} regions and checked how fluxes at 8, 24, and 160~\mkm{} change during their expansion. While they found that the electron number density decreases along with increasing size of the \hii{} regions in agreement with theoretical expectations, there was not any particular trend for the IR fluxes. They proposed that an absence of continuous growth in the masses of the neutral envelopes around the \hii{} regions with time is related to non-uniform gas distribution in the parental molecular clouds, followed by leaks of the ionizing photons through the discontinuities. Recently, \citet{Topchieva_2019} showed using a sample of 32 embedded \hii{} regions, that the morphology of the PDRs around them is likely 3D (spherical) rather than 2D (plane-like). The scenario proposed agrees with 3D simulations of expanding \hii{} regions in turbulent magnetized media \citep[e.g. by][]{Arthur_2011} or hydrodynamical simulations by \citet{Walch_2015}, for \hii{} regions in fractal molecular clouds. Recent analysis of the {\it Herschel} images of RCW~120 by \citet{Marsh_2019} demonstrates significant differences in the spatial distributions of warm and cold dust. The former is localized near the ionizing star in a confined volume, but cold dust is distributed in a clumpy manner in a broad area around the \hii{} region.

We conclude the paper with an idea that due to different optical depth of dust emission the bubbling galactic disk found by {\it Spitzer} becomes a disk full of clumps organized around young massive stars on the {\it Herschel} and APEX~ATLASGAL images.

\begin{figure*}
\includegraphics[width=\columnwidth]{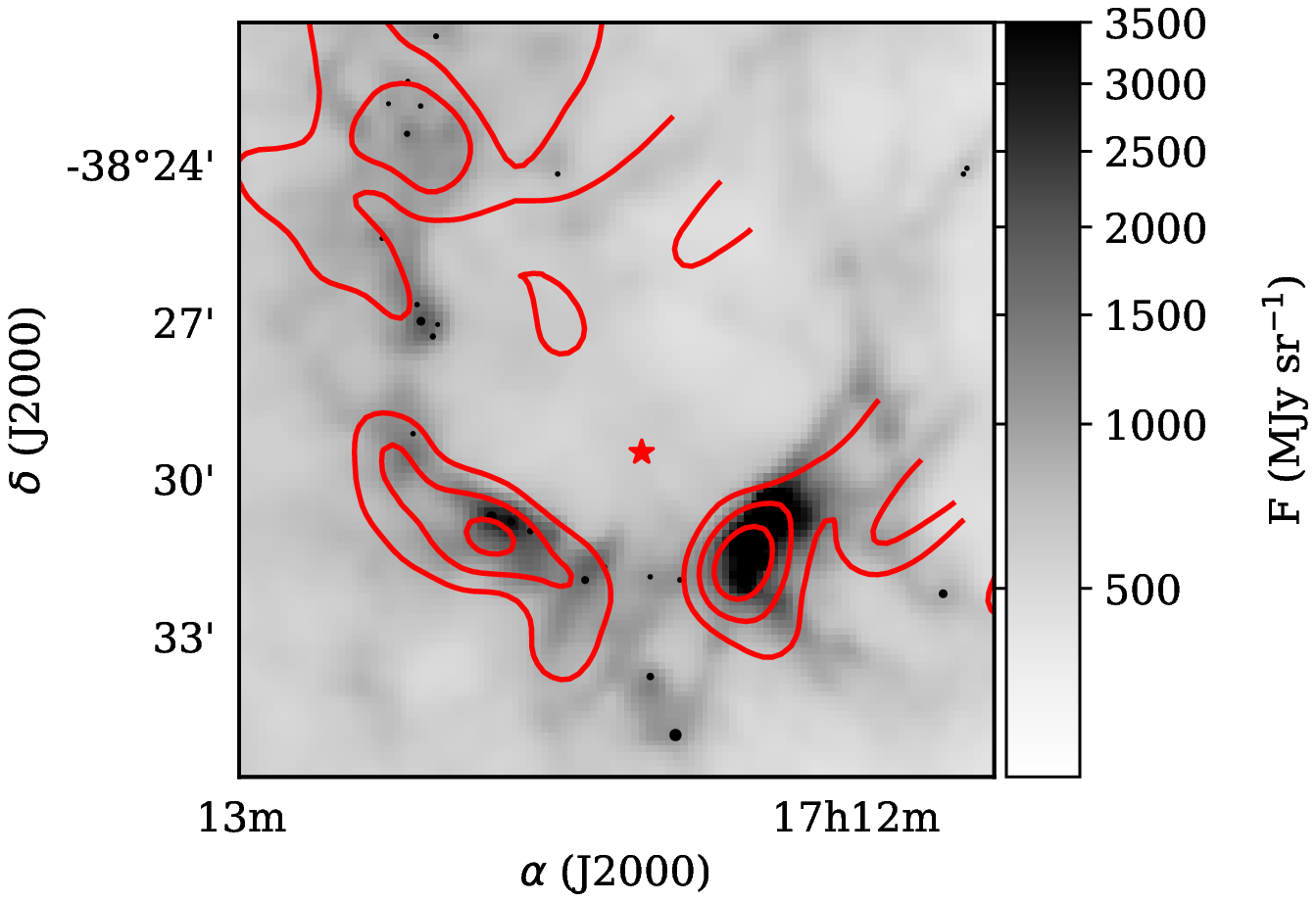}
\includegraphics[width=\columnwidth]{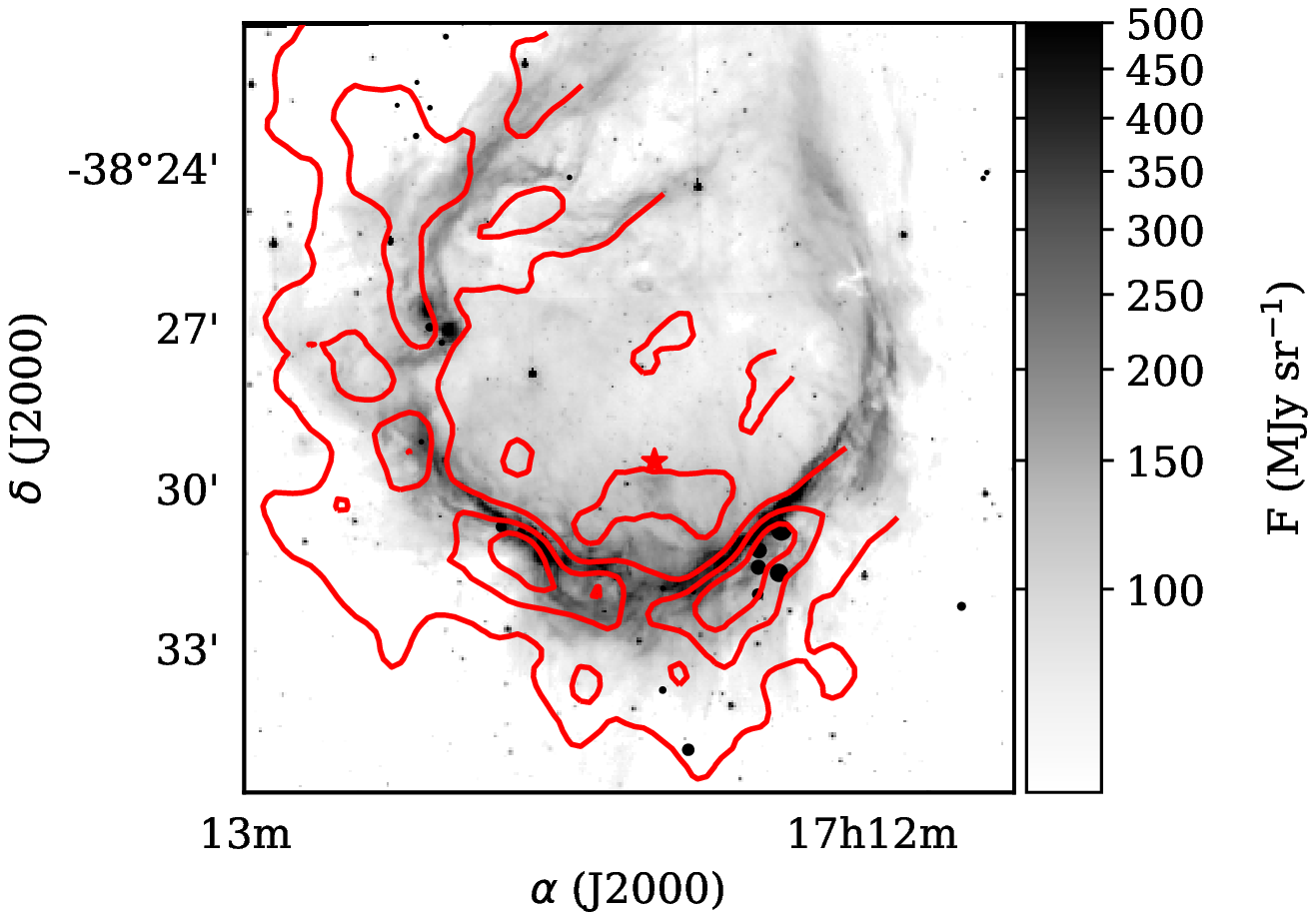}
\caption{Left: \co(2--1) integrated intensity (red contours) overlaid on {\it Spitzer} image at 8~\mkm{}. The contours are for 4.5, 20.0 and 30.0 K~\kms{}. Black circles show locations of YSOs described by \citet{Figueira_2017}. The position of the ionizing star of RCW~120 is marked as a red star. Right: \cvo(2--1) integrated intensity (red contours) overlaid on {\it Herschel} image at 350~\mkm{}. The contours corresponds to 4.5, 7.0 and 9.5 K~\kms{}.}
\label{fig:opticaldepth}
\end{figure*}

\section{Conclusions}

\begin{itemize}
\item In this work, we presented observations of RCW~120 in lines of CO isotopologues, and analysed these observations (along with archival data) to put constraints on the geometry of RCW~120. Using results of numerical modelling and radiative transfer simulations, we show that the integrated intensities of \co(2--1) and \cvo(2--1) lines in the \hii{} region and its environment cannot be reproduced within the framework of a spherical shell model. However, the observational data are consistent with a model of a toroidal dense molecular shell around RCW~120 seen face-on.

\item The \co(2--1) and \co(3--2) lines show deep self-absorption, which cannot be explained even by the torus model, and require additional absorption in foreground cold gas. Extinction maps at 2~\mkm{}, and also infrared images at longer wavelengths, show that RCW~120 is a part of a large-scale absorbing cloud 11.3~pc in size in the plane of the sky. An extended foreground cloud with a gas number density of about 50~cm$^{-3}$ can fit the self-absorption profiles of the \co{} lines.

\item There is a spatial coincidence of optically thick gas and dust tracers, such as \co(2--1) line emission and 8~\mkm{} emission of PAHs. The neutral envelope of RCW~120 looks almost continuous in these tracers. The same is true for a pair of optically thin dust emission in the far infrared (160--1100~\mkm{}) and with \cvo(2--1) line emission. The optically thin emission demonstrates discontinuities in the neutral envelope, where strong ultraviolet radiation leaks the ionized gas volume. 

\end{itemize}

\section*{Acknowledgements}

We are thankful to D.~A.~Kovaleva, V.~V.~Akimkin for fruitful discussions of RCW~120 and also to staff of Onsala Space Observatory for their care for technical details of the O-083.F-9311A-2009 project.

M.~S.~Kirsanova and Ya.~N.~Pavlyuchenkov were partly funded by the Russian Foundation for Basic Research, research project 18-32-20049. 

S.~V.~Salii and A.~M.~Sobolev work was supported in part by the Ministry of Education and Science (the basic part of the State assignment, RK no. AAAA-A17-117030310283-7) and by the Act no. 211 of the Government of the Russian Federation, agreement no. 02.A03.21.0006.

This research is based on observations with the Atacama Pathfinder EXperiment (APEX) telescope. APEX is a collaboration between the Max Planck Institute for Radio Astronomy, the European Southern Observatory, and the Onsala Space Observatory.

The ATLASGAL project is a collaboration between the Max-Planck-Gesellschaft, the European Southern Observatory (ESO) and the Universidad de Chile. It includes projects E-181.C-0885, E-078.F-9040(A), M-079.C-9501(A), M-081.C-9501(A) plus Chilean data.

This research has made use of NASA's Astrophysics Data System Bibliographic Services; SIMBAD database, operated at CDS, Strasbourg, France~\citep{Wenger_2000};  Aladin web page~\citep{2000A&AS..143...33B}; Astropy, a community-developed core Python package for Astronomy~\citep{Astropy_2013}; APLpy, an open-source plotting package for Python \citep[(http://aplpy.github.com)][]{APLpy_2012}.

\bibliographystyle{mnras}
\bibliography{expansion} 

\begin{thebibliography}{}
\makeatletter
\relax
\def\mn@urlcharsother{\let\do\@makeother \do\$\do\&\do\#\do\^\do\_\do\%\do\~}
\def\mn@doi{\begingroup\mn@urlcharsother \@ifnextchar [ {\mn@doi@}
  {\mn@doi@[]}}
\def\mn@doi@[#1]#2{\def\@tempa{#1}\ifx\@tempa\@empty \href
  {http://dx.doi.org/#2} {doi:#2}\else \href {http://dx.doi.org/#2} {#1}\fi
  \endgroup}
\def\mn@eprint#1#2{\mn@eprint@#1:#2::\@nil}
\def\mn@eprint@arXiv#1{\href {http://arxiv.org/abs/#1} {{\tt arXiv:#1}}}
\def\mn@eprint@dblp#1{\href {http://dblp.uni-trier.de/rec/bibtex/#1.xml}
  {dblp:#1}}
\def\mn@eprint@#1:#2:#3:#4\@nil{\def\@tempa {#1}\def\@tempb {#2}\def\@tempc
  {#3}\ifx \@tempc \@empty \let \@tempc \@tempb \let \@tempb \@tempa \fi \ifx
  \@tempb \@empty \def\@tempb {arXiv}\fi \@ifundefined
  {mn@eprint@\@tempb}{\@tempb:\@tempc}{\expandafter \expandafter \csname
  mn@eprint@\@tempb\endcsname \expandafter{\@tempc}}}

\bibitem[\protect\citeauthoryear{{Akimkin}, {Kirsanova}, {Pavlyuchenkov}  \&
  {Wiebe}}{{Akimkin} et~al.}{2015}]{Akimkin_2015}
{Akimkin} V.~V.,  {Kirsanova} M.~S.,  {Pavlyuchenkov} Y.~N.,   {Wiebe} D.~S.,
  2015, \mn@doi [\mnras] {10.1093/mnras/stv187}, \href
  {http://esoads.eso.org/abs/2015MNRAS.449..440A} {449, 440}

\bibitem[\protect\citeauthoryear{{Akimkin}, {Kirsanova}, {Pavlyuchenkov}  \&
  {Wiebe}}{{Akimkin} et~al.}{2017}]{Akimkin_2017}
{Akimkin} V.~V.,  {Kirsanova} M.~S.,  {Pavlyuchenkov} Y.~N.,   {Wiebe} D.~S.,
  2017, \mn@doi [\mnras] {10.1093/mnras/stx797}, \href
  {http://esoads.eso.org/abs/2017MNRAS.469..630A} {469, 630}

\bibitem[\protect\citeauthoryear{{Anderson} et~al.,}{{Anderson}
  et~al.}{2010}]{Anderson_2010}
{Anderson} L.~D.,  et~al., 2010, \mn@doi [\aap] {10.1051/0004-6361/201014657},
  \href {http://adsabs.harvard.edu/abs/2010A%26A...518L..99A} {518, L99}

\bibitem[\protect\citeauthoryear{{Anderson}, {Bania}, {Balser}  \&
  {Rood}}{{Anderson} et~al.}{2011}]{Anderson_2011}
{Anderson} L.~D.,  {Bania} T.~M.,  {Balser} D.~S.,   {Rood} R.~T.,  2011,
  \mn@doi [\apjs] {10.1088/0067-0049/194/2/32}, \href
  {http://adsabs.harvard.edu/abs/2011ApJS..194...32A} {194, 32}

\bibitem[\protect\citeauthoryear{{Anderson}, {Zavagno}, {Barlow},
  {Garc{\'{\i}}a-Lario}  \& {Noriega-Crespo}}{{Anderson}
  et~al.}{2012}]{anderson_12}
{Anderson} L.~D.,  {Zavagno} A.,  {Barlow} M.~J.,  {Garc{\'{\i}}a-Lario} P.,
  {Noriega-Crespo} A.,  2012, \mn@doi [\aap] {10.1051/0004-6361/201117640},
  \href {http://esoads.eso.org/abs/2012A%26A...537A...1A} {537, A1}

\bibitem[\protect\citeauthoryear{{Anderson}, {Bania}, {Balser}, {Cunningham},
  {Wenger}, {Johnstone}  \& {Armentrout}}{{Anderson}
  et~al.}{2014}]{Anderson_2014}
{Anderson} L.~D.,  {Bania} T.~M.,  {Balser} D.~S.,  {Cunningham} V.,  {Wenger}
  T.~V.,  {Johnstone} B.~M.,   {Armentrout} W.~P.,  2014, \mn@doi [\apjs]
  {10.1088/0067-0049/212/1/1}, \href
  {http://adsabs.harvard.edu/abs/2014ApJS..212....1A} {212, 1}

\bibitem[\protect\citeauthoryear{{Anderson} et~al.,}{{Anderson}
  et~al.}{2015}]{Anderson_2015}
{Anderson} L.~D.,  et~al., 2015, \mn@doi [\apj] {10.1088/0004-637X/800/2/101},
  \href {http://esoads.eso.org/abs/2015ApJ...800..101A} {800, 101}

\bibitem[\protect\citeauthoryear{{Arthur}, {Henney}, {Mellema}, {de Colle}  \&
  {V{\'a}zquez-Semadeni}}{{Arthur} et~al.}{2011}]{Arthur_2011}
{Arthur} S.~J.,  {Henney} W.~J.,  {Mellema} G.,  {de Colle} F.,
  {V{\'a}zquez-Semadeni} E.,  2011, \mn@doi [\mnras]
  {10.1111/j.1365-2966.2011.18507.x}, \href
  {http://adsabs.harvard.edu/abs/2011MNRAS.414.1747A} {414, 1747}

\bibitem[\protect\citeauthoryear{{Astropy Collaboration} et~al.,}{{Astropy
  Collaboration} et~al.}{2013}]{Astropy_2013}
{Astropy Collaboration} et~al., 2013, \mn@doi [\aap]
  {10.1051/0004-6361/201322068}, \href
  {http://adsabs.harvard.edu/abs/2013A%26A...558A..33A} {558, A33}

\bibitem[\protect\citeauthoryear{{Beaumont} \& {Williams}}{{Beaumont} \&
  {Williams}}{2010}]{Beaumont_2010}
{Beaumont} C.~N.,  {Williams} J.~P.,  2010, \mn@doi [\apj]
  {10.1088/0004-637X/709/2/791}, \href
  {http://adsabs.harvard.edu/abs/2010ApJ...709..791B} {709, 791}

\bibitem[\protect\citeauthoryear{{Belitsky} et~al.,}{{Belitsky}
  et~al.}{2006}]{2006SPIE.6275E..0GB}
{Belitsky} V.,  et~al., 2006, in Society of Photo-Optical Instrumentation
  Engineers (SPIE) Conference Series. p. 62750G, \mn@doi{10.1117/12.671383}

\bibitem[\protect\citeauthoryear{{Bisbas} et~al.,}{{Bisbas}
  et~al.}{2015}]{Bisbas_2015}
{Bisbas} T.~G.,  et~al., 2015, \mn@doi [\mnras] {10.1093/mnras/stv1659}, \href
  {http://adsabs.harvard.edu/abs/2015MNRAS.453.1324B} {453, 1324}

\bibitem[\protect\citeauthoryear{{Bock}, {Large}  \& {Sadler}}{{Bock}
  et~al.}{1999}]{Bock_1999}
{Bock} D.~C.-J.,  {Large} M.~I.,   {Sadler} E.~M.,  1999, \mn@doi [\aj]
  {10.1086/300786}, \href {http://adsabs.harvard.edu/abs/1999AJ....117.1578B}
  {117, 1578}

\bibitem[\protect\citeauthoryear{{Bonnarel} et~al.,}{{Bonnarel}
  et~al.}{2000}]{2000A&AS..143...33B}
{Bonnarel} F.,  et~al., 2000, \mn@doi [\aaps] {10.1051/aas:2000331}, \href
  {https://ui.adsabs.harvard.edu/abs/2000A&AS..143...33B} {143, 33}

\bibitem[\protect\citeauthoryear{{Cardelli}, {Clayton}  \& {Mathis}}{{Cardelli}
  et~al.}{1989}]{Cardelli_1989}
{Cardelli} J.~A.,  {Clayton} G.~C.,   {Mathis} J.~S.,  1989, \mn@doi [\apj]
  {10.1086/167900}, \href
  {https://ui.adsabs.harvard.edu/\#abs/1989ApJ...345..245C} {345, 245}

\bibitem[\protect\citeauthoryear{{Churchwell} et~al.,}{{Churchwell}
  et~al.}{2006}]{Churchwell_2006}
{Churchwell} E.,  et~al., 2006, \mn@doi [\apj] {10.1086/507015}, \href
  {http://adsabs.harvard.edu/abs/2006ApJ...649..759C} {649, 759}

\bibitem[\protect\citeauthoryear{{Churchwell} et~al.,}{{Churchwell}
  et~al.}{2007}]{Churchwell_2007}
{Churchwell} E.,  et~al., 2007, \mn@doi [\apj] {10.1086/521646}, \href
  {http://adsabs.harvard.edu/abs/2007ApJ...670..428C} {670, 428}

\bibitem[\protect\citeauthoryear{{Condon}, {Cotton}, {Greisen}, {Yin},
  {Perley}, {Taylor}  \& {Broderick}}{{Condon} et~al.}{1998}]{Condon_1998}
{Condon} J.~J.,  {Cotton} W.~D.,  {Greisen} E.~W.,  {Yin} Q.~F.,  {Perley}
  R.~A.,  {Taylor} G.~B.,   {Broderick} J.~J.,  1998, \mn@doi [\aj]
  {10.1086/300337}, \href {http://esoads.eso.org/abs/1998AJ....115.1693C} {115,
  1693}

\bibitem[\protect\citeauthoryear{{Deharveng}, {Zavagno}  \&
  {Caplan}}{{Deharveng} et~al.}{2005}]{Deharveng_2005}
{Deharveng} L.,  {Zavagno} A.,   {Caplan} J.,  2005, \mn@doi [\aap]
  {10.1051/0004-6361:20041946}, \href
  {http://adsabs.harvard.edu/abs/2005A%26A...433..565D} {433, 565}

\bibitem[\protect\citeauthoryear{{Deharveng}, {Zavagno}, {Schuller}, {Caplan},
  {Pomar{\`e}s}  \& {De Breuck}}{{Deharveng} et~al.}{2009}]{Deharveng_09}
{Deharveng} L.,  {Zavagno} A.,  {Schuller} F.,  {Caplan} J.,  {Pomar{\`e}s} M.,
    {De Breuck} C.,  2009, \mn@doi [\aap] {10.1051/0004-6361/200811337}, \href
  {http://adsabs.harvard.edu/abs/2009A%26A...496..177D} {496, 177}

\bibitem[\protect\citeauthoryear{{Deharveng} et~al.,}{{Deharveng}
  et~al.}{2010}]{2010A&A...523A...6D}
{Deharveng} L.,  et~al., 2010, \mn@doi [\aap] {10.1051/0004-6361/201014422},
  \href {http://adsabs.harvard.edu/abs/2010A%26A...523A...6D} {523, A6}

\bibitem[\protect\citeauthoryear{{Elmegreen} \& {Lada}}{{Elmegreen} \&
  {Lada}}{1977}]{Elmegreen_1977}
{Elmegreen} B.~G.,  {Lada} C.~J.,  1977, \mn@doi [\apj] {10.1086/155302}, \href
  {http://adsabs.harvard.edu/abs/1977ApJ...214..725E} {214, 725}

\bibitem[\protect\citeauthoryear{{Everett} \& {Churchwell}}{{Everett} \&
  {Churchwell}}{2010}]{2010ApJ...713..592E}
{Everett} J.~E.,  {Churchwell} E.,  2010, \mn@doi [\apj]
  {10.1088/0004-637X/713/1/592}, \href
  {http://adsabs.harvard.edu/abs/2010ApJ...713..592E} {713, 592}

\bibitem[\protect\citeauthoryear{{Figueira} et~al.,}{{Figueira}
  et~al.}{2017}]{Figueira_2017}
{Figueira} M.,  et~al., 2017, \mn@doi [\aap] {10.1051/0004-6361/201629379},
  \href {http://esoads.eso.org/abs/2017A%26A...600A..93F} {600, A93}

\bibitem[\protect\citeauthoryear{{Figueira}, {Bronfman}, {Zavagno}, {Louvet},
  {Lo}, {Finger}  \& {Rod{\'o}n}}{{Figueira} et~al.}{2018}]{Figueira_2018}
{Figueira} M.,  {Bronfman} L.,  {Zavagno} A.,  {Louvet} F.,  {Lo} N.,  {Finger}
  R.,   {Rod{\'o}n} J.,  2018, \mn@doi [\aap] {10.1051/0004-6361/201832930},
  \href {http://adsabs.harvard.edu/abs/2018A%26A...616L..10F} {616, L10}

\bibitem[\protect\citeauthoryear{{Gaia Collaboration}}{{Gaia
  Collaboration}}{2018}]{2018yCat.1345....0G}
{Gaia Collaboration} 2018, VizieR Online Data Catalog, \href
  {https://ui.adsabs.harvard.edu/#abs/2018yCat.1345....0G} {p. I/345}

\bibitem[\protect\citeauthoryear{{Goldsmith} \& {Langer}}{{Goldsmith} \&
  {Langer}}{1999}]{Goldsmith_1999}
{Goldsmith} P.~F.,  {Langer} W.~D.,  1999, \mn@doi [\apj] {10.1086/307195},
  \href {http://adsabs.harvard.edu/abs/1999ApJ...517..209G} {517, 209}

\bibitem[\protect\citeauthoryear{{Hanaoka} et~al.,}{{Hanaoka}
  et~al.}{2019}]{2019PASJ...71....6H}
{Hanaoka} M.,  et~al., 2019, \mn@doi [\pasj] {10.1093/pasj/psy126}, \href
  {http://adsabs.harvard.edu/abs/2019PASJ...71....6H} {71, 6}

\bibitem[\protect\citeauthoryear{{Hosokawa} \& {Inutsuka}}{{Hosokawa} \&
  {Inutsuka}}{2006}]{Hosokawa_2006}
{Hosokawa} T.,  {Inutsuka} S.-i.,  2006, \mn@doi [\apj] {10.1086/504789}, \href
  {http://adsabs.harvard.edu/abs/2006ApJ...646..240H} {646, 240}

\bibitem[\protect\citeauthoryear{{Juvela} \& {Montillaud}}{{Juvela} \&
  {Montillaud}}{2016}]{Juvela_2016}
{Juvela} M.,  {Montillaud} J.,  2016, \mn@doi [\aap]
  {10.1051/0004-6361/201425112}, \href
  {http://adsabs.harvard.edu/abs/2016A%26A...585A..38J} {585, A38}

\bibitem[\protect\citeauthoryear{{Kirsanova}, {Wiebe}  \&
  {Sobolev}}{{Kirsanova} et~al.}{2009}]{Kirsanova_2009}
{Kirsanova} M.~S.,  {Wiebe} D.~S.,   {Sobolev} A.~M.,  2009, \mn@doi [\arep]
  {10.1134/S106377290907004X}, \href
  {http://adsabs.harvard.edu/abs/2009ARep...53..611K} {53, 611}

\bibitem[\protect\citeauthoryear{{Leger} \& {Puget}}{{Leger} \&
  {Puget}}{1984}]{Leger_1984}
{Leger} A.,  {Puget} J.~L.,  1984, \aap, \href
  {http://adsabs.harvard.edu/abs/1984A%26A...137L...5L} {137, L5}

\bibitem[\protect\citeauthoryear{{Mackey}, {Haworth}, {Gvaramadze}, {Mohamed},
  {Langer}  \& {Harries}}{{Mackey} et~al.}{2016}]{Mackey_2016}
{Mackey} J.,  {Haworth} T.~J.,  {Gvaramadze} V.~V.,  {Mohamed} S.,  {Langer}
  N.,   {Harries} T.~J.,  2016, \mn@doi [\aap] {10.1051/0004-6361/201527569},
  \href {http://adsabs.harvard.edu/abs/2016A%26A...586A.114M} {586, A114}

\bibitem[\protect\citeauthoryear{{Mangum} \& {Shirley}}{{Mangum} \&
  {Shirley}}{2015}]{Mangum_2015}
{Mangum} J.~G.,  {Shirley} Y.~L.,  2015, \mn@doi [\pasp] {10.1086/680323},
  \href {http://adsabs.harvard.edu/abs/2015PASP..127..266M} {127, 266}

\bibitem[\protect\citeauthoryear{{Marsh} \& {Whitworth}}{{Marsh} \&
  {Whitworth}}{2019}]{Marsh_2019}
{Marsh} K.~A.,  {Whitworth} A.~P.,  2019, \mn@doi [\mnras]
  {10.1093/mnras/sty3186}, \href
  {http://adsabs.harvard.edu/abs/2019MNRAS.483..352M} {483, 352}

\bibitem[\protect\citeauthoryear{{Martins}, {Schaerer}  \& {Hillier}}{{Martins}
  et~al.}{2005}]{Martins_2005}
{Martins} F.,  {Schaerer} D.,   {Hillier} D.~J.,  2005, \mn@doi [\aap]
  {10.1051/0004-6361:20042386}, \href
  {http://adsabs.harvard.edu/abs/2005A%26A...436.1049M} {436, 1049}

\bibitem[\protect\citeauthoryear{{Martins}, {Pomar{\`e}s}, {Deharveng},
  {Zavagno}  \& {Bouret}}{{Martins} et~al.}{2010}]{Martins_10}
{Martins} F.,  {Pomar{\`e}s} M.,  {Deharveng} L.,  {Zavagno} A.,   {Bouret}
  J.~C.,  2010, \mn@doi [\aap] {10.1051/0004-6361/200913158}, \href
  {http://adsabs.harvard.edu/abs/2010A%26A...510A..32M} {510, A32}

\bibitem[\protect\citeauthoryear{{McClure-Griffiths}, {Dickey}, {Gaensler},
  {Green}, {Haverkorn}  \& {Strasser}}{{McClure-Griffiths}
  et~al.}{2005}]{McClure-Griffiths_2005}
{McClure-Griffiths} N.~M.,  {Dickey} J.~M.,  {Gaensler} B.~M.,  {Green} A.~J.,
  {Haverkorn} M.,   {Strasser} S.,  2005, \mn@doi [\apjs] {10.1086/430114},
  \href {http://adsabs.harvard.edu/abs/2005ApJS..158..178M} {158, 178}

\bibitem[\protect\citeauthoryear{{Ochsendorf}, {Verdolini}, {Cox}, {Bern{\'e}},
  {Kaper}  \& {Tielens}}{{Ochsendorf} et~al.}{2014}]{dustwave}
{Ochsendorf} B.~B.,  {Verdolini} S.,  {Cox} N.~L.~J.,  {Bern{\'e}} O.,  {Kaper}
  L.,   {Tielens} A.~G.~G.~M.,  2014, \mn@doi [\aap]
  {10.1051/0004-6361/201423545}, \href
  {http://adsabs.harvard.edu/abs/2014A%26A...566A..75O} {566, A75}

\bibitem[\protect\citeauthoryear{{Pavlyuchenkov} \& {Shustov}}{{Pavlyuchenkov}
  \& {Shustov}}{2004}]{Pavlyuchenkov:2004}
{Pavlyuchenkov} Y.~N.,  {Shustov} B.~M.,  2004, \mn@doi [Astronomy Reports]
  {10.1134/1.1704676}, \href
  {https://ui.adsabs.harvard.edu/\#abs/2004ARep...48..315P} {48, 315}

\bibitem[\protect\citeauthoryear{{Pavlyuchenkov}, {Wiebe}, {Shustov},
  {Henning}, {Launhardt}  \& {Semenov}}{{Pavlyuchenkov}
  et~al.}{2008}]{Pavlyuchenkov:2008}
{Pavlyuchenkov} Y.,  {Wiebe} D.,  {Shustov} B.,  {Henning} T.,  {Launhardt} R.,
    {Semenov} D.,  2008, \mn@doi [\apj] {10.1086/592564}, \href
  {https://ui.adsabs.harvard.edu/\#abs/2008ApJ...689..335P} {689, 335}

\bibitem[\protect\citeauthoryear{{Pavlyuchenkov}, {Kirsanova}  \&
  {Wiebe}}{{Pavlyuchenkov} et~al.}{2013}]{Pavlyuchenkov_2013}
{Pavlyuchenkov} Y.~N.,  {Kirsanova} M.~S.,   {Wiebe} D.~S.,  2013, \mn@doi
  [\arep] {10.1134/S1063772913070056}, \href
  {http://adsabs.harvard.edu/abs/2013ARep...57..573P} {57, 573}

\bibitem[\protect\citeauthoryear{{Raga}, {Cant{\'o}}  \&
  {Rodr{\'{\i}}guez}}{{Raga} et~al.}{2012}]{Raga_2012}
{Raga} A.~C.,  {Cant{\'o}} J.,   {Rodr{\'{\i}}guez} L.~F.,  2012, \rmxaa, \href
  {http://adsabs.harvard.edu/abs/2012RMxAA..48..149R} {48, 149}

\bibitem[\protect\citeauthoryear{{Robitaille} \& {Bressert}}{{Robitaille} \&
  {Bressert}}{2012}]{APLpy_2012}
{Robitaille} T.,  {Bressert} E.,  2012, {APLpy: Astronomical Plotting Library
  in Python}, Astrophysics Source Code Library (\mn@eprint {ascl} {1208.017})

\bibitem[\protect\citeauthoryear{{Robitaille}, {Churchwell}, {Benjamin},
  {Whitney}, {Wood}, {Babler}  \& {Meade}}{{Robitaille}
  et~al.}{2012}]{Robitaille_2012}
{Robitaille} T.~P.,  {Churchwell} E.,  {Benjamin} R.~A.,  {Whitney} B.~A.,
  {Wood} K.,  {Babler} B.~L.,   {Meade} M.~R.,  2012, \mn@doi [\aap]
  {10.1051/0004-6361/201219073}, \href
  {http://adsabs.harvard.edu/abs/2012A%26A...545A..39R} {545, A39}

\bibitem[\protect\citeauthoryear{{Russeil}}{{Russeil}}{2003}]{Russeil_03}
{Russeil} D.,  2003, \mn@doi [\aap] {10.1051/0004-6361:20021504}, \href
  {http://adsabs.harvard.edu/abs/2003A%26A...397..133R} {397, 133}

\bibitem[\protect\citeauthoryear{{Samal}, {Deharveng}, {Zavagno}, {Anderson},
  {Molinari}  \& {Russeil}}{{Samal} et~al.}{2018}]{2018A&A...617A..67S}
{Samal} M.~R.,  {Deharveng} L.,  {Zavagno} A.,  {Anderson} L.~D.,  {Molinari}
  S.,   {Russeil} D.,  2018, \mn@doi [\aap] {10.1051/0004-6361/201833015},
  \href {http://adsabs.harvard.edu/abs/2018A%26A...617A..67S} {617, A67}

\bibitem[\protect\citeauthoryear{{S{\'a}nchez-Cruces},
  {Castellanos-Ram{\'{\i}}rez}, {Rosado}, {Rodr{\'{\i}}guezGonz{\'a}lez}  \&
  {Reyes-Iturbide}}{{S{\'a}nchez-Cruces} et~al.}{2018}]{Sanchez-Cruces_2018}
{S{\'a}nchez-Cruces} M.,  {Castellanos-Ram{\'{\i}}rez} A.,  {Rosado} M.,
  {Rodr{\'{\i}}guezGonz{\'a}lez} A.,   {Reyes-Iturbide} J.,  2018, \rmxaa,
  \href {http://adsabs.harvard.edu/abs/2018RMxAA..54..375S} {54, 375}

\bibitem[\protect\citeauthoryear{{Schuller} et~al.,}{{Schuller}
  et~al.}{2009}]{Schuller_2009}
{Schuller} F.,  et~al., 2009, \mn@doi [\aap] {10.1051/0004-6361/200811568},
  \href {https://ui.adsabs.harvard.edu/\#abs/2009A&A...504..415S} {504, 415}

\bibitem[\protect\citeauthoryear{{Schuller} et~al.,}{{Schuller}
  et~al.}{2017}]{Schuller_2017}
{Schuller} F.,  et~al., 2017, \mn@doi [\aap] {10.1051/0004-6361/201628933},
  \href {http://esoads.eso.org/abs/2017A%26A...601A.124S} {601, A124}

\bibitem[\protect\citeauthoryear{{Sellgren}}{{Sellgren}}{1984}]{Sellgren_1984}
{Sellgren} K.,  1984, \mn@doi [\apj] {10.1086/161733}, \href
  {http://adsabs.harvard.edu/abs/1984ApJ...277..623S} {277, 623}

\bibitem[\protect\citeauthoryear{{Simpson} et~al.,}{{Simpson}
  et~al.}{2012}]{Simpson_2012}
{Simpson} R.~J.,  et~al., 2012, \mn@doi [\mnras]
  {10.1111/j.1365-2966.2012.20770.x}, \href
  {http://adsabs.harvard.edu/abs/2012MNRAS.424.2442S} {424, 2442}

\bibitem[\protect\citeauthoryear{{Spitzer}}{{Spitzer}}{1978}]{Spitzer_1978}
{Spitzer} L.,  1978, {Physical processes in the interstellar medium}.
New York Wiley-Interscience, 1978.~333 p., \mn@doi{10.1002/9783527617722}

\bibitem[\protect\citeauthoryear{{Topchieva}, {Wiebe}, {Kirsanova}  \&
  {Krushinskii}}{{Topchieva} et~al.}{2017}]{Topchieva_2017}
{Topchieva} A.~P.,  {Wiebe} D.~S.,  {Kirsanova} M.~S.,   {Krushinskii} V.~V.,
  2017, \mn@doi [Astronomy Reports] {10.1134/S1063772917120083}, \href
  {http://adsabs.harvard.edu/abs/2017ARep...61.1015T} {61, 1015}

\bibitem[\protect\citeauthoryear{{Topchieva}, {Wiebe}  \&
  {Kirsanova}}{{Topchieva} et~al.}{2018a}]{Topchieva_2018}
{Topchieva} A.,  {Wiebe} D.,   {Kirsanova} M.~S.,  2018a, \mn@doi [Research in
  Astronomy and Astrophysics] {10.1088/1674-4527/18/8/91}, \href
  {http://adsabs.harvard.edu/abs/2018RAA....18...91T} {18, 091}

\bibitem[\protect\citeauthoryear{{Topchieva}, {Kirsanova}  \&
  {Sobolev}}{{Topchieva} et~al.}{2018b}]{Topchieva_2018ion}
{Topchieva} A.~P.,  {Kirsanova} M.~S.,   {Sobolev} A.~M.,  2018b, \mn@doi
  [Astronomy Reports] {10.1134/S1063772918110082}, \href
  {http://adsabs.harvard.edu/abs/2018ARep...62..764T} {62, 764}

\bibitem[\protect\citeauthoryear{{Topchieva}, {Akimkin}  \&
  {Smirnov-Pinchukov}}{{Topchieva} et~al.}{2019}]{Topchieva_2019}
{Topchieva} A.~P.,  {Akimkin} V.~V.,   {Smirnov-Pinchukov} G.~V.,  2019, arXiv
  e-prints, \href {https://ui.adsabs.harvard.edu/abs/2019arXiv190511077T} {p.
  arXiv:1905.11077}

\bibitem[\protect\citeauthoryear{{Tremblin} et~al.,}{{Tremblin}
  et~al.}{2014}]{Tremblin_14}
{Tremblin} P.,  et~al., 2014, \mn@doi [\aap] {10.1051/0004-6361/201322700},
  \href {http://adsabs.harvard.edu/abs/2014A%26A...564A.106T} {564, A106}

\bibitem[\protect\citeauthoryear{{Vassilev} et~al.,}{{Vassilev}
  et~al.}{2008}]{2008A&A...490.1157V}
{Vassilev} V.,  et~al., 2008, \mn@doi [\aap] {10.1051/0004-6361:200810459},
  \href {http://adsabs.harvard.edu/abs/2008A%26A...490.1157V} {490, 1157}

\bibitem[\protect\citeauthoryear{{Walch}, {Whitworth}, {Bisbas}, {Hubber}  \&
  {W{\"u}nsch}}{{Walch} et~al.}{2015}]{Walch_2015}
{Walch} S.,  {Whitworth} A.~P.,  {Bisbas} T.~G.,  {Hubber} D.~A.,
  {W{\"u}nsch} R.,  2015, \mn@doi [\mnras] {10.1093/mnras/stv1427}, \href
  {https://ui.adsabs.harvard.edu/\#abs/2015MNRAS.452.2794W} {452, 2794}

\bibitem[\protect\citeauthoryear{{Wenger} et~al.,}{{Wenger}
  et~al.}{2000}]{Wenger_2000}
{Wenger} M.,  et~al., 2000, \mn@doi [\aaps] {10.1051/aas:2000332}, \href
  {http://adsabs.harvard.edu/abs/2000A%26AS..143....9W} {143, 9}

\bibitem[\protect\citeauthoryear{{Wilson}}{{Wilson}}{1999}]{Wilson_1999}
{Wilson} T.~L.,  1999, \mn@doi [Reports on Progress in Physics]
  {10.1088/0034-4885/62/2/002}, \href
  {http://adsabs.harvard.edu/abs/1999RPPh...62..143W} {62, 143}

\bibitem[\protect\citeauthoryear{{Zavagno}, {Pomar{\`e}s}, {Deharveng},
  {Hosokawa}, {Russeil}  \& {Caplan}}{{Zavagno} et~al.}{2007}]{Zavagno_2007}
{Zavagno} A.,  {Pomar{\`e}s} M.,  {Deharveng} L.,  {Hosokawa} T.,  {Russeil}
  D.,   {Caplan} J.,  2007, \mn@doi [\aap] {10.1051/0004-6361:20077474}, \href
  {http://adsabs.harvard.edu/abs/2007A%26A...472..835Z} {472, 835}

\bibitem[\protect\citeauthoryear{{Zavagno} et~al.,}{{Zavagno}
  et~al.}{2010}]{Zavagno_2010}
{Zavagno} A.,  et~al., 2010, \mn@doi [\aap] {10.1051/0004-6361/201014623},
  \href {http://adsabs.harvard.edu/abs/2010A%26A...518L..81Z} {518, L81}

\bibitem[\protect\citeauthoryear{{van der Tak}, {Black}, {Sch{\"o}ier},
  {Jansen}  \& {van Dishoeck}}{{van der Tak} et~al.}{2007}]{radexpaper}
{van der Tak} F.~F.~S.,  {Black} J.~H.,  {Sch{\"o}ier} F.~L.,  {Jansen} D.~J.,
   {van Dishoeck} E.~F.,  2007, \aap, 468, 627

\makeatother
\end{thebibliography}

\bsp	
\label{lastpage}
\end{document}